\documentclass[twocolumn,superscriptaddress,amsfont,amssymb,amsmath, showpacs,balancelastpage,nofootinbib,aps,prd,10pt]{revtex4-2}
\usepackage{graphicx,longtable,mathrsfs,color,array}
\usepackage[unicode=true,pdfusetitle,
bookmarks=true,bookmarksnumbered=true,bookmarksopen=true,bookmarksopenlevel=1,
breaklinks=false,pdfborder={0 0 0},backref=false,colorlinks=true]{hyperref}
\hypersetup{citecolor=blue,filecolor=blue,linkcolor=blue,urlcolor=blue,pdfauthor={Name}
}
\usepackage[dvipsnames]{xcolor}
\usepackage{amssymb,amsmath,mathtools,mathrsfs,enumitem}
\usepackage{epsfig,subfigure,placeins,float}
\usepackage{booktabs,longtable,multirow}
\usepackage{exscale,relsize}
\usepackage[normalem]{ulem}
\usepackage[T1]{fontenc}
\usepackage[utf8]{inputenc}
\usepackage{enumerate}
\usepackage{times, mathptmx}
\usepackage{tikz}
\usepackage{aas_macros}
\usepackage{tabularx}
\usepackage{tabularray}
\usepackage{diagbox}
\usepackage{multirow}
\usepackage{pifont}

\newcolumntype{M}[1]{>{\centering\arraybackslash}m{#1}}
\usepackage{hhline}
\usepackage{makecell}
\usepackage{cleveref}
\usepackage{siunitx}
\usepackage{multirow}
\usepackage{booktabs}

\usetikzlibrary{arrows,positioning,decorations.markings,decorations.pathmorphing,calc}

\newcommand{\be}{\begin{equation}}
\newcommand{\ee}{\end{equation}}
\newcommand{\ba}{\begin{align}}
\newcommand{\ea}{\end{align}}

\newcommand{\comment}[1]{}

\newcolumntype{C}[1]{>{\centering\let\newline\\\arraybackslash\hspace{0pt}}m{#1}}

\newcommand{\contwo}{c_2}
\newcommand{\conthree}{c_3}
\newcommand{\approxcon}{\beta}

\newcommand{\nocontentsline}[3]{}
\newcommand{\tocless}[2]{\bgroup\let\addcontentsline=\nocontentsline#1{#2}\egroup}

\def\mpl{M_{\rm Pl}}

\newcommand{\red}[1]{#1}

\newcommand{\jnh}[1]{}

\definecolor{hyperref}{RGB}{026,028,087}

\def\gsim{ \lower .75ex \hbox{$\sim$} \llap{\raise .27ex \hbox{$>$}} }
\def\lsim{ \lower .75ex \hbox{$\sim$} \llap{\raise .27ex \hbox{$<$}} }

\newlength{\stheight}
\newcommand\textst[1][fu-grey]{
\ifmmode\setlength{\stheight}{+1.0ex}
\else\setlength{\stheight}{+0.5ex}
\fi
\bgroup\markoverwith{\textcolor{#1}{\rule[\the\stheight]{2pt}{1.0pt}}}\ULon
} 
 
\newcommand{\textins}[2][fu-grey]{
\ifmmode\mathcolor{#1}{#2}
\else\textcolor{#1}{#2}\@\,
\fi
}
\graphicspath{{./Plots/}}
\allowdisplaybreaks

\tikzstyle{vecArrow} = [thick, decoration={markings,mark=at position
1 with {\arrow[semithick]{open triangle 60}}},
double distance=1.4pt, shorten >= 5.5pt,
preaction = {decorate},
postaction = {draw,line width=1.4pt, white,shorten >= 4.5pt}]

\newcommand{\order}[1]{{\cal O}\left(#1\right)}

\begin{document}

\setcounter{tocdepth}{5}
\title{Stable black hole solutions with cosmological hair}

\author{Laurens Smulders}
\affiliation{Department of Physics \& Astronomy, University College London, London, WC1E 6BT, U.K}

\author{Johannes Noller}
\affiliation{Department of Physics \& Astronomy, University College London, London, WC1E 6BT, U.K}
\affiliation{Institute of Cosmology \& Gravitation, University of Portsmouth, Portsmouth, PO1 3FX, U.K.}

\begin{abstract}
Dynamical dark energy theories generically introduce a time-dependent field that causes the accelerated expansion of the Universe on large scales. When embedding black hole solutions in such a cosmological spacetime, this time dependence naturally gives rise to cosmological hair, i.e. the local black hole physics is no longer controlled by just the mass and spin of the black hole, but also impacted by the dark energy field. However, known such solutions are unstable. Focusing on the cubic Galileon as a concrete and illustrative example, we discuss the restrictions imposed on physical solutions by their regularity and stability in detail. We explicitly derive solutions that both recover the desired cosmological long-range behaviour and give rise to well-behaved, regular and stable, short-range dynamics around black holes. We show how the nature of the scalar hair around these local black hole solutions encodes cosmological information, highlighting novel and tantalising prospects of directly probing cosmological dynamics with black hole observations. 
\end{abstract}

\date{\today}
\maketitle

\tableofcontents

\section{Introduction} \label{sec:Intro}

While no-hair theorems have a long history \cite{Israel:1967wq,Carter:1971zc,Bekenstein:1971hc, Ruffini:1971bza, Bekenstein:1972ky,Robinson:1975bv, Bekenstein:1996pn,Hui:2012qt,Sotiriou:2013qea,Capuano:2023yyh,Yazadjiev:2025ezx, Maselli:2015yva} , they can be violated. Where in General Relativity (GR) the phenomenology associated with black holes is controlled by at most three parameters -- mass, spin and charge -- in theories introducing novel gravitational physics, additional parameters, i.e. `hair', can control the dynamics. This novel gravitational physics typically arises in the form of additional fields coupling to the usual tensorial degrees of freedom of GR. Scalar-tensor theories, where an additional scalar degree of freedom is introduced, are a minimal modification in this sense and have therefore unsurprisingly been a workhorse for understanding beyond-GR black hole solutions. 

Cosmology provides some of the most compelling theoretical and observational clues suggesting that physics beyond GR -- or, in a cosmological context, beyond $\Lambda{}$CDM -- may be present. In particular, the presence of dynamical dark energy, i.e. dark energy different from a pure cosmological constant $\Lambda$, would generically introduce new degrees of freedom with non-trivial time dependence and many candidate scalar-tensor theories, where the scalar corresponds to dynamical dark energy, exist -- see \cite{Copeland:2006wr,Clifton:2011jh,Babichev:2013vji,Joyce:2014kja,Koyama:2015vza,Kobayashi:2019hrl} for reviews. Unlike for static scalar profiles, for which there are strong no-hair theorems even for scalar-tensor theories \cite{Bekenstein:1996pn,Hui:2012qt,Sotiriou:2013qea,Capuano:2023yyh,Yazadjiev:2025ezx, Maselli:2015yva}, such time-dependent degrees of freedom (whether dark energy-related or not) are known to generically give rise to hairy solutions, see e.g. \cite{Jacobson:1999vr,Babichev:2013cya,Lara:2025hqh}. However, while several such hairy solutions are known to exist at the background level, where the stability of such known black hole solutions (and hence the behaviour of perturbations around known background solutions) has been fully investigated, they have been found to be unstable, with the instability typically associated with even parity modes in black hole perturbation theory \cite{Babichev:2013cya,Kobayashi:2014eva,Babichev:2016kdt,BenAchour:2018dap,Motohashi:2019sen,Charmousis:2019vnf,deRham:2019gha,Takahashi:2020hso,Khoury:2020aya,Takahashi:2021bml}.\footnote{
Note that a number of other known hairy solutions exist, for which stability has yet to be fully probed. These include the solutions proposed in \cite{Babichev:2016kdt,BenAchour:2018dap,Bakopoulos:2023fmv}, for which odd parity stability has been demonstrated in \cite{Sirera:2024ghv,Kobayashi:2025evr,Charmousis:2025xug}, with even-parity stability remaining unexplored. Using a different approach, partial stability  criteria (so-called `proto-stability') have been identified and demonstrated to hold for the hairy solutions found in \cite{Lara:2025hqh}.
}
In this context, a key challenge is therefore to establish whether a {\it stable} black hole solution in such scalar-tensor theories can be identified, which is embedded in a cosmological spacetime whose expansion is accelerated by dynamical dark energy as described by the scalar degree of freedom present. Or, alternatively, whether/in what sense the instability of solutions investigated until now points at an underlying no-go result.

The stability of black hole solutions in the presence of timelike scalar profiles has predominantly been investigated for so-called stealth solutions, i.e. solutions where the metric background corresponds to a GR solution (e.g. a Schwarzschild-de Sitter spacetime) and the hair purely arises via the existence of a non-trivial scalar profile. Indeed, requiring the existence of such stealth solutions already places constraints on the underlying scalar-tensor theories \cite{Kobayashi:2025evr}. However, the generic instabilities encountered when investigating such solutions motivate going beyond this ansatz and exploring theories where also the metric solution is deformed away from GR solutions and displays hair. In this paper we do so by investigating the cubic Galileon \cite{Nicolis:2008in} in detail, for which the existence of non-GR black hole solutions is known \cite{Babichev:2013cya,Babichev:2016fbg}. In this paper we discuss challenges in constructing such regular solutions, building on the analysis of \cite{Babichev:2025ric}, as well as the stability properties of the solutions we find. We identify generic obstacles to finding a stable and regular solution in the context of static black hole solutions, but show how these can be evaded as well. In particular, we explicitly derive novel, stable and regular solutions that are quasi-stationary (i.e. not static, but slowly varying).
\\

{\bf Outline}:
The paper is organised as follows. In section \ref{sec:CosmologicalEmbedding} we review the cubic Galileon and the cosmological/long-range limit of static black hole solutions in this theory. In section \ref{sec:BlackHoleSolutions} we then derive approximate analytical short-range solutions for such black holes, discuss what the branching structure of the general solution implies for how short- and long-range solutions can be connected, and then construct full numerical solutions across the full range of scales. Black hole perturbations around these solutions are then discussed in section \ref{sec:BlackHolePerturbations}, where we carry out a detailed stability analysis for the identified solutions. We find that solutions generically display instabilities in the scalar field around the short-range black hole solution. In section \ref{sec:QuasiStationaryBlackHoles} we therefore relax the initial assumption of a purely static black hole solution and allow the introduction of a mild time dependence, arguably a rather natural extension given the scalar background already displays time dependence. We repeat the analysis of the previous sections and show that regular and stable local black hole solutions with the desired cosmological long-range behaviour exist in this setup, therefore identifying a candidate dark energy theory (the cubic Galileon) with a well-behaved short-range black hole solution. We finally conclude in section \ref{sec:Conclusions} before collecting additional relevant details in the appendices. 
\red{More details on the calculations behind the results presented in this paper can be found in the accompanying GitHub repository~\cite{stable-black-holes-cosmological-hair}. }
Throughout we work in \textit{natural} units, where $c=\hbar=1$.

\section{Black Holes in a Cubic Galileon Universe: The cosmological limit} \label{sec:CosmologicalEmbedding}

Galileon scalar-tensor theories \cite{Nicolis:2008in} have been one of the primary workhorses in improving our understanding of scalar-tensor theories in both cosmological and strong gravity regimes. They capture some of the essential features of broad classes of scalar-tensor theories (e.g. of the Horndeski type \cite{Horndeski:1974wa,Deffayet:2011gz}), possessing non-linear interactions that give rise to self-accelerating solutions on larger scales \cite{Silva:2009km,DeFelice:2010pv} as well as a screening mechanism on smaller scales \cite{Nicolis:2008in,Burrage:2010rs,Babichev:2013usa}. In addition, the shift symmetry these models come endowed with protects them from radiative corrections -- see \cite{Luty:2003vm,Nicolis:2004qq,deRham:2010eu,Burrage:2010cu,Burrage:2011bt,deRham:2014wfa, Pirtskhalava:2015nla} and also \cite{deRham:2012ew,Goon:2016ihr,Saltas:2016nkg,Noller:2018eht,Heisenberg:2019wjv,Heisenberg:2020cyi,Goon:2020myi} for recent extensions of these arguments (also in other cosmologically relevant scalar-tensor theories). The shift symmetry further ensures that the aforementioned cosmological large scale behaviour comes in the form of analytically tractable attractor solutions \cite{DeFelice:2010pv}. Galileons are therefore a key `lamppost' theory to illustrate the behaviour of generic scalar-tensor theories with such features.

\subsection{The cubic Galileon} \label{sec:CubGal}

The cubic Galileon is the simplest non-trivial Galileon theory and therefore indeed one of the simplest scalar-tensor theories with non-trivial, higher-derivative and non-linear interactions. Its action is given by
\begin{align}
    \label{eq:CubicGalileonAction}
    S=\int d^4x\sqrt{-g}\left[
        \frac{\mpl^2}{2}R
        -\contwo (\partial\varphi)^2
        -\frac{\conthree}{\Lambda_3^3}\Box\varphi(\partial\varphi)^2
    \right],
\end{align}
where in addition to the scalar $\varphi$ we have also included a standard kinetic term for the metric $g_{\mu\nu}$, the Ricci scalar $R$. $g$ is the determinant of the metric, $\mpl$ is the reduced Planck mass, and $\Box \equiv \nabla_\mu\nabla^\mu$ is the d'Alembertian operator. $\contwo$ and $\conthree$ are dimensionless parameters and $\Lambda_3$ is an in principle arbitrary mass scale, but we will typically set it to its standard cosmological value such that $\Lambda_3^3 = \mpl H_0^2$, where $H_0$ is the value taken by the Hubble constant today. We normalise the Galileon field $\varphi$ in such a way that $\contwo=\pm\frac{1}{2}$, where we will discuss the relevance and background-dependence of the sign of this coefficient more below.

For Galileon theories, standard no-hair theorems \cite{Bekenstein:1971hc, Bekenstein:1972ky, Bekenstein:1995un} do not apply due to the (higher) derivative nature of its interactions. However, it can be shown that an alternative no-hair theorem exists \cite{Hui:2012qt}. This alternative theorem relies on the background configuration of the Galileon scalar field being purely radial, so can in particular be violated by introducing a time dependence in the scalar profile. When this time dependence is linear it will not propagate into the equations of motion due to the shift symmetry of the action and we can still find static solutions for the metric with such an ansatz.

Above we mentioned a sign choice for the $\contwo$ coefficient. On a flat Minkowski background, the canonical choice for this is $\contwo = \frac{1}{2}$, while the opposite (negative) sign leads to a ghost instability against scalar perturbations around this spacetime and a homogeneous background scalar field. Therefore, one cannot find stable black hole solutions with long-distance Minkowski asymptotes for this choice of $\contwo$. However, cosmological long-distance asymptotes are of course not Minkowski in nature, but instead we expect to recover expanding Friedmann-Robertson-Walker (FRW) behaviour in the large distance limit. A good proxy for this is requiring a de Sitter spacetime at large distances, simplifying the calculation and keeping it analytically tractable. As such we will consider black holes embedded in a cosmological deSitter-like spacetime, with the accelerated expansion at large scales driven by the scalar field. In other words, we are interested in a setup where the additional degree of freedom represented by the scalar plays the role of dark energy and will investigate what the consequences of this are for dynamics near the black hole. For the specific case of self-accelerating (long distance) solutions for the Galileon, stability on cosmological scales can be obtained even for the `wrong' sign of $\contwo$. In fact, for the cubic Galileon the existence of cosmological self-accelerating solutions mandates a negative $c_2$, while for general Galileons the same requirement is imposed by observational constraints for such a cosmological solution \cite{Barreira:2013jma}, highlighting that flat Minkowski vacua are disconnected from cosmologically preferred background solutions for Galileons -- a feature also representative of a much wider class of scalar-tensor theories \cite{Traykova:2021hbr}.

\subsection{Cosmological solution} \label{sec:CosmologicalSolution}

We will now focus on static non-rotating black hole solutions with de Sitter asymptotics for the cubic Galileon \eqref{eq:CubicGalileonAction}, closely following \cite{Babichev:2016fbg}. We consider the most general spherically symmetric, static form of the metric and a scalar field with a purely radial profile augmented with a linear time dependence. In spherical polar coordinates $(t,r,\theta,\phi)$ this ansatz then takes the form
\begin{align}
    ds^2 &= -h(r)dt^2+\frac{dr^2}{f(r)}+r^2d\Omega^2, \nonumber
    \\
    \varphi(t,r) &= qt+q\int dr\frac{\Xi(r)}{h(r)}.
    \label{eq:Ansatz}
\end{align}
Here, $d\Omega^2=d\theta^2+\sin^2\theta d\phi^2$ is the angular line element, while $h$, $f$ and $\Xi$ are functions of radius and $q$ is a dimensionful constant, to be determined through the equations of motion. Note that the appearance of $h$ in the $r$-integral is for convenience and follows the notation of \cite{Babichev:2016fbg} -- we could equally have written $\varphi(t,r) = qt+q\Psi(r)$, where the radial dependence would instead be encoded in the function $\Psi(r)$, cf. \cite{Babichev:2013cya}.

We find the equations of motion by varying \eqref{eq:CubicGalileonAction} with respect to the inverse metric and solve $\frac{\delta S}{\delta g^{\mu\nu}}=0$. After substituting in the ansatz we have nonzero $(tt)$, $(tr)$, $(rr)$, $(\theta\theta)$ and $(\phi\phi)$ components. The $(\theta\theta)$ and $(\phi\phi)$ components are proportional to each other due to spherical symmetry. They are also redundant due to Bianchi identities. Secondly, we can vary $S$ with respect to the scalar field to find another equation of motion. This equation can be written as a conservation equation by calculating the Noether current $J^\mu$ due to the shift symmetry $\varphi\rightarrow\varphi+\epsilon$:
\begin{align}
    \label{eq:ShiftSymmetryCurrent}
    J^\mu
    =2\contwo\partial^\mu\varphi
    +2\frac{\conthree}{\Lambda_3^3}\partial^\mu\varphi\Box\varphi-\frac{\conthree}{\Lambda_3^3}\partial^\mu\left(\partial\varphi\right)^2.
\end{align}
The scalar equation then gives $\nabla_\mu J^\mu=0$, which by substituting in the ansatz gives $\partial_r\left(r^2\sqrt{\frac{h}{f}}J^r\right)=0$. However, the $(tr)$ equation from variation with respect to the metric gives $J^r=0$, so the scalar equation is also redundant. We are then left with three independent equations to solve for the three free functions in the ansatz:
\begin{align}
     \frac{\conthree q^2}{\Lambda_3^3}(r^4h)'\frac{f}{h}\Xi^2-\frac{\conthree}{\Lambda_3^3} q^2r^4h'+2\contwo q r^4h\Xi &=0,
     \nonumber\\
     \contwo q^2 r^2\left(\frac{f}{h}\Xi^2-1\right)+\mpl^2rfh'+\mpl^2h(f-1) &=0,
     \nonumber\\
    \frac{q^2\left(\frac{f}{h}\Xi^2-1\right)}{\mpl^2rh^2}\left[\contwo r^2\sqrt{\frac{h}{f}}+\frac{\conthree q}{\Lambda_3^3}\left(r^2\sqrt{\frac{f}{h}}\Xi\right)'\right] 
         &= \left(\sqrt{\frac{f}{h}}\right)',
         \label{eq:EoMsDimensionful}
\end{align}
where a prime indicates differentiation with respect to $r$. The first equation is the $(tr)$ equation, while the second and third equations are linear combinations of the $(tt)$, $(tr)$ and $(rr)$ equations. These equations permit a de Sitter-like solution with
\begin{align}
    \label{eq:CosmologicalSolution}
    f(r)=h(r)=1-H^2r^2,
    &&
    \Xi(r)=-\frac{\contwo}{3\conthree}\frac{\Lambda_3^3}{q}r,
    \nonumber\\
    H^4=-\frac{\contwo^3}{27 \conthree^2}\frac{\Lambda_3^6}{\mpl^2},
    &&
    q^2=q_0^2 \equiv \frac{\contwo^2}{9\conthree^2}\frac{\Lambda_3^6}{H^2},
\end{align}
where $H$ is a constant Hubble parameter and we note that setting $\Lambda_3^3 = \mpl H_0^2$ indeed recovers $H \sim H_0$, as expected, with $\{\contwo,\conthree\}$ affecting the precise proportionality constant relating them. We can make this solution explicitly homogeneous by transforming to Friedmann coordinates. There are several possible coordinate transformations that give an explicitly homogeneous metric. If we require that there is no change of sign between the Schwarzschild coordinate time $t$ and cosmic time $\tau$ (i.e. $\left({\partial\tau}/{\partial t}\right)_r>0$) we are restricted to
\begin{align}
    \label{eq:FriedmannCoordinates}
    \tau=t\pm\frac{1}{2H}\ln\left|1-(Hr)^2\right|,
    &&
    \rho=re^{\mp Ht}\left|1-(Hr)^2\right|^{-1/2}.
\end{align}
The remaining $\pm$-signs are fixed by requiring homogeneity of the scalar field, which further depends on the sign of $q$. Specifically, for $q=\pm q_0 \equiv \pm\ ({\contwo}/{3\conthree})({\Lambda_3^3}/{H})$, we require the corresponding upper or lower sign in equation \eqref{eq:FriedmannCoordinates}. We then find for the metric and scalar field
\begin{align}
    ds^2=-d\tau^2+e^{\pm2H\tau}\left(d\rho^2+\rho^2d\Omega^2\right),&&
    \varphi=\pm q_0\tau.
\end{align}
We see that for the lower sign, we have a shrinking scale factor. Therefore, in order to obtain the physically interesting solution of an expanding universe in the frame where the scalar field is explicitly homogeneous we require $q=+q_0$. To summarise, this then leaves us with the fully sign-fixed set of coordinate transformations and metric and scalar background solutions
\begin{align}
    \label{eq:FriedmannCoordinatesSummary}
    \tau &=t+\frac{1}{2H}\ln\left|1-(Hr)^2\right|,
    \quad \rho =re^{- Ht}\left|1-(Hr)^2\right|^{-1/2}, \nonumber \\
    ds^2 & =-d\tau^2+e^{2H\tau}\left(d\rho^2+\rho^2d\Omega^2\right),
    \quad \varphi = q_0\tau.
\end{align}

\subsection{Long range approximation} \label{sec:LongRangeLimit}

Having established the existence of asymptotic de Sitter solutions, it is interesting to compute the leading order corrections to this in the long-range limit due to the presence of the black hole. In \cite{Babichev:2016fbg} such long-range solutions were found by expanding the background functions about $\frac{1}{r}$ as follows
\begin{align}
    h(r) &= \sum_{n=-2}^{\infty}\frac{c_n^{(h)}}{r^n},
    &f(r) &= \sum_{n=-2}^{\infty}\frac{c_n^{(f)}}{r^n},
    &\Xi(r) &= \sum_{n=-1}^{\infty}\frac{c_n^{(\Xi)}}{r^n}.
\end{align}
Solving this order by order one finds
\begin{align}
    \label{eq:LongRangeSolutions}
    h(r)&=-H^2r^2+1+\frac{\mu}{r}+\order{\frac{1}{r^2}},
    \nonumber\\
    f(r)&=-H^2r^2+1-\frac{1}{6}\left(1-\frac{q^2}{q_0^2}\right)+\frac{\mu}{r}+\order{\frac{1}{r^2}},
    \nonumber\\
    \Xi(r)&=-\frac{\contwo}{3\conthree}\frac{\Lambda_3^3}{q}\left(r-\frac{1}{2H^2r}\left(1-\frac{q^2}{q_0^2}\right)-\frac{\mu}{2H^2r^2}+\order{\frac{1}{r^3}}\right),
\end{align}
where we notice from \eqref{eq:Ansatz} that $\varphi$ is related to the integral of $\Xi$ and we therefore go to one order higher in the above expansion for $\Xi$. Here, $H$ and $q_0$ are given by equation \eqref{eq:CosmologicalSolution} and $\mu$ is an arbitrary integration constant. We see that the leading order terms indeed correspond to the cosmological solution discussed above, while higher order terms represent the desired corrections due to the presence of a black hole embedded in this cosmology. One might naively expect the $1/r$ correction terms shown explicitly in \eqref{eq:LongRangeSolutions} to be the leading corrections to a de Sitter solution close to the cosmological horizon $r_c$. However, since \eqref{eq:LongRangeSolutions} is an expansion not about $r_c$ but about $r=\infty$, the terms we have omitted can give significant contributions close to the  cosmological horizon, depending on their coefficients. For example, in the case where $1-(q/q_0)^2\sim\order{1}$, the omitted terms for $f$ and $h$ must be at least $\order{1}$ as well, since $f$ and $h$ must cross zero at the same point. Related to this, we will see below that for homogeneity around the cosmological horizon we require $q$ close to $q_0$ and will discuss the nature of corrections away from de Sitter on large scales in more detail when discussing explicit solutions in section \ref{sec:BlackHoleSolutions}.

It is worth highlighting that $q$ is in principle arbitrary for this solution and not necessarily equal to $q_0$. However, requiring homogeneity of physical observables on length scales $r\sim H^{-1}$ implies $q\sim q_0$. Consider for example $(\partial\varphi)^2$:
\begin{align}
    (\partial\varphi)^2=-q_0^2\left[1+\frac{1}{6H^2r^2}\left(1-\frac{q^2}{q_0^2}\right)+\order{\frac{1}{r^3}}\right].
    \label{eq:PartialPhiLongRange}
\end{align}
This has an inhomogeneous part that decays with a characteristic length scale $r_\text{hom}$, where we define
\begin{align}
    \label{eq:rHom}
    r_\text{hom} \equiv H^{-1}\left|\frac{q_0^2-q^2}{q_0^2}\right|^{\frac{1}{2}}.
\end{align}
To retain homogeneity on cosmological scales, we require $r_\text{hom}\ll H^{-1}$, recovering the bound found in \cite{Babichev:2025ric}, which we can here express as\footnote{
If the onset of inhomogeneities, i.e. $r$-dependence, on smaller scales is controlled by the $1/(H^2r^2)$ term in \eqref{eq:PartialPhiLongRange}, then we also expect $r_\text{hom}\gg r_s$, with $r_s$ the Schwarzschild radius of the black hole. Combining this condition with the one found above one then finds
\begin{align}
    \label{eq: q q0 relation}
    H^2r_s^2\ll\left|\frac{q_0^2-q^2}{q_0^2}\right|\ll1.
\end{align}
For example, taking $H=H_0=67.7(\text{km}/\text{s})/\text{Mpc}$, we find $q_0^2=7.4\times10^{-11}\text{eV}^4$. This is as expected in the cosmological limit, where one would expect $\dot\phi \sim \Lambda_2^2 \sim \mpl H_0 \sim (10^{-3} \rm eV)^2$, which matches well with the recovered value for $q_0$. Considering a black hole merger remnant with a $\sim10^6M_\odot$ mass typical for LISA detections, we then have
\begin{align}
    \frac{q^2}{q_0^2} &= 1 +\epsilon, \quad\quad \text{where} \quad\quad 4.7\times10^{-34} \ll\epsilon\ll 1.
\end{align}
} 
\begin{align}
    \label{eq:qVariation}
    \frac{q^2}{q_0^2} &= 1 +\epsilon, \quad\quad \text{where} \quad\quad \epsilon\ll 1.
\end{align}

\section{Approximate analytical and exact numerical black hole solutions} \label{sec:BlackHoleSolutions} 

Having discussed cosmological and approximate long-range solutions in the previous section, we now wish to explore the nature of short-range solutions, i.e. solutions in the vicinity of the black hole. We will first do so analytically, writing down approximate short-range solutions and finally using these to find numerical solutions that cover the full range of scales and hence connect the long and short-range limits.

\subsection{Parameter hierarchies and approximate black hole solutions} \label{sec:ShortRangeLimit}

Finding exact analytical solutions to the equations of motion \eqref{eq:EoMsDimensionful} is challenging, but luckily one can find highly accurate approximate solutions leveraging the fact that there is a significant hierarchy of scales for the solutions we consider. Namely, cosmological and black hole scales are separated by approximately 20 orders of magnitude, which indeed is the approximate hierarchy separating the $H_0$ and $\Lambda_3$ scales. Proceeding in such an effective field theory spirit, by requiring that our black holes are embedded in a realistic cosmology giving rise to these scales, we can therefore deduce the orders of magnitude of all relevant parameters in the action and use this to identify the (significantly simplified) leading order contributions to the equations of motion. In order to efficiently read off scale hierarchies, it will be useful to recast the equations of motion in dimensionless form. To this end we use the following dimensionless parameters as defined in \cite{Babichev:2016fbg}:
\begin{align}
    \alpha_1 &\equiv \frac{\conthree}{\contwo}\frac{q}{\Lambda_3^3 r_0},
    &\alpha_2 &\equiv -2\contwo\frac{q^2r_0^2}{\mpl^2},
    & x &\equiv\frac{r}{r_0}.
    \label{eq:DimensionlessParameters}
\end{align}
Here, $r_0$ is some for now arbitrary length scale that we will later fix to the Schwarzschild radius of the black hole. In terms of these dimensionless quantities, the full equations of motion \eqref{eq:EoMsDimensionful} can be rewritten as
\begin{align}
    \label{eq:EoMsDimensionless1}
    \alpha_1(x^4h)'\frac{f}{h}\Xi^2+2x^4h\Xi-\alpha_1x^4h'&=0,
    \\
    \label{eq:EoMsDimensionless2}
    \alpha_2x^2\left[1-\frac{f}{h}\Xi^2\right]+2xfh'+2h(f-1)&=0,
    \\
    \label{eq:EoMsDimensionless3}
    \left[1-\frac{f}{h}\Xi^2\right]\left[\alpha_2x^2\sqrt{\frac{h}{f}}+\alpha_1\alpha_2\left(x^2\sqrt{\frac{f}{h}}\Xi\right)'\right]&=2xh^2\left(\sqrt{\frac{f}{h}}\right)',
\end{align}
where in this case a prime denotes differentiation with respect to $x$.

It is now straightforward to identify hierarchically suppressed contributions to the equations of motion, which are in particular encoded in the $\alpha_i$ parameters in \eqref{eq:DimensionlessParameters}. As discussed above, the model parameters must be such that the long-range solutions in equation \eqref{eq:LongRangeSolutions} have $H\sim H_0$ and $q\sim q_0$ due to equation \eqref{eq:qVariation}. Consistent with our earlier discussion, equation \eqref{eq:CosmologicalSolution} then implies that indeed $\Lambda_3^3\sim H_0^2\mpl$ and $q\sim \Lambda_2^2 \sim H_0\mpl$. For the $\alpha_i$ parameters we then have $\alpha_1 \sim (H_0 r_s)^{-1}$ and $\alpha_2 \sim (H_0 r_s)^2$, where we have fixed $r_0$ to the Schwarzschild radius $r_s$ of the black hole, i.e.  $H_0r_0 = H_0 r_s$, which will be parametrically suppressed given the large hierarchy between cosmological and black hole scales. Indeed, defining a cosmological horizon $r_c$, we have $r_c\approx 1/H_0$ (with small corrections due to the presence of the black hole). We can then write the $\alpha_i$ parameters in the more instructive form
\begin{align}
\label{eq:AlphaHierarchy}
    \alpha_1 &\sim \frac{r_c}{r_s} \gg 1, &\alpha_2 &\sim \left(\frac{r_s}{r_c}\right)^2 \ll 1.
\end{align}
For example, for a typical LVK-merger remnant of $\sim50M_\odot$, $H_0 r_s \approx {r_s}/{r_c} \sim 10^{-22}$, while for $\sim10^6M_\odot$ masses typical for LISA detections $H_0r_s\approx r_s/r_c\sim10^{-17}$, so we recover the approximate 20 orders of magnitude separation discussed above. Naturally such strong hierarchies in the parameters will allow us to dramatically simplify the equations of motion when considering the region close to the black hole.

Focusing on this short-range regime, we are therefore looking for an approximate solution in the region near the horizon. We have set $r_0=r_s$, which means that the horizon is defined to be at $x=1$.\footnote{
Note, we are using $r_s$ to refer to the position of the horizon and do not a priori set $r_s = 2 G M$.} 
In the region around the horizon we therefore have $x\sim\order{1}$. We are interested in solutions where the scalar field is screened locally, such that we expect our solution to be similar to the Schwarzschild solution. For such solutions, $f(x)$ and $h(x)$, as well as their derivatives are $\order{1}$ away from the horizon.\footnote{
Of course, this is not the case at the horizon where $f(x),h(x)\propto(x-1)$. This will lead to additional suppression or enhancement of terms in the equations of motion. The approximate solutions we find will be accurate as long as this suppression/enhancement is of a smaller order than $r_s/r_c$: $|x-1|\gg r_s/r_c$. This error will therefore be at most a small change in the position of the horizon of order $r_s/r_c$ and finds its origin in the fact that the relative error in the solution naturally diverges where the solution approaches zero.
}
We can use equation \eqref{eq:EoMsDimensionless1} to determine $\order{\Xi}$. We see that if $\order{\Xi}$ is large, the first term will dominate. If it is small the third term will dominate. In either of these two cases, no solutions close to the Schwarzschild one can be obtained. Therefore, we are interested in solutions with $\Xi\sim\order{1}$. This means that the only suppression or enhancement of terms in the equations \eqref{eq:EoMsDimensionless1}--\eqref{eq:EoMsDimensionless3} comes from the hierarchies of the $\alpha_i$ parameters. If we expand the equations in terms of the small parameter $r_s/r_c \ll 1$ and only keep the leading order terms, we find these simplify significantly to
\begin{align}
\label{eq:ShortRangeSimpleEOM}
    (x^4h)'\frac{f}{h}\Xi^2-x^4h'&=0,
    \nonumber\\
    xfh'+h(f-1)&=0,
    \nonumber\\
    2xh^2\left(\sqrt{\frac{f}{h}}\right)'&=0.
\end{align}
This system of equations has a unique non-trivial solution
\begin{align}
    \label{eq:ShortRangeSolutions}
    f(x)&=1-\frac{1}{x},
    \quad
    h(x)=\approxcon \left(1-\frac{1}{x}\right),
    \nonumber\\
    &\Xi_\pm(x)=\pm\text{sign}(\alpha_1)\sqrt{\frac{\approxcon}{4x-3}},
\end{align}
where we have fixed one integration constant by requiring that $f(1)=h(1)=0$, since this is where we defined the horizon to be. Since the first equation is quadratic in $\Xi$, there exist two branches of opposite sign. In naming these branches we have introduced $\text{sign}(\alpha_1)$ for later convenience when we relate these short-range solutions to the full ones. While we have found these solutions for the region where $x=\order{1}$, a more detailed analysis of the relative errors, which we relegate to appendix \ref{sec:AppendixShortRangeLimitAccuracy}, shows that these approximate solutions will be accurate as long as $x\ll(r_s/r_c)^{-2/3}$. Note that this radius has the same order of magnitude as the Vainshtein radius found in \cite{Babichev:2012re, Babichev:2025ric}.

The approximate short-range solution is ultimately controlled by just one free parameter, namely an integration constant which we denote $\approxcon$. At this stage we cannot express this parameter in terms of the underlying theory parameters in the action \eqref{eq:CubicGalileonAction}, but this can in principle be done by matching this short-range solution to the correct cosmological behaviour on large scales.\footnote{
For the Schwarzschild(-deSitter) solution in GR (with cosmological constant), this integration constant is also present, but fixed to $\approxcon=1$ by requiring flat (deSitter) spacetime as $x\rightarrow\infty$. We essentially aim to do the same thing, but since our short-range solutions are not accurate all the way to the cosmological limit, the connection between the two limits is more complicated and does not necessarily imply $\approxcon=1$.
}
We will discuss this point in detail below.

While we cannot fix $\approxcon$ at this stage, it is worth estimating how a small deviation away from a Schwarschild short-range solution might arise parametrically. Considering the dimensionless parameters controlling the full set of equations of motion \eqref{eq:EoMsDimensionless1}-\eqref{eq:EoMsDimensionless3}, the parametric freedom is primarily encoded in $\alpha_1$ and $\alpha_2$. Since the only $\order{1}$ combination of these parameters is $\alpha_1^2\alpha_2$ (and powers thereof), in order to obtain a non-negligible parameter dependence of $\approxcon$, it is natural to expect the leading order dependence of this parameter on $\alpha_1$ and $\alpha_2$ to ultimately (when matching to the long-range solution) be a function of this combination, so  $\approxcon=\approxcon(\alpha_1^2\alpha_2)$.\footnote{
This is the case if one implicitly assumes the functional dependence to be polynomial. However, for other dependencies (e.g. logarithmic), parameters of  $\order{r_s/r_c}$ could still contribute to $\approxcon$ at leading order. Nevertheless, in \cite{Smulders:2026bya} we find $\beta$ numerically for a range of parameter values and confirm that it indeed only depends on $\alpha_1^2\alpha_2$ at the leading order.
}
In addition, we have
\begin{align}
    \alpha_1^2\alpha_2=\frac{2}{3}\left(\frac{q}{q_0}\right)^4\approx\frac{2}{3}-\frac{4}{3}\frac{q_0^2-q^2}{q_0^2}.
\end{align}
Using equation \eqref{eq:qVariation}, we see that $\alpha_1^2\alpha_2\approx\frac{2}{3}$. Although corrections to this must be small, they can in principle still be several orders of magnitude higher than $r_s/r_c$. Therefore, we could see deviations from $\approxcon\left(\frac{2}{3}\right)$ that do not drop out at the leading order in $r_s/r_c$. Physically, deviations in $\approxcon$ represent local black hole hair. However, note that $\approxcon$ can be removed {\it locally} by rescaling $t\rightarrow t/\sqrt{\approxcon}, q \rightarrow q\sqrt{\beta}$, which means that $\approxcon$ cannot be determined using local observables only.\footnote{
We thank Shinji Mukohyama for discussions on this point.
} 
Crucially, the expansion rate of the universe on large scales defines a reference time-normalisation -- see equation \eqref{eq:FriedmannCoordinatesSummary} which defines the Schwarzschild time coordinate $t$ uniquely from the cosmic time $\tau$. This makes $\approxcon$ an observable parameter when considering the embedding of the black hole in an expanding universe.

\subsection{Branching structure of the scalar solution} \label{sec:BranchingStructure}

We now have analytic approximations for the short- and long-range behaviours of $f$, $h$ and $\Xi$. For the intermediate region however, analytic solutions are not readily available and, mimicking the approach of \cite{Babichev:2016fbg}, we will therefore resort to numerical integration -- see section \ref{sec:NumericalSolutions}. Nevertheless, before diving head first into the numerics, we can make some useful observations about the structure that the solutions need to have in the intermediate region in order to match the short- and long-range limits.

We can partly characterize this structure by the roots of equation \eqref{eq:EoMsDimensionless1}. Since it is quadratic in $\Xi$ two branches will exist for the scalar solution. This is described in \cite{Babichev:2025ric} and already apparent from our short-range solution \eqref{eq:ShortRangeSolutions}. We can write \eqref{eq:EoMsDimensionless1} as
\begin{align}
    \label{eq: A, B, C defs}
    A\Xi^2+B\Xi+C=0,
\end{align}
where $A=\alpha_1(x^4h)'\frac{f}{h}$, $B=2x^4h$ and $C=-\alpha_1x^4h'$. We then define the two branches of $\Xi$ in terms of $f(x)$, $h(x)$ and $x$ as
\begin{align}
\label{eq:BranchDefinitions}
    \Xi_\pm &=\frac{-B\pm\sqrt{\Delta}}{2A},
     &\Delta &\equiv B^2-4AC.
\end{align}

We are looking for solutions with cosmological asymptotes such that $f(x),h(x)\rightarrow-H^2r_s^2x^2$ and $\Xi(x)\rightarrow-\frac{x}{3\alpha_1}$ as $x\rightarrow\infty$ -- see equation \eqref{eq:LongRangeSolutions}. Using these cosmological asymptotes for $f$ and $h$ to calculate $\Xi_\pm$ to leading order in $x$ we find
\begin{align}
    \Xi_\pm(x)=-\frac{x}{6\alpha_1}\mp\frac{x}{6\alpha_1}+\order{x^0},
\end{align}
from which it is clear that $\Xi$ must follow the $\Xi_+$ branch as $x\rightarrow\infty$ in order to recover the correct cosmological behaviour. More specifically, it can be shown that for any kind of cosmological horizon to occur (where $f$ and $h$ cross zero from above), $\Xi$ must follow the $\Xi_+$ branch (see appendix \ref{sec:AppendixCosmologicalHorizonBranch}). Similarly, we can use the short-range solutions from equation \eqref{eq:ShortRangeSolutions} to evaluate $\Xi_\pm$ to leading order in $r_s/r_c$ and find the $\Xi_\pm$ labels defined here are the same as their short-range equivalents defined in equation \eqref{eq:ShortRangeSolutions}.

From the above it can be deduced that a solution that follows the $\Xi_-$ branch near the black hole horizon can only have cosmological asymptotes if a branch switch occurs in the intermediate region. For continuous solutions a switch of branch can only occur at \textit{branch points} where the two branches intersect, i.e. $\frac{\sqrt{\Delta}}{2A}=0$. However, {whether a solutions switches branch at a certain branch point is predetermined by the initial conditions} if we require that the solution for $\Xi(x)$ remains smooth. Smoothness of $\Xi(x)$ is important, since through the equations of motion the metric functions $f(x)$ and $h(x)$ can only be smooth if $\Xi(x)$ is smooth. A non-smooth $f(x)$ or $h(x)$ generically leads to divergences in for example the Ricci or Kretschmann scalars, indicating an unphysical spacetime.\footnote{
Smoothness here means that all derivatives of the functions are finite and continuous. However, one could try to violate this assumption by introducing a discontinuity at some very high derivative. This would still lead to divergences in (very high derivatives of) physical observables, but these might not be strictly observable in a practical sense.
}
We will now elaborate on this point.

Given some smooth $f(x)$ and $h(x)$, the individual $\Xi_\pm$ branches are not necessarily smooth. This is because $\sqrt{\Delta}$ is not necessarily smooth across branch points. However, two smooth solutions which are combinations of the $\Xi_\pm$ branches through switches at branch points will always exist. We can write a \textit{smooth} solution for $\Xi(x)$ as
\begin{align}
    \Xi_D(x)=\frac{-B+D}{2A},
\end{align}
where $D(x)$ is some smooth function satisfying $D^2=\Delta$. The sign of $D(x)$ determines what branch the solution follows at a certain $x$. At branch points $D(x)$ must be zero. Suppose this zero is of multiplicity $m$ at some $x=x_0$. This means that in some neighbourhood of $x_0$ we can write
\begin{align}
    D(x)=\bar{D}(x)(x-x_0)^m
    \Rightarrow \Delta(x)=\bar{D}^2(x)(x-x_0)^{2m},
\end{align}
where $\bar{D}(x_0)\neq0$. For odd multiplicity zeros, $D(x)$ will change sign and a switch of branch will occur, while there is no branch switching for even multiplicity zeros.

Take now the solutions for a certain set of parameters $t=(\alpha_1,\alpha_2,\approxcon)$: $f_t(x)$, $h_t(x)$ and $\Xi_t(x)$. We can consider the paths through parameter space along which $f_t$, $h_t$ and $\Xi_t$ smoothly deform, i.e. $f_{t+\delta t}(x)-f_t(x)\rightarrow 0$ as $\delta t \rightarrow 0$, etc. We expect this to happen when the solutions are \textit{analytic} in our region of interest ($x>1$) in addition to being smooth.\footnote{
We know that the short range approximations will undergo smooth deformations from \eqref{eq:ShortRangeSolutions}. Suppose we have some $x_0$ in the vicinity of the black hole where the function will smoothly deform. The function being analytic then means that we can Taylor expand to show that the deformation is smooth at any point:
\begin{align}
    f_{t+\delta t}(x)-f_t(x)
    =\sum_{n=0}^\infty\frac{(x-x_0)^n}{n!}\left(f_{t+\delta t}^{(n)}(x_0)-f_t^{(n)}(x_0)\right)
    \rightarrow0,
\end{align}
where this analyticity is a feature we expect for physical theories.
}
$A_t$, $B_t$, $C_t$ and $D_t$ will then smoothly deform as well. We can consider the possibility of moving along such a path between solutions with a different number of odd multiplicity branch points. Crucially, odd multiplicity branch points are topologically protected: we cannot locally form a single odd multiplicity branch point without altering the solution for $D_t$ at an arbitrary distance since it involves a change of sign. This means we can only smoothly create and annihilate odd multiplicity branch points in pairs. Two solutions that differ in the number of odd multiplicity branch points by an odd amount therefore cannot be connected by smooth paths through the parameter space. Therefore, such solutions must lie in disconnected regions of the parameter space, separated by regions where analytic solutions do not exist at all. 

For example, if we numerically find some solution with an odd number of branch switches, and want to find a solution with an even number of branch switches, we cannot find this by smoothly changing our input parameters without passing through regions in the parameter space where analytic solutions do not exist. 

\subsection{Numerical solutions} \label{sec:NumericalSolutions}

For our black hole solutions, we now have long-range cosmological asymptotes from \eqref{eq:LongRangeSolutions} and short-range near-horizon approximations from \eqref{eq:ShortRangeSolutions}. The constant $\approxcon$ in the short-range solutions can in principle be fixed by requiring that the short-range solution connects to the correct cosmological limit. To do this, we need to know the full solution. However, in the intermediate region we do not have a readily available analytic solution, so we here turn to numerically investigating this regime instead, following an approach that is analogous to those in \cite{Babichev:2016fbg} and \cite{Emond:2019myx}.

To solve for the background numerically, we use the dimensionless set of equations \eqref{eq:EoMsDimensionless1}-\eqref{eq:EoMsDimensionless3}. This is an index-2 system of differential-algebraic equations (DAE)\footnote{
The index refers to the minimum number of times we need to differentiate the equations to obtain a system of ordinary differential equations -- see e.g. \cite{doi:10.1137/1.9781611971224,hairer_numerical_1989} for thorough discussions of DAEs.
} 
and we therefore reduce the index of the system to $0$ by differentiating the first two equations. Subsequently, three equations remain containing $h$, $f$, $\Xi$, $h'$, $f'$, $\Xi'$ and $h''$. We will start the integration at $x = 1$, so a full set of initial conditions consists of specifications for $h(1)$, $f(1)$, $\Xi(1)$ and $h'(1)$. However, the initial conditions must be consistent with the two underlying undifferentiated equations. This leaves only two initial conditions that we can freely fix: we choose $h(1)$ and $h'(1)$. To have a horizon at $r_0$, i.e. $x=1$, we need to set $h(1)=0$. This essentially fixes $r_0=r_s$. $h'(1)$ is then fixed through the \textit{shooting method}, i.e. by varying $h'(1)$ until we find the value that gives the correct long-range cosmological behaviour after numerical integration. Note that $h'(1)$ essentially encodes the value of $\approxcon$, since from the approximate solutions we have $h'(1)\approx\approxcon$. In practice, due to the coordinate singularity at $x=1$, the numerical integration breaks down there and we have to start the integration slightly outside the horizon \cite{Emond:2019myx}. We then use as initial condition for $h$: $h(1+\delta x)=h'(1+\delta x)\delta x$ and $h'(1+\delta x)$ is still fixed via the shooting method. Although $f(1+\delta x)$ and $\Xi(1+\delta x)$ will be fixed by the equations of motion, we do have one further freedom to choose either of the two branches for $\Xi$ initially (which may or may not switch once integrating further out in the solution). We then numerically integrate out to large $x$ for several values of $h'(1+\delta x)$ and find the initial value that most closely matches the cosmological limit. 
\begin{figure*}
    \includegraphics[width=\textwidth]{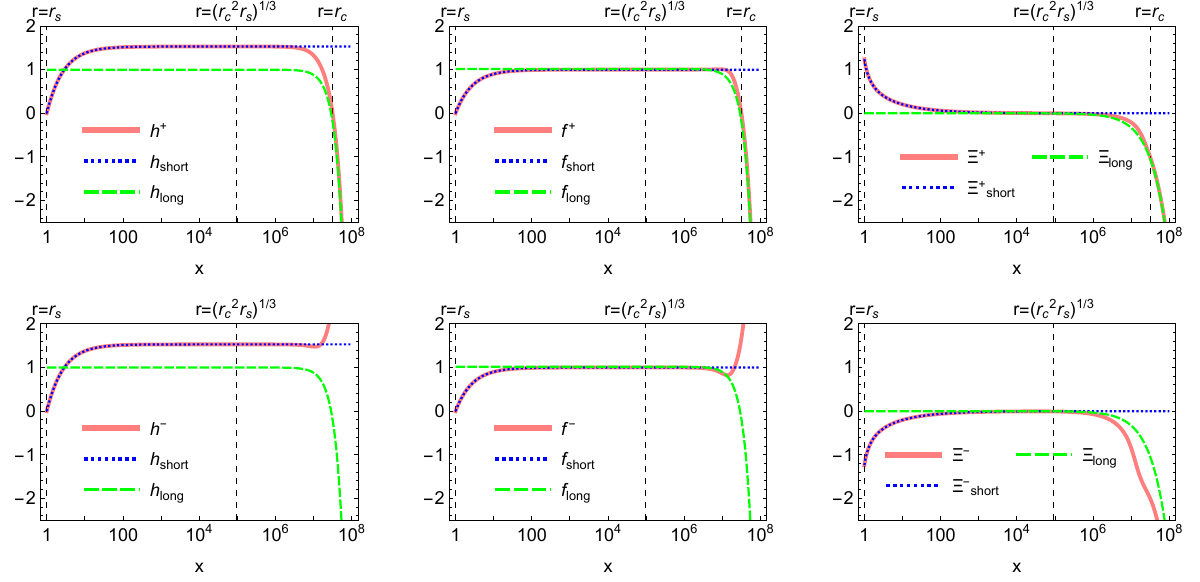}
    \caption{
    Here we show numerical solutions to \eqref{eq:EoMsDimensionless1}-\eqref{eq:EoMsDimensionless3} for a specific example case, where we have chosen $\alpha_1=10^7$ and $\alpha_2=8\times10^{-15}$ \eqref{eq:DimensionlessParameters} (see main text for a discussion of the chosen values). Going from left to right, we compare the full numerical solutions for $h(x),f(x)$ and $\Xi(x)$ together with their analytical short- and long-range approximations (\ref{eq:ShortRangeSolutions},\ref{eq:LongRangeSolutions}), where $x \equiv r/r_s$. $x = 1$ and $x = r_c/r_s$ denote black hole and cosmological horizons, respectively. $x=(r_c/r_s)^{2/3}$ is the scale where the short-range solution is generically expected to become inaccurate (see appendix \ref{sec:AppendixShortRangeLimitAccuracy}). 
    {\bf Top row}: Here we start the evolution in the $\Xi_+$ branch \eqref{eq:BranchDefinitions} at short scales and integrate out to large $x$. The one parameter freedom in this solution, captured by the parameter $\approxcon$ in \eqref{eq:ShortRangeSolutions}, is then fixed via the shooting method: scanning over initial conditions of $h'$ to match the solution to the desired (cosmological) long-range solution. We find an excellent fit for $\approxcon=1.53$. 
    {\bf Bottom row}: Here we instead start in the $\Xi_-$ branch on short scales. No value of $\approxcon$ could be found here to connect to the desired cosmological limit. Instead the solution on large scales generically diverges away from that desired limit, as is illustrated here with a solution for $\approxcon = 1.53$ in this branch. In section \ref{sec:BranchingStructure} we discuss how this can be understood in terms of the overall branching structure of solutions across all $x$ values. 
    }
    \label{fig:NumericalSolutions}
\end{figure*}

Having laid out the numerical procedure above, it is instructive to consider some concrete examples in detail. We will choose the following values for the $\alpha_i$ parameters: $\alpha_1=10^7$ and $\alpha_2=8\times10^{-15}$. Here it is important to highlight that, in a realistic scenario -- in other words, when substituting cosmologically motivated values for the relevant coefficients into \eqref{eq:DimensionlessParameters} -- the hierarchy between $\alpha_1$ and $\alpha_2$ would be more extreme, namely $\alpha_1\sim10^{20}$ and $\alpha_2\sim10^{-40}$. However, with such vastly different coefficients in the equations, the maximum size of floating point numbers becomes an obstacle for numerical computations in practice. Similar to the approach employed by \cite{Babichev:2016fbg}, we therefore choose the above example values that satisfy \eqref{eq:AlphaHierarchy} and retain all the relevant qualitative features we want to illustrate. Figure \ref{fig:NumericalSolutions} shows the numerical solutions for $f(x)$, $h(x)$, $\Xi(x)$ and their long- and short-range approximations for these parameter values. In the top row we have chosen to start in the $\Xi_+$ branch and found the initial condition for $h'$ such that the solutions approach their cosmological limits for large $x$. We performed a least-squares fit on the short-range part of the solution to determine the parameter $\approxcon$ and find an excellent fit in the $x\ll(r_c/r_s)^{2/3}$ regime for $\approxcon=1.53$. 

Crucially, and closely related to our choice of illustrative $\alpha_i$ discussed above, we expect the leading order behaviour of the short-range solutions to depend only on the combination $\alpha_1^2\alpha_2$ (as described in section \ref{sec:ShortRangeLimit}). Therefore, we can safely assume that the exact orders of magnitude of $\alpha_1$ and $\alpha_2$ do not matter for studying the short-range behaviour, as long as $\alpha_1\gg1$ and $\alpha_2\sim\alpha_1^{-2}$. We have verified this in practice: varying $\alpha_1$ and $\alpha_2$ while keeping $\alpha_1^2\alpha_2$ constant does not change the overall qualitative form of the solutions or the value of $\approxcon$ that we find. Changing the value of $\alpha_1^2\alpha_2$ does change the value of $\approxcon$ and we find that $\alpha_1^2\alpha_2=\frac{2}{3}$ (i.e. $q=q_0$) corresponds to $\approxcon=1$, recovering a Schwarzschild metric background in the short-range approximation -- see figure 1 in \cite{Smulders:2026bya} for the detailed results showing this. Since $\approxcon$ is the only parameter affecting the form of the approximate local solutions~\eqref{eq:ShortRangeSolutions}, we can conclude that our use of reduced hierarchies is justified for deriving the results in this paper.

\begin{figure}
    \centering
    \includegraphics[width=\linewidth]{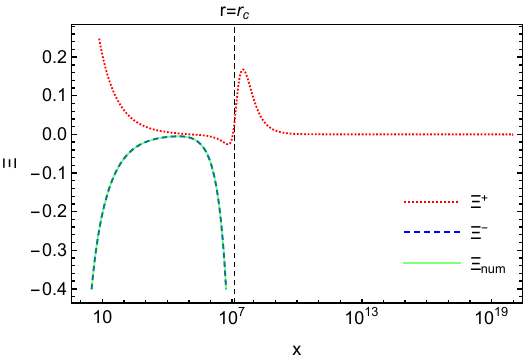}
    \caption{
    Here we show the numerical solution for $\Xi(x)$ from the bottom row of figure \ref{fig:NumericalSolutions}, integrated up to large $x$, as well as the $\Xi_\pm(x)$ branches calculated from the solutions for $f(x)$ and $h(x)$. The solution fully follows the $\Xi_-$ branch as expected and we find no branch points anywhere in this range.
    }
    \label{fig:MinusBranchSolutionBranchPoints}
\end{figure}

We have not found any numerical solutions that follow the $\Xi_-$ branch 
in the short-range regime and connect to the desired cosmological long-range solution. The bottom row of figure \ref{fig:NumericalSolutions} shows an example of an attempt at finding such a solution, using the same parameters as in the top row. As is visible in the figure, the solutions closely follow the near-horizon approximations on small scales, but do not match with the desired cosmological limit on large scales. In order to connect to the desired asymptotic behaviour on cosmological scales, there needs to be a branch switch in between the short- and long-range limits, as described in section \ref{sec:BranchingStructure}. Yet the solution shown remains on the $\Xi_-$ branch throughout -- see figure \ref{fig:MinusBranchSolutionBranchPoints}. To find a solution with the `correct' cosmological asymptotes, we would have to change the parameters in such a way that an odd number of odd multiplicity branch points forms in between the short- and long-range limits. As described in section \ref{sec:BranchingStructure} an odd number of odd multiplicity branch points cannot be formed continuously, so the only way to smoothly deform into a physically relevant solution is to move an odd multiplicity branch point from elsewhere in the solution to the intermediate region. It can be shown that branch points cannot pass horizons (appendix \ref{sec:AppendixCosmologicalHorizonBranch}), so the odd multiplicity branch point would have to come from large $x$. We have integrated the solution shown in the bottom row of figure \ref{fig:NumericalSolutions} up to very large $x$ (up to $x\approx10^{20}$, while the would-be cosmological horizon is around $x\approx10^{7}$) and not found any branch points that could be moved to lower $x$.
This is shown in figure \ref{fig:MinusBranchSolutionBranchPoints} where we have plotted the numerical solution for $\Xi(x)$ as well as the $\Xi_\pm(x)$ branches calculated from the numerical solutions for $f(x)$ and $h(x)$. This therefore suggests that this solution cannot be smoothly deformed into a cosmologically relevant solution. The parameters for which cosmologically relevant solutions exist would therefore have to lie in a region of parameter space that is disconnected from the current region. 

Searching for such a disconnected region, we have run the numerical integration for a range of input values $(\alpha_1,\alpha_2,\beta)=(a_1\times10^6,a_2\times10^{-12},a_3)$, while starting the integration in the $\Xi_-$ branch. Recall that $\alpha_1\sim r_c/r_s$ and $\alpha_2\sim (r_s/r_c)^2$ while we only expect the value of the combination $\alpha_1^2\alpha_2$ to affect the qualitative behaviour of solutions. We therefore probe physically relevant solutions by making a representative choice for $\alpha_1$ and $\alpha_2$, such that $a_1$ and $a_2$ are $\order{1}$. Similarly, $\approxcon=1$ corresponds to a Schwarzschild-like solution on black hole scales so we expect $a_3\sim\order{1}$ as well for black holes resembling GR ones. To investigate the resulting parameter sapce, we construct a three-dimensional grid with each of $(a_1,a_2,a_3)$ running from $0.01$ to $10$ with a spacing of $1$ and secondly a narrower grid from $0.5$ to $1.5$ and a spacing of $0.1$. For none of these input values, a numerical solution that is in the correct ($\Xi_+$) branch at large $x$ is recovered.

These empirical finds suggest that 1) there is no region of parameter space (with the hierarchies required for the cosmological expansion) that is disconnected from the parameters we have used in our attempted solution, and hence 2) a short-range solution in the $\Xi_-$ branch {\it cannot} be connected to the long-range cosmological asymptotes in the $\Xi_+$ branch.

\begin{figure*}
    \centering
    \includegraphics[width=\textwidth]{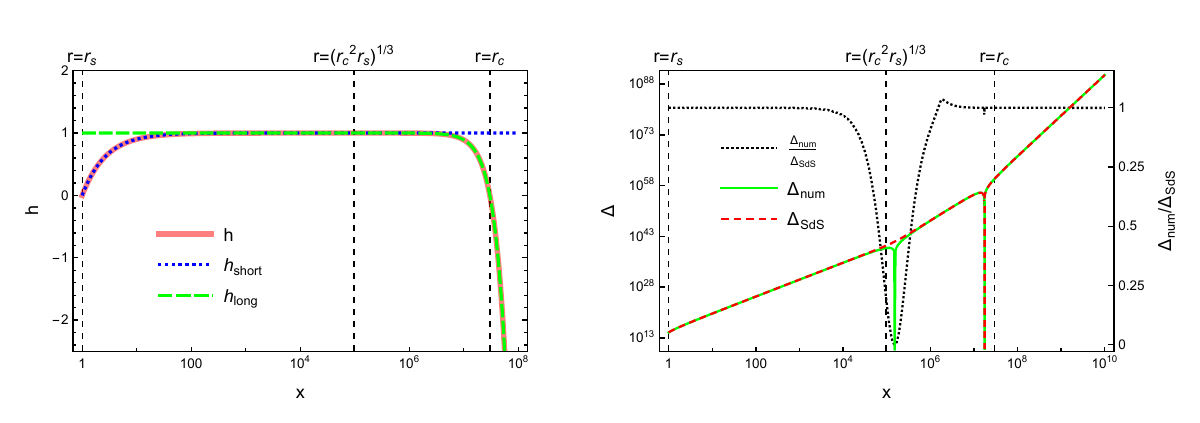}
    \caption{
    Here we show a numerical solution for the specific case highlighted at the end of section \ref{sec:NumericalSolutions}, namely a choice of $\alpha_i$ such that $q = q_0$ for which $\approxcon = 1$ is a solution and hence the respective limits of a Schwarzschild-de Sitter spacetime are recovered at short and long distances. We show such a numerical solution, as well as the corresponding small and large scale analytic approximations for the example values $\alpha_1=10^7$, $\alpha_2=\frac{2}{3}\times10^{-14}$ here.   
    {\bf Left plot}: Given the asymptotic analytic behaviour, one may expect that Schwarzschild-de Sitter is a good fit throughout all $x$ values. Here we see that this is indeed appears to be the case when looking at the overall form of the full numerical solution. However, while the solution for the metric superficially resembles Schwarzschild-deSitter, there are important differences in the branching structure that can be understood when looking at the discriminant $\Delta$ \eqref{eq:BranchDefinitions}.
    {\bf Right plot}: Here we illustrate the branching structure of the solution by plotting the discriminant $\Delta$ for the numerical solution shown in the left plot, $\Delta_{\rm num}$, as well as for a true Schwarzschild-deSitter solution, $\Delta_{\rm SdS}$. At small and large scales, the numerical solution closely follows the Schwarzschild-deSitter solution. In particular we recover a common branch point close to $r_c$. However, on intermediate scales the two solutions deviate from each other, with the crucial difference that the numerical solution has an additional (odd multiplicity) branch point. This induces an additional switch of branch, which means that in the numerical solution the short-range $\Xi_+$ branch connects to the cosmological limit instead of the $\Xi_-$ branch.
    }
    \label{fig:SdSLikeSolution}
\end{figure*}

Finally, we wish to highlight an interesting special case, namely fixing $q=q_0$. In this case one finds $\approxcon\approx1$, i.e. approximate Schwardschild behaviour on small scales. From equation \eqref{eq:LongRangeSolutions} we find that $q=q_0$ gives de Sitter-like behaviour on large scales. This kind of behaviour can be explained by Vainshtein screening on small scales and $\varphi$ taking on the role of a cosmological constant on large scales, as argued in \cite{Babichev:2025ric,Babichev:2012re}. One may therefore assume that such a solution remains close to Schwarzschild-deSitter throughout all scales. However, this assumption seems to fail on intermediate scales, where the analytical short- and long-range approximations discussed do not apply, leading to important qualitative differences in the overall solution. In the left-hand panel of figure \ref{fig:SdSLikeSolution} we have plotted the numerical solution for $h(x)$ as well as the short- and long-range approximations for $\alpha_1=10^7$ and $\alpha_2=\frac{2}{3}\times10^{-14}$. Although the solution naively seems to closely follow Schwarzschild-deSitter all the way through, a different picture arises when we compare the discriminant calculated from this numerical solution to the one calculated from a pure Schwarzschild-deSitter solution as is shown in the right-hand panel. As expected, the discriminants closely overlap in the short- and long-range limits, sharing an odd-multiplicity branch point close to $r=r_c$. However, in the intermediate range there are significant differences. Crucially, the numerical solution has an additional odd multiplicity branch point which induces an additional branch switch. This is why the solution described here must follow the $\Xi_+$ branch in the vicinity of the black hole, unlike the Schwarzschild-deSitter solutions conjectured in \cite{Babichev:2025ric,Babichev:2012re}.

\section{Black Hole Perturbations and Stability Criteria} \label{sec:BlackHolePerturbations}

Having discussed background solutions in detail in the previous sections, we will now investigate perturbations around this solution. We will focus on perturbations in the vicinity of the black hole where black hole perturbation theory will apply. The main aim of the calculation in this section will be to understand the stability properties of solutions. 

\subsection{Schematic quadratic actions} \label{sec:QuadraticActionsSchematic}

We begin by perturbing the metric and scalar field about their static backgrounds \eqref{eq:Ansatz}:
\begin{align}
    g_{\mu\nu}=\bar{g}_{\mu\nu}+\delta g_{\mu\nu{}},
    &&
    \varphi=\bar{\varphi}+\delta\varphi,
\end{align}
where $\bar{g}_{\mu\nu}$ and $\bar{\varphi}$ are the metric and scalar backgrounds while $\delta g_{\mu\nu}$ and $\delta\varphi$ are their perturbations, respectively. We then substitute these perturbations into the action \eqref{eq:CubicGalileonAction} and expand to second order in the perturbations to find the quadratic action. Note that no assumption about working in the short-distance regime has been made so far. 

In addition to the dimensionless radial coordinate $x \equiv r/r_0$, it will be useful to introduce a dimensionless time coordinate $y=t/r_0$. In these coordinates, the background metric and scalar field have the form
\begin{align}
    \bar{g}_{\mu\nu}dx^\mu dx^\nu&=r_0^2\left[-h(x)dy^2+\frac{dx^2}{f(x)}+x^2d\Omega^2\right],
    \nonumber\\
    \bar{\varphi}&=qr_0\left[y+\int dx\frac{\Xi(x)}{h(x)}\right].
\end{align}
We can now perturb each term in the action \eqref{eq:CubicGalileonAction}, find all of its contributions to the quadratic action and assess what the leading order terms are for each type of contribution. It is at this point that we will again focus on the short-range regime, where $r/r_c \ll 1$. As in the previous section, the existence of such large hierarchies means we can find excellent simplified approximations to the quadratic action in the vicinity of black holes. Specifically, this means we again take $r_0 = r_s$. In the dimensionless coordinates $(x,y)$, partial derivatives will then not contribute any orders of magnitude, since our approximate solutions and their derivatives are all $\order{1}$ in the region of interest. Any background scalar fields left in the quadratic action for perturbations will contribute $qr_s$, while the remaining metric background factors can only contribute some factor of $r_s$, which must be such that the overall term is dimensionless. 

In order to make the interactions structure more explicit, we will also canonically normalise the perturbations such that their kinetic terms are $\order{1}$. We do so by re-defining the perturbations as follows 
\begin{align}
    \label{eq:NormalisedPerturbations}
    g_{\mu\nu}=\bar{g}_{\mu\nu}+\frac{r_s}{\mpl}\delta g_{\mu\nu},&&
    \varphi=\bar{\varphi}+\sqrt{\frac{\Lambda_3^3}{|\conthree q|r_s}}\delta\varphi.
\end{align}
where we recall that $q\sim H_0\mpl$ and $\Lambda^3_3\sim H_0^2\mpl$, while $\contwo$ and $\conthree$ are both $\order{1}$. Using these normalisations and keeping only the leading order in $H_0 r_s \approx r_s/r_c$ of every type of contribution to the quadratic action (i.e. we now zoom in on the short-distance regime), we find the following schematic result 
\begin{align}
    \label{eq:QuadraticActionSchematic}
    S^{(2)}_{\text{leading order}}\sim\int d^4x&\biggl[
        \delta g^2 \color{black} 
        + \delta g \partial\delta g \color{black} 
        +  (\partial\delta g)^2 \color{black}
        \biggr.\nonumber\\&\biggl.
        +  (\partial\delta\varphi)^2 \color{black} 
        +  \partial\delta\varphi\partial^2\delta\varphi \color{black}
        \biggr.\nonumber\\&\biggl.
        +  (r_s/r_c)^{\frac{1}{2}}\delta g\partial\delta\varphi \color{black} 
        \biggr.\nonumber\\&\biggl.
        + (r_s/r_c)^{\frac{1}{2}}\partial\delta g\partial\delta\varphi \color{black} 
    \biggr].
\end{align}
In this schematic representation we have omitted all indices, but it should be understood that $\delta g$ represents the metric perturbation with both indices lowered. Also note that we have integrated by parts to remove terms such as $\delta g \partial^2\delta g$. Note that terms in the first line arise from perturbing the Einstein-Hilbert term, while the remaining contributions shown all come from the cubic Galileon interaction. The standard scalar kinetic term only provides sub-leading corrections to the terms shown. We stress that in deriving the form of \eqref{eq:QuadraticActionSchematic} we have only isolated the leading order contributions to each type of term (e.g. to $(\partial\delta g)^2$ or to $\delta g\partial\delta\varphi$), but not assessed the relative importance of different terms yet.

Several additional useful observations can already be made from this schematic form of the leading order quadratic action. Firstly, the terms coupling the scalar and metric perturbations are suppressed by a factor $\sqrt{r_s/r_c}$. Therefore, at leading order, the scalar and metric perturbations decouple near the black hole horizon. This is an important difference to e.g. the coupling of perturbations in Horndeski-type scalar-tensor theories around a Schwarzschild metric background with no (background) scalar hair \cite{Tattersall:QNMsHorn}.\footnote{
Note that this remains true for general stationary metric and radial scalar backgrounds, i.e. scalar and metric perturbations generically couple around such backgrounds in the even sector \cite{Kobayashi:2014wsa}.
} 
We can demonstrate this decoupling more precisely by explicitly demixing the scalar and metric perturbations in the full quadratic action through a shift in the metric perturbation. This shift will then be explicitly suppressed near the black hole horizon -- we show this explicitly in appendix \ref{sec:AppendixScalarDecoupling}. Secondly, as already implied above, the scalar kinetic ($\contwo$) term does not contribute to any leading order term in the quadratic action, and instead the leading order contribution to the kinetic term of scalar perturbations comes from the cubic Galileon $\conthree$ term (as a consequence of Vainshtein screening). This has the important consequence that the sign of $\contwo$ will not matter for stability of perturbations near the black hole horizon. Finally, since only the Einstein-Hilbert term contributes to the metric part of the quadratic action at the leading order, the cubic Galileon terms only affect the metric quasi-normal modes indirectly through their effect on the background metric solution. Therefore, using a stealth solution for the metric\footnote{
`Stealth solution' is standard nomenclature for a metric solution that exactly resembles a GR solution in this context, while the scalar may follow some non-trivial background solution.
} 
would not lead to modifications away from standard GR behaviour at leading order in the dynamics of metric perturbations.

In section \ref{sec:QuadraticActionsFull} we will use our approximate background solutions \eqref{eq:ShortRangeSolutions} to find the quadratic Lagrangians for the perturbations. In the above discussion we focused on the approximate behaviour of perturbations in the short-distance regime (i.e. near the black hole horizon).  However, in section \ref{sec:ShortRangeLimit} we remarked that the approximate solutions we found are accurate for a significantly larger range, so a natural question is up to which radius the leading order quadratic action \eqref{eq:QuadraticActionSchematic} found here is applicable. A more involved analysis which we relegate to appendix \ref{sec:AppendixScalarDecoupling} shows that \eqref{eq:QuadraticActionSchematic} including the demixing of $\delta g$ and $\delta\varphi$ in the leading order is applicable as long as $x\ll(r_c/r_s)^{2/3}$. Note that this is exactly the same range for which our approximate background solutions are accurate.

\subsection{Modified quadratic actions in the Regge-Wheeler gauge} \label{sec:QuadraticActionsFull}

As described above, we substitute the perturbed metric and scalar field from equation \eqref{eq:NormalisedPerturbations} into the action and expand to second order in the perturbations to find the quadratic action. Next, we expand the perturbations in terms of spherical harmonics and use the gauge freedom to fix to the \textit{Regge-Wheeler gauge} \cite{ReggeWheeler,Maggiore:2018sht}.
In this gauge, the perturbations have the forms
\begin{widetext}
\begin{align}
    \label{eq:ReggeWheelerGauge}
    \delta g_{\mu\nu}dx^\mu dx^\nu&=
    \sum_{\ell=0}^{\infty}\sum_{m=-\ell}^{\ell}\left[
        h(x)dy^2H_{\ell m}^{(0)}(y,x)
        +\frac{dx^2}{f(x)}H_{\ell m}^{(2)}(y,x)
    \right]Y_{\ell m}(\theta,\phi),
    \nonumber\\
    &+\sum_{\ell=1}^{\infty}\sum_{m=-\ell}^{\ell}\left[
        2dydxH_{\ell m}^{(1)}(y,x)
        -2dyd\theta\frac{1}{\sin\theta}h_{\ell m}^{(0)}(y,x)\partial_\phi
        +2dyd\phi\sin\theta h_{\ell m}^{(0)}(y,x)\partial_\theta
    \right]Y_{\ell m}(\theta,\phi),
    \nonumber\\
    &+\sum_{\ell=2}^{\infty}\sum_{m=-\ell}^{\ell}\left[
        x^2d\Omega^2K_{\ell m}(y,x)
        -2dxd\theta\frac{1}{\sin\theta}h_{\ell m}^{(1)}(y,x)\partial_\phi
        +2dxd\phi\sin\theta h_{\ell m}^{(1)}(y,x)\partial_\theta
    \right]Y_{\ell m}(\theta,\phi),
    \nonumber\\\nonumber\\
    \delta\varphi&=\sum_{\ell=0}^{\infty}\sum_{m=-\ell}^{\ell}\delta\varphi_{\ell m}(y,x)Y_{\ell m}(\theta,\phi).
\end{align}
\end{widetext}
Here, $h_{\ell m}^{(0)}$ and $h_{\ell m}^{(1)}$ are fields describing metric perturbations of odd parity, while $H_{\ell m}^{(0)}$, $H_{\ell m}^{(1)}$, $H_{\ell m}^{(2)}$ and $K_{\ell m}$ describe metric perturbations of even parity. The (even parity) scalar perturbations are (with a slight abuse of notation) described by the fields $\delta\varphi_{\ell m}$. $Y_{\ell m}$ are the usual spherical harmonic functions. For $\ell\geq2$ the perturbations are fully gauge fixed, while some additional gauge freedom remains for $\ell=1$ and $\ell=0$. These additional gauge freedoms and the gauge transformations leading to equation \eqref{eq:ReggeWheelerGauge} are described in more detail in appendix \ref{sec:AppendixReggeWheelerGauge}. Here, we will focus on the $\ell\geq2$ part of the quadratic action for the metric perturbation. The monopole ($\ell=0$) and dipole ($\ell=1$) metric perturbations contain no dynamical degrees of freedom so are not of interest to us here. We describe these in more detail in appendix \ref{sec:AppendixMonopoleDipole}. For the scalar perturbations, there are dynamical degrees of freedom in the monopole and dipole perturbations and the discussion in this section will apply to all $\ell$ for the scalar perturbations.

We then substitute the approximate solutions for the background fields and the Regge-Wheeler perturbations into the quadratic action and focus on the leading order part in the short-range limit as discussed above.  Fields with different spherical harmonic numbers separate due to the spherical symmetry of the background which means that we can integrate over the angular dependence. We observe that the remaining quadratic action separates into three decoupled contributions:
\begin{align}
    \mathcal L_{\text{leading order}}^{(2)}=\mathcal L_{\text{odd}}^{(2)}+\mathcal L_{\text{even}}^{(2)}+\mathcal L_{\text{scalar}}^{(2)}.
\end{align}
Each contribution describes a single dynamical degree of freedom. We describe these degrees of freedom with the fields $w_{\ell m}(y,x)$, $\psi_{\ell m}(y,x)$ (for $\ell\geq2$) and $\delta\varphi_{\ell m}(y,x)$ (for all $\ell$). These are related to our original Regge-Wheeler fields by
\begin{align}
    w_{\ell m}&=\dot{h}_{\ell m}^{(1)}-h_{\ell m}'^{(0)}+\frac{2}{x}h_{\ell m}^{(0)},
    \nonumber\\
    \psi_{\ell m}&=H_{\ell m}^{(2)}+\frac{(\ell+2)(\ell-1)x+3}{2(x-1)}K_{\ell m}-x K_{\ell m}'
    \nonumber\\
    \delta\varphi_{\ell m}&=\delta\varphi_{\ell m},
\end{align}
where a dot denotes a partial derivative with respect to our dimensionless time coordinate $y$ and primes denote partial derivatives with respect to the dimensionless radial coordinate $x$. In terms of these fields, the contributions to the quadratic action are
\begin{widetext}
\begin{align}
    \label{eq:QuadraticActionsFull}
    \frac{4\sqrt{\approxcon}(\ell+2)(\ell-1)}{\ell(\ell+1)}\mathcal L_{\text{odd}}^{(2)}
    &=\frac{x^3}{\approxcon(x-1)}\dot{w}^2
    -x(x-1)w'^2
    -\frac{\ell(\ell+1)x-4}{x}w^2,
    \nonumber\\\nonumber\\
    \frac{\sqrt{\approxcon}\ell(\ell+1)}{(\ell+2)(\ell-1)}\mathcal L_{\text{even}}^{(2)}
    &=\frac{x^3(x-1)}{\left((\ell+2)(\ell-1)x+3\right)^2}\dot{\psi}^2
    -\approxcon\frac{x(x-1)^3}{\left((\ell+2)(\ell-1)x+3\right)^2}\psi'^2
    \nonumber\\\nonumber\\
    &-\approxcon\frac{(x-1)\left(3-\left(\ell(\ell+1)+4\right)x+\ell(\ell+1)(\ell+2)(\ell-1)x(x-1)\right)}{\left((\ell+2)(\ell-1)x+3\right)^3}\psi^2,
    \nonumber\\\nonumber\\
    \mathcal L_{\text{scalar}}^{(2)}
    &=\mp\frac{x^2(6x-5)(2x-3)}{\approxcon(x-1)^2(4x-3)^{\frac{3}{2}}}\dot{\delta\varphi}^2
    -2\text{sign}(\alpha_1)\frac{x}{\sqrt{\approxcon}(x-1)}\dot{\delta\varphi}\delta\varphi'
    \pm\sqrt{4x-3}\delta\varphi'^2
    \nonumber\\\nonumber\\
    &
    \pm2\ell(\ell+1)\frac{2x-3}{x(4x-3)^{\frac{3}{2}}}\delta\varphi^2,
\end{align}
\end{widetext}
where we have omitted the sums over $\ell$ and $m$ as well as the angular indices of the fields for simplicity. We show the quadratic actions where the $\pm$-signs indicate the $\Xi_\pm$ branch, with the upper sign for the $\Xi_+$ background and the lower sign for the $\Xi_-$ background. Note that this quadratic action, which is valid near the black hole horizon to the leading order in $r_s/r_c$, only depends on the theory parameters through the value of $\approxcon$, the sign of $\alpha_1$ and the branch of the local solution for $\Xi$. We show the derivation of the quadratic actions in \eqref{eq:QuadraticActionsFull} in detail in appendix \ref{sec:AppendixQuadraticActions}.

\subsection{Stability against perturbations of the black hole background} \label{sec:Stability}

The conventional method of studying (in)stabilities is to calculate the Hamiltonian density and checking that it is positive-definite everywhere. Doing so directly from our starting points \eqref{eq:QuadraticActionsFull}, we here obtain apparent tachyon instabilities for the even metric and scalar perturbations as well as apparent ghost and/or gradient instability for the scalar perturbations. However, this way of identifying ghost and gradient instabilities is not robust and the apparent instabilities we find might be simply due to a poor choice of coordinates \cite{Babichev:2018uiw,Sawicki:2024ryt}. We can explain this as follows: for a minimally coupled field $a$, the kinetic part of the quadratic action will be $-\frac{1}{2}g^{\mu\nu}\partial_\mu a\partial_\nu a$. We can generalise this to write the kinetic part of the quadratic action for e.g. $w$ as $-\frac{1}{2}S^{\mu\nu}\partial_\mu w\partial_\nu w$. We then obtain a set of \textit{effective} metrics ($S^{-1}_{\mu\nu}$) that govern the propagation of the perturbations. For the propagation problem to be well defined, we need timelike and spacelike coordinates that are timelike and spacelike with respect to every effective metric simultaneously. This is covered in more detail in appendix \ref{sec:AppendixHamiltonianBoundedness}. In our $(y,x)$ coordinates $\partial_x$ is not spacelike with respect to the effective metric for the scalar perturbations. We can fix this by defining a new timelike coordinate $\tilde{y}$:
\begin{align}
    \tilde{y}=y\pm\text{sign}(\alpha_1)\int dx\frac{x}{\sqrt{\approxcon}(x-1)\sqrt{4x-3}},
\end{align}
which also happens to be the timelike coordinate for which $\mathcal L_{\text{scalar}}^{(2)}$ is diagonal. With this coordinate, $\partial_x$ is spacelike with respect to both the spacetime metric and the effective metric for scalar perturbations in the region outside the black hole horizon. Using this coordinate, the Hamiltonian can be written
\begin{widetext}
\begin{align}   
    \frac{4\sqrt{\approxcon}(\ell+2)(\ell-1)}{\ell(\ell+1)}\mathcal H_{\text{odd}}^{(2)}
    &=\frac{4x^3}{\approxcon(4x-3)}(\partial_{\tilde{y}}w)^2
    +x(x-1)w'^2
    +\frac{\ell(\ell+1)x-4}{x}w^2,
    \nonumber\\\nonumber\\
    \frac{\sqrt{\approxcon}\ell(\ell+1)}{(\ell+2)(\ell-1)}\mathcal H_{\text{even}}^{(2)}
    &=\frac{4x^3(x-1)^2}{\left((\ell+2)(\ell-1)x+3\right)^2(4x-3)}(\partial_{\tilde{y}}\psi)^2
    +\approxcon\frac{x(x-1)^3}{\left((\ell+2)(\ell-1)x+3\right)^2}\psi'^2
    \nonumber\\\nonumber\\
    &+\approxcon\frac{(x-1)\left(3-\left(\ell(\ell+1)+4\right)x+\ell(\ell+1)(\ell+2)(\ell-1)x(x-1)\right)}{\left((\ell+2)(\ell-1)x+3\right)^3}\psi^2,
    \nonumber\\\nonumber\\
    \mathcal H_{\text{scalar}}^{(2)}
    &=\mp\frac{12x^2}{\approxcon(4x-3)^{\frac{3}{2}}}(\partial_{\tilde{y}}\delta\varphi)^2
    \mp\sqrt{4x-3}\delta\varphi'^2
    \mp2\ell(\ell+1)\frac{2x-3}{x(4x-3)^{\frac{3}{2}}}\delta\varphi^2.
\end{align}
\end{widetext}
In these coordinates, the kinetic part of the Hamiltonian is positive-definite for the $\Xi_-$ background. There is therefore no ghost or gradient instability in the region outside the black hole horizon for this background. However, for the $\Xi_+$ background there remains a ghost instability in the scalar modes, as was also shown in \cite{Babichev:2012re}. This is a true instability that cannot be removed through a coordinate transformation, since there exists no direction which is timelike with respect to all effective metrics in this case. We show this in detail in appendix \ref{sec:AppendixHamiltonianBoundedness}.
Note from the quadratic Lagrangian for scalar perturbations \eqref{eq:QuadraticActionsFull}, that the effective metric for these perturbations does not have a horizon at $r=r_s$.\footnote{
There will most likely be a horizon for scalar perturbations at some $r\leq3r_s/4$, but this lies outside the region where our quadratic actions are accurate.
} 
Scalar perturbations can therefore travel past the black hole horizon and we thus need to check the stability of scalar perturbations in the black hole interior as well. Extending our analysis down to the minimum radius where our quadratic actions are accurate ($r\approx3r_s/4$), we find that the $\Xi_-$ background has a ghost instability behind the black hole horizon -- see appendix \ref{sec:AppendixHamiltonianBoundedness} for details.\footnote{We thank Gilles Esposito-Far\`ese for discussions on this point.}
Therefore, neither branch is stable against scalar perturbations in the vicinity of the black hole.\footnote{
Furthermore, apparent tachyon instabilities remain in the even metric and scalar modes since the $\psi^2$ and $\delta\varphi^2$ terms in the Hamiltonian are not positive-definite. However, this is the case for the metric modes when expressed in these variables even in GR. A thorough analysis of the stability as in \cite{Wald:1979lth} in this case shows that the background (despite the apparent tachyonic instability) is in fact stable against even parity metric perturbations. For the scalar perturbations a more involved analysis would be required to study their tachyon stability.
}

\section{Quasi-stationary black hole solutions} \label{sec:QuasiStationaryBlackHoles}

In the previous section we found that the background considered there is unstable in the short-range limit. This suggests that, starting with the spherically symmetric and static ansatz \eqref{eq:Ansatz}, no physical and self-consistent solution exists that yields a self-accelerating (de Sitter-like) cosmology at large scales and describes a stable black hole on short scales.\footnote{
Note that the stability against scalar perturbations has already been studied for the $r_s \ll r \ll \left(r_sr_c^2\right)^{1/3}$ regime in \cite{Babichev:2012re}, uncovering the ghost instability in the $\Xi_+$ branch, but naturally (given the $r$--regime) not the instability behind the black hole horizon in the $\Xi_-$ branch.
}
This motivates exploring whether the ansatz \eqref{eq:Ansatz} may have been too restrictive. Indeed, in view of the results of the previous sections and given the inherent time dependence of the scalar field $\varphi$, it is natural to suspect that a self-consistent solution requires going beyond a purely static ansatz for the metric. In this section we therefore explore the possibility of finding stable solutions when relaxing the requirement for static black holes, instead considering \textit{quasi-stationary} solutions with a suppressed time dependence induced by a non-zero shift-symmetry current $J^r$ \eqref{eq:ShiftSymmetryCurrent}. This approach is further motivated by earlier results including a suppressed time dependence to find well-behaved black hole solutions in other theories. For example, \cite{DeFelice:2022xvq} shows that strong coupling in quadratic DHOST theories \cite{deRham:2019gha} can be avoided by introducing a \textit{Scordatura} term \cite{Motohashi:2019ymr} into the action. \cite{DeFelice:2022qaz} shows that in such a theory, no stealth solutions exist, but approximately stealth ones with a small time dependence on Hubble timescales do. Meanwhile, \cite{Rosen:2017dvn} shows that issues with a singular horizon and non-physical asymptotes can be overcome by introducing a time dependence with a characteristic time scale of the inverse graviton mass. Finally, while this paper was being completed, \cite{Lara:2025hqh} derived stability and regularity bounds for black holes with cosmological asymptotes in $K$-essence theories with a scalar Gauss-Bonnet term, which include a lower bound for $J^r$, inducing a time-dependence in the metric -- see Eq. \eqref{eq:ConservedCurrent} and \eqref{eq:CharacteristicTimeScale}.

\subsection{Non-zero shift-symmetry current} \label{sec:NonZeroShiftSymmetryCurrent}

Before incorporating explicit time dependence in our ansatz, it is instructive to briefly reconsider the static equations of motion \eqref{eq:EoMsDimensionful}. As described in section \ref{sec:CosmologicalSolution}, the first equation in \eqref{eq:EoMsDimensionful}, i.e. the $(tr)$ equation, can be written as $J^r=0$. Therefore, a non-zero shift-symmetry current cannot be included while satisfying this equation. Let us for a moment put this equation aside and instead consider the dependent but less constraining equation of motion for the scalar ($\frac{\delta S}{\delta\varphi}$), which can be written as a current conservation
\begin{align}
\label{eq:ConservedCurrent}
    \partial_r\left(\sqrt{-g}J^r\right)&=0
    &\Rightarrow& &r^2\sqrt{\frac{h}{f}}J^r&=\alpha_{BH}\mpl r_s,
\end{align}
where the current is as given in \eqref{eq:ShiftSymmetryCurrent}. We have introduced a new (dimensionless) constant $\alpha_{BH}$, following the discussion in \cite{Babichev:2025ric}. Replacing the $(tr)$ equation with the scalar equation for now, our equations of motion in dimensionless form are
\begin{align}
    \label{eq:EoMsQuasiStationary}
    \alpha_1(x^4h)'\frac{f}{h}\Xi^2+2x^4h\Xi-\alpha_1x^4h'&=\alpha_4 x^2h\sqrt{\frac{h}{f}},
    \nonumber\\
    \alpha_2x^2\left[1-\frac{f}{h}\Xi^2\right]+2xfh'+2h(f-1)&=0,
    \nonumber\\
    \left[1-\frac{f}{h}\Xi^2\right]\left[\alpha_2x^2\sqrt{\frac{h}{f}}+\alpha_1\alpha_2\left(x^2\sqrt{\frac{f}{h}}\Xi\right)'\right]&
    \nonumber\\
    -2xh^2\left(\sqrt{\frac{f}{h}}\right)'&=0,
\end{align}
where we have introduced a new dimensionless parameter $\alpha_4$
\begin{align}
\label{eq:alpha4Def}
    \alpha_4\equiv\alpha_{BH}\frac{\mpl r_s}{\contwo q r_0^2}\sim\alpha_{BH}\frac{r_c}{r_0}\frac{r_s}{r_0}.
\end{align}
A non-zero $J^r$ adds a source term to the first (scalar) equation of motion controlled by $\alpha_4$. As we will show in sections \ref{sec:BlackHoleSolutionsQuasiStationary} and \ref{sec:StabilityQuasiStationary}, this changes the background scalar profile and, most importantly, stabilizes the $\Xi_+$ branch.

Of course, the above treatment has not been fully self-consistent so far. Solutions to \eqref{eq:EoMsQuasiStationary} will not be solutions to the full static equations of motion for $\alpha_4\neq0$, since the $(tr)$ equation that we replaced by the $\frac{\delta S}{\delta\varphi}$ equation of motion \eqref{eq:EoMsDimensionless1} will not be satisfied. However, as we will show in what follows, allowing for time dependence in the metric ansatz introduces additional terms containing time derivatives to all equations. It is then possible to find quasi-stationary solutions, where the time dependence is such that the full time-dependent form of the $(tr)$ equation is satisfied, while the time-dependent terms in the other equations are suppressed such that \eqref{eq:EoMsQuasiStationary} is \textit{approximately} satisfied. And so, as we will show in detail below, \eqref{eq:EoMsQuasiStationary} will turn out to capture the leading order contributions to the equations of motion in this case.

\subsection{Time-dependent equations of motion} \label{sec:TimeDependentEoMs}

As outlined above, we will now investigate solutions where the metric background incorporates a time dependence driven by a non-zero shift-symmetry current. To do this, we change our ansatz \eqref{eq:Ansatz} to consider the most general spherically symmetric (but {\it not} static) metric and scalar field. The metric will then take the same form as in the static case, with the exception that $h$ and $f$ now depend on time. We add a general time dependence to the scalar field on top of the linear dependence by making $\Psi(r)$ time-dependent as well.\footnote{
Note that in principle one could try an ansatz where the scalar field only has linear time dependence as in \eqref{eq:Ansatz}, but an investigation of the time dependence in the vicinity of the black hole showed that no solutions consistent with this ansatz and our assumptions about the orders of magnitude of the solution exist. The time dependence of the black hole metric therefore induces a more complicated time dependence in the scalar field as well. This is described in more detail in appendix \ref{sec:AppendixScalarAnsatzTimeDependence}.
}
We then have
\begin{align}
    \label{eq:TimeDependentAnsatz}
    ds^2&=-h(t,r)dt^2+\frac{dr^2}{f(t,r)}+r^2d\Omega^2,
    \nonumber\\
    \varphi&=q t +q\Psi(t,r), \quad \text{where} \quad
    \Psi(t,r) \equiv \int dr\frac{\Xi(t,r)}{h(t,r)}.
\end{align}

Substituting this into the equations of motion, we find the same structure as described in section \ref{sec:CosmologicalSolution}. 
We can take linear combinations of the $(tt)$, $(tr)$ and $(rr)$ equations of motion due to variation with respect to the inverse metric and write these (in dimensionless form) as \eqref{eq:EoMsDimensionless1}-\eqref{eq:EoMsDimensionless3}, with newly added source-terms on the right-hand side due to time dependence:
\begin{align}
    \label{eq:EoMsTimeDependentFull}
    \alpha_1(x^4h)'\frac{f}{h}\Xi^2+2x^4h\Xi-\alpha_1x^4h'&=\mathcal{E}_1^{(t)},
    \nonumber\\
    \alpha_2x^2\left[1-\frac{f}{h}\Xi^2\right]+2xfh'+2h(f-1)&=\mathcal{E}_2^{(t)},
    \nonumber\\
    \left[1-\frac{f}{h}\Xi^2\right]\left[\alpha_2x^2\sqrt{\frac{h}{f}}+\alpha_1\alpha_2\left(x^2\sqrt{\frac{f}{h}}\Xi\right)'\right]&
    \nonumber\\
    -2xh^2\left(\sqrt{\frac{f}{h}}\right)'&=\mathcal{E}_3^{(t)}.
\end{align}
The $\mathcal{E}_i^{(t)}$ terms encoding new contributions related to the non-static metric solution are cumbersome, but we give their leading orders in the vicinity of the black hole here:
\begin{align}
    \label{eq:TimeDependentSourceTerms}
    \mathcal{E}_1^{(t)}
    &=\frac{2}{\alpha_2}x^3h^2\frac{\dot{f}}{f}
    +\dots,
    \nonumber\\
    \mathcal{E}_2^{(t)}
    &=-2 x \Xi \dot{f}
    +\dots,
    \nonumber\\
    \mathcal{E}_3^{(t)}
    &=4x\Xi\dot{f}
   +\dots,
\end{align}
where we have omitted contributions suppressed by factors of $r_s/r_c$ through the $\alpha_1$ and $\alpha_2$ parameters. If all time-derivatives vanish, $\mathcal{E}_i^{(t)}=0$ and we recover the static case. The equation of motion from variation with respect to the scalar can again be written as a conservation of the shift-symmetry current ($\nabla_\mu J^\mu=0$) 
given in \eqref{eq:ShiftSymmetryCurrent}. Note that with our new ansatz, $J^t$ and $J^r$ will contain additional terms compared to the static case due to time-derivatives of the metric from e.g. $\Box\varphi$. We can write the conservation equation in components as
\begin{align}
    \partial_t\left(\sqrt{-g}J^t\right)+\partial_r\left(\sqrt{-g}J^r\right)=0.
\end{align}

\subsection{\label{sec:BlackHoleSolutionsQuasiStationary}Quasi-stationary background solutions}

We will now consider \textit{quasi-stationary} black hole solutions. Near the black hole horizon, these solutions will have some characteristic time-scale over which they vary $\Gamma^{-1}$ (e.g. $\dot{f}=\partial_yf\sim\Gamma r_s f$). We require that $\Gamma$ is large enough such that the $(tr)$ equation is satisfied, yet small enough such that the time-dependent terms in all other equations of motion are sufficiently suppressed such that \eqref{eq:EoMsQuasiStationary} is approximately satisfied. We will show the compatibility of these two bounds near the black hole horizon, where we can use that $f$, $h$, $\Xi$ and their radial derivatives are $\order{1}$ since this is the case for the static solutions and we are looking for quasi-stationary solutions resembling these (or at least having the same order of magnitude).

Firstly, we look at the $(tr)$ equation (i.e. the first equation of \eqref{eq:EoMsTimeDependentFull}). From \eqref{eq:EoMsQuasiStationary} we know that due to the non-zero shift-symmetry current the left-hand side will be $\sim\alpha_4\sim \alpha_{BH} r_c/r_s$ -- to derive this note that $\alpha_{BH}$ is given in \eqref{eq:alpha4Def}, while we empirically know from above that $x$ and $h$ are ${\cal O}(1)$ in the static solution, which we remain close to due to quasi-stationarity. From \eqref{eq:TimeDependentSourceTerms} we deduce that ${\mathcal E}_1^{(t)}\sim(r_c/r_s)^2\Gamma r_s$, where we recall that $\alpha_2 \sim (r_s/r_c)^2$ and we again use that $x \sim {\cal O}(1) \sim h$. Therefore, in order for the time dependence to compensate the non-vanishing part of the left-hand side of the equation near the black hole horizon we require
\begin{align}
    \label{eq:CharacteristicTimeScale}
    \Gamma\sim\frac{\alpha_{BH}}{r_c},
\end{align}
as is also found in \cite{Babichev:2025ric} by considering the energy-momentum tensor of the scalar field. Similarly, requiring that the time-dependent terms in the remaining three equations are suppressed -- recall we are looking for quasi-stationary solutions where the impact of the time-dependent terms is present where required in \eqref{eq:EoMsTimeDependentFull}, but otherwise minimal -- we find an upper bound
\begin{align}
    \label{eq:AlphaBHBound}
    |\alpha_{BH}|\ll r_c/r_s.
\end{align}

Since we are interested in solutions resembling the GR Schwarzschild solution (and therefore our previous static solution), we require that the accretion ($\alpha_4$) term in the scalar equation of motion is at most of the same order as the other terms.\footnote{
If it were much larger, it would be the single dominant term in the equation giving $h(x)=0$ or $f(x)\rightarrow\infty$ at leading order. This could be overcome if $\Xi(x)\gg1$, but we will not consider that case here since we are interested in background solutions resembling the static ones.
}
It can then at most contribute to the leading order of the equation, giving $\alpha_4\lesssim\alpha_1$. Note that in this case, the accretion term can only contribute to the leading order of the equations near the black hole horizon due to the $x^{-2}$ suppression compared to the other terms. We therefore do not expect the presence of this term to affect the branching structure in the intermediate regime. Thus, one should not expect that including time dependence will sufficiently alter the conclusions for the static case and allow the existence of a regular solution that connects the $\Xi_-$ branch near the black hole horizon to the desired cosmological limit. However, the accretion term can affect the background solutions and hence the stability analysis near the black hole, so its inclusion can potentially stabilise the short-range $\Xi_+$ branch. We will now investigate this possibility further.

Given the above discussion, we are particularly interested in choices of $\alpha_4$ for which the accretion term contributes to the leading order of the equations of motion near the black hole. We therefore take $\alpha_4\sim\alpha_1$. Using \eqref{eq:CharacteristicTimeScale} we find $\Gamma\sim H$, so the black hole evolves on very long cosmological time scales of the order of a Hubble time, which is indeed much longer than the typically relevant black hole time scales. In this case $\alpha_{BH}\sim 1$ so it satisfies the bound given by \eqref{eq:AlphaBHBound}, required for quasi-stationarity of the solutions.\footnote{
Strictly speaking this bound only guarantees that the time dependence is sufficiently suppressed near the black hole horizon. However, due to the $r^{-2}$ dependence of the shift-symmetry current, it is natural to expect its effect to be even further suppressed at larger scales. In appendix \ref{sec:AppendixTimeDependenceSuppression} we confirm this suppression across all length scales explicitly for the numerical solution we find in section \ref{sec:FullStableQuasiStationaryBlackHoleSolution}.
} 

We then solve the quasi-stationary background equations \eqref{eq:EoMsQuasiStationary} at the leading order and find the same approximate background solutions for $f$ and $h$ as before in \eqref{eq:ShortRangeSolutions}, but an altered solution for $\Xi$:
\begin{align}
    \label{eq:XiShortRangeQuasiStationarySolution}
    \Xi_\pm(x)=\pm\text{sign}(\alpha_1)\sqrt{\frac{\approxcon}{4x-3}\left(1+\frac{\alpha_4\sqrt{\approxcon}(x-1)}{\alpha_1 x}\right)},
\end{align}
with the important difference compared to the static case that the integration constants $\approxcon=\approxcon(t)$ and $r_s=r_s(t)$ (which is hidden in $x$, $\alpha_1$ and $\alpha_4$) are now functions of time. By considering the leading order of the time-dependent $(tr)$ equation 
\begin{align}
    \label{eq:RadiusEvolution}
    \frac{1}{r_s}\frac{d r_s}{dt}\approx-\frac{\alpha_2\alpha_4}{2r_s\sqrt{\approxcon}}\sim\frac{\alpha_{BH}}{r_c}.
\end{align}
So $r_s$ would evolve on the expected time-scale $\Gamma$ and for a negative (positive) value of $\alpha_4$ the black hole would grow (shrink). For a negative value of $\alpha_4$ the time dependence can therefore be thought of as accretion onto the black hole of energy associated with the scalar field. Additionally, there is no constraint on $\dot{\approxcon}$ at leading order so for simplicity we take $\approxcon$ to be constant. Studying the accuracy of the approximate solution in the same way as for the static case -- see appendix \ref{sec:AppendixShortRangeLimitAccuracy} -- we find again that these approximate solutions are accurate as long as $x\ll(r_c/r_s)^{2/3}$.

Finally, note that for $1+\alpha_4\sqrt{\approxcon}/\alpha_1<0$, $\Xi$ becomes complex for
\begin{align}
    x>\left(1+\frac{\alpha_1}{\alpha_4\sqrt{\approxcon}}\right)^{-1}.
\end{align}
So to ensure a physical, real scalar solution, we require that this region lies outside the regime of validity of the approximate solutions and hence
\begin{align}
    1+\frac{\alpha_1}{\alpha_4\sqrt{\approxcon}}\ll (r_s/r_c)^{2/3}
\end{align}
in the case where $1+(\alpha_4/\alpha_1) \sqrt{\approxcon}<0$. Regardless of the sign of this term, we find a similar but more general bound resulting from the stability analysis below.

\subsection{Stability of quasi-stationary black hole solutions} \label{sec:StabilityQuasiStationary}

Including the accretion of scalar shift-symmetry charge gives solutions with (suppressed) time dependence. The radial parts of the solutions for $f$ and $h$ are identical to those in the static case. However, the solution for the scalar background $\Xi$ is altered depending on the value of $\alpha_4$. We saw earlier that at leading order, the quadratic actions for the metric perturbations do not depend on the scalar background. Therefore, in the time-dependent case considered here the quadratic actions for the even and odd metric perturbations will be identical to those in \eqref{eq:QuadraticActionsFull}. However, the quadratic action for scalar perturbations does change, which will modify the stability criteria.

We follow the same procedure as in section \ref{sec:Stability}, now using our altered background, to find the leading order quadratic action for the scalar perturbations, as well as the Hamiltonian:
\begin{widetext}
\begin{align}
    \label{eq:ScalarQuadraticActionQuasiStationary}
    \mathcal L_{\text{scalar}}^{(2)}
    &=\mp\frac{x^2}{\approxcon(4x-3)^{3/2}(x-1)^2}\left(12(x-1)^2\xi(x)-\frac{4x-3}{\xi(x)}\right)\dot{\delta\varphi}^2
    -2\lambda\frac{x}{\sqrt{\approxcon}(x-1)}\dot{\delta\varphi}\delta\varphi'
    \nonumber\\\nonumber\\
    &\pm\sqrt{4x-3}\xi(x)\delta\varphi'^2
    \pm\ell(\ell+1)\frac{(4x^2-6x+3)\xi(x)^2-(4x-3)}{x(x-1)(4x-3)^{3/2}\xi(x)}\delta\varphi^2,
    \nonumber\\\nonumber\\
    \mathcal H_{\text{scalar}}^{(2)}
    &=\mp\frac{x^2}{\approxcon(4x-3)^{3/2}(x-1)^2}\left(12(x-1)^2\xi(x)-\frac{4x-3}{\xi(x)}\right)\dot{\delta\varphi}^2
    \mp\sqrt{4x-3}\xi(x)\delta\varphi'^2
    \nonumber\\\nonumber\\
    &\mp\ell(\ell+1)\frac{(4x^2-6x+3)\xi(x)^2-(4x-3)}{x(x-1)(4x-3)^{3/2}\xi(x)}\delta\varphi^2,
    \nonumber\\\nonumber\\
    \xi(x)&=\sqrt{1+\frac{\alpha_4\sqrt{\beta}}{\alpha_1}\frac{x-1}{x}}.
\end{align}
\end{widetext}
Where the $\pm$ signs indicate the $\Xi_\pm$ branch and $\lambda=\text{sign}(\alpha_1)$. This reduces to the static case for $\alpha_4\rightarrow0$. We want to investigate the effect of time dependence on the stability of the $\Xi_+$ branch, since this is the branch that is required in the vicinity of the black hole to connect to the cosmological asymptotes on large scales. Again, we see that in these coordinates, the Hamiltonian is not bounded from below for any value of $\alpha_4$. However, as in section \ref{sec:Stability} this is due to using the `wrong' coordinates. A thorough analysis of the stability (see appendix \ref{sec:AppendixStabilityQuasiStationary}) shows that there is a stable region between $x=1$ and some $x=x_{\text{max}}$ for $\alpha_4\sqrt{\approxcon}/\alpha_1\leq3$, where
\begin{align}
    \label{eq:xMax}
    x_{\text{max}}=\frac{(3+2\sqrt{3})\sqrt{3-\frac{\alpha_4\sqrt{\approxcon}}{\alpha_1}}+\sqrt{3}\left(3+2\sqrt{3}-\frac{\alpha_4\sqrt{\approxcon}}{\alpha_1}\right)}{2\left(1+\frac{\alpha_4\sqrt{\approxcon}}{\alpha_1}\right)}.
\end{align}
Since this stability analysis is only accurate for $x\ll(r_c/r_s)^{2/3}$, we require that $x_\text{max}\gsim (r_c/r_s)^{2/3}\gg1$. This can be achieved by taking $\alpha_4\sqrt{\approxcon}/\alpha_1$ close to $-1$. In that case we find
\begin{align}
    x_\text{max}\sim\left(1+\frac{\alpha_4\sqrt{\approxcon}}{\alpha_1}\right)^{-1}.
\end{align}
So for stability against scalar perturbations in the region where our analysis is valid we require
\begin{align}
    \label{eq:StabilityConstraint}
    \frac{\alpha_4\sqrt{\approxcon}}{\alpha_1}=-1+\epsilon, \quad \text{where} \quad |\epsilon|\lsim(r_s/r_c)^{2/3}
\end{align}
Note that in order for the long-range metric to describe an expanding universe in the frame where the scalar field is explicitly homogeneous we require $q=+q_0$ (see section \ref{sec:CosmologicalSolution}), which implies $\alpha_1>0$. Therefore, in this stable case $\alpha_4<0$, and it then follows from \eqref{eq:RadiusEvolution} that the time dependence indeed corresponds to accretion of energy onto the black hole. Furthermore, due to the smallness of $\epsilon$ this stability requirement very tightly constrains the value of $\alpha_4$ in terms of $\beta$ and $\alpha_1$. Lastly note that although equation~\eqref{eq:StabilityConstraint} remains valid as the black hole evolves, the resulting $\alpha_4$ has a small time dependence since this stability requirement is defined for $r_0=r_s$ and $r_s$ is slowly growing in time -- in other words, as the black hole evolves due to accretion, the required accretion rate ($dr_s/dt$) for stabilisation evolves with it -- see equation~\eqref{eq:RadiusEvolution}.

\begin{figure}
    \centering
    \includegraphics[width=\linewidth]{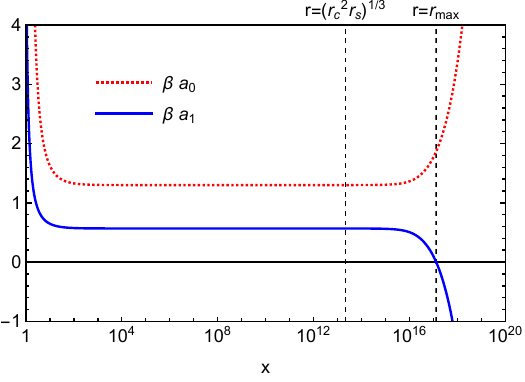}
    \caption{The coefficients $a_0$ and $a_1$ in the kinetic part of the Hamiltonian density \eqref{eq:ExplicitlyStableHamiltonianQuasiStationary} in explicitly stable coordinates for a specific value of the accretion rate ($\alpha_4\sqrt{\approxcon}/\alpha_1=-1+10^{-16}$). For boundedness from below these must be positive. This is the case up to some maximum radius, $r_\text{max}=r_s x_\text{max}$ with $x_\text{max}$ defined in \eqref{eq:xMax}. Therefore, for this choice of $\alpha_4$, $r_\text{max}\gg(r_c^2r_s)^{1/3}$, the radius up to which our stability analysis with approximate short-range solutions is accurate, which means the approximate solutions in the $\Xi_+$ branch are stable in the region that we have studied. Note that fixing $\alpha_4\sqrt{\beta}/\alpha_1=-1$ takes $r_\text{max}\rightarrow\infty$.}
    \label{fig:ExplicitlyStableHamiltonianCoefficients}
\end{figure}

To further demonstrate the stability in this case we can transform to coordinates $(\tilde{y},\tilde{x})$ in which the kinetic part of the Hamiltonian is explicitly bounded from below:
\begin{align}
    \label{eq:ExplicitlyStableCoordinatesQuasiStationary}
    &\tilde{y} = y - \int dx \gamma(x),\quad
    \tilde{x} = y - \int dx \delta(x),
    \nonumber\\\nonumber\\
    \gamma(x) &= \frac{\lambda x}{2\sqrt{\beta}(4x-3)(x-1)\xi}
    \left[\left(\left((4+2\sqrt{3}\right)x
    \right.\right.
    \nonumber\\
    &\quad\quad\quad\quad\quad\quad\quad\left.\left.
    -\left(3+2\sqrt{3}\right)\right)\xi-\sqrt{4x-3}\right]
    \nonumber\\\nonumber\\
    \delta(x) &= \frac{\lambda x}{2\sqrt{\beta}(4x-3)(x-1)\xi}
    \left[-\left(\left((4+2\sqrt{3}\right)x
    \right.\right.
    \nonumber\\
    &\quad\quad\quad\quad\quad\quad\quad\left.\left.
    -\left(3+2\sqrt{3}\right)\right)\xi-\sqrt{4x-3}\right].
\end{align}
In these coordinates, we find a quadratic Hamiltonian density
\begin{align}
    \label{eq:ExplicitlyStableHamiltonianQuasiStationary}
    \mathcal{H}^{(2)}_{\text{scalar,stable}} 
    = a_0(\partial_{\tilde{y}} \delta\varphi)^2
    +a_1(\partial_{\tilde{x}} \delta\varphi)^2
    +a_2\delta\varphi^2,
\end{align}
where the $a_i$ coefficients are cumbersome and given in appendix \ref{sec:AppendixStabilityQuasiStationary}. To demonstrate the boundedness of the kinetic part of the Hamiltonian we plot $a_0$ and $a_1$ in figure \ref{fig:ExplicitlyStableHamiltonianCoefficients} for the example value of $\alpha_4\sqrt{\approxcon}/\alpha_1=-1+10^{-16}$. As expected, we see that these coefficients are positive and therefore the kinetic part of the Hamiltonian is bounded from below up to some $r_\text{max}=r_s x_\text{max}$, where $x_\text{max}$ is given by \eqref{eq:xMax}. For this choice of $\alpha_4$, this $r_\text{max}$ is out of the region where our stability analysis is accurate, so the $\Xi_+$ branch is stable against scalar perturbations in this region. 

In closing this subsection, it is worth mentioning that we have focused on investigating the stability of solutions in the short-range regime here, while the long-range cosmological solution is already known to be free of ghost and gradient instabilities \cite{Kobayashi:2009wr,DeFelice:2011bh}. Further constraints on the model may be obtainable by considering stability bounds from the intermediate regime, which we have analysed numerically here, so this would be interesting to investigate further in the future.

\subsection{Stable accreting black hole solutions with cosmological asymptotes} \label{sec:FullStableQuasiStationaryBlackHoleSolution}

In the previous sections we have shown that in a fully static scenario, the background in the vicinity of the black hole is unstable to scalar perturbations near the black hole horizon. In this section, we have described how a \textit{quasi-stationary} solution driven by accretion of scalar shift-symmetry charge onto the black hole affects the scalar background near the black hole horizon and we showed that for specific values of the accretion parameter $\alpha_4$ the $\Xi_+$ branch becomes stable in the range that we can describe using our approximate solutions. It then remains to be shown that such a stable quasi-stationary solution following the $\Xi_+$ branch near the black hole horizon can be connected to the cosmological large scale asymptotes. To this end, we again turn to numerical integration, as in section \ref{sec:NumericalSolutions}.
\begin{figure*}
    \includegraphics[width=\textwidth]{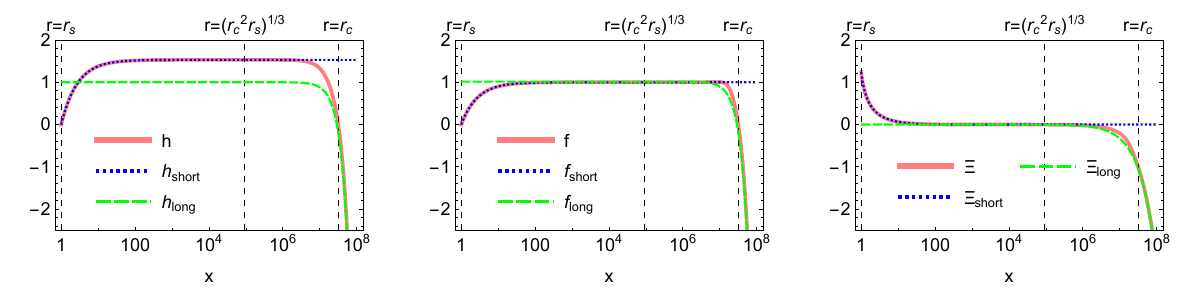}
    \caption{The quasi-stationary numerical solution obtained using the shooting method described in section \ref{sec:NumericalSolutions} for $\alpha_1=10^7$, $\alpha_2=8\times10^{-15}$, $\alpha_4=-8084527.4$ and $\Xi$ starting off in the $\Xi_+$ branch. We find that the solution connects to the correct cosmological asymptotes at large scales for $\beta=1.5299975$. For these values, the solution is stable against perturbations in the short-range regime.}
    \label{fig:NumericalSolutionQuasiStationary}
\end{figure*}
Figure \ref{fig:NumericalSolutionQuasiStationary} shows an example of a numerical solution for $\alpha_1=10^7$, $\alpha_2=8\times10^{-15}$ and $\alpha_4=-8084527.4$ with $\Xi$ following the $\Xi_+$ branch near the black hole horizon and connecting to the correct cosmological behaviour at large scales. Fitting the approximate solutions \eqref{eq:ShortRangeSolutions} and \eqref{eq:XiShortRangeQuasiStationarySolution} to this numerical solution, we find excellent agreement for $\approxcon \approx 1.5299975$. We have given a large number of digits for these parameters in order to demonstrate that the constraint \eqref{eq:StabilityConstraint} is satisfied. We have
\begin{align}
    r_s/r_c=\left(\frac{\alpha_2}{54 \alpha_1^2}\right)^{1/4}&\approx3.49\times 10^{-8},
    \nonumber\\
    \frac{\alpha_4\sqrt{\approxcon}}{\alpha_1}&=-1+\epsilon,
    \nonumber\\
    \epsilon\approx5\times 10^{-9}&\ll(r_s/r_c)^{2/3}\approx10^{-5}.
\end{align}
So this solution is indeed stable against scalar perturbations over the range that we have studied. As intended, we have therefore found a regular and stable solution that follows the $\Xi_+$ branch close to the black hole, with well-behaved perturbations and dynamics in this short-distance regime, while successfully connecting to the desired cosmological behaviour -- since there are no branch points for this solution, it stays on the (cosmological) $\Xi_+$ branch throughout.

\section{Conclusions} \label{sec:Conclusions}

In this paper we have investigated black hole solutions in scalar-tensor theories in the presence of a timelike scalar, as cosmologically motivated by theories of dynamical dark energy. The existence of hairy solutions in such theories is well-known \cite{Jacobson:1999vr,Babichev:2013cya}, but known such solutions suffer from instabilities, typically encountered when investigating even parity perturbations \cite{Babichev:2013cya,Kobayashi:2014eva,Babichev:2016kdt,BenAchour:2018dap,Motohashi:2019sen,Charmousis:2019vnf,deRham:2019gha,Takahashi:2020hso,Khoury:2020aya,Takahashi:2021bml}. In this paper we specifically focused on the cubic Galileon as a candidate scalar-tensor theory, which captures the qualitative long-range (dark energy) behaviour representative of large classes of scalar-tensor theories. We investigated both the existence and regularity of background solutions in this theory, as well as the stability properties of these solutions,  extending the work of \cite{Babichev:2016fbg, Babichev:2025ric,Babichev:2012re}. Our key findings are:
\begin{itemize}
    \item We derived asymptotic analytical solutions for the self-accelerating cubic Galileon. In particular, in section \ref{sec:BlackHoleSolutions} we found novel analytic solutions in the short-range regime, i.e. in the vicinity of black holes.
    \item We investigated how the analytically tractable short- and long-range solutions can be connected to build a full numerical solution covering all scales. We paid special attention to the branching structure of the scalar solution, showing why for static black hole solutions, a short-range solution in the $\Xi_-$ branch can likely not be connected to the cosmologically relevant long-range solution.
    \item In section \ref{sec:BlackHolePerturbations} we investigated the stability of the static black hole solution \eqref{eq:ShortRangeSolutions} found in section \ref{sec:BlackHoleSolutions} to perturbations, crucially finding that the background in the vicinity of the black hole is unstable against scalar perturbations. This means we cannot identify any stable and static black hole solution that connects to the cosmologically relevant long-range solution.
    \item Motivated by this, we investigated time-dependent black hole solutions, where the black hole time dependence is nevertheless suppressed to allow for the existence of quasi-stationary solutions. We find background solutions for this case, which crucially alter the stability properties of perturbations significantly. As a result we identify 
    (quasi-stationary) black hole solutions that connect to the desired cosmological long-range behaviour and are stable to perturbations in the vicinity of the black hole -- see section \ref{sec:QuasiStationaryBlackHoles} for details. It will be interesting to explore whether full stability -- also covering the intermediate regime -- can be proven for this theory in the future.
\end{itemize}

Overall, the key result is therefore that we have identified a representative dynamical (scalar-tensor) dark energy theory, which gives rise to hairy {\it and} (locally) stable black holes and therefore opens the door to testing dark energy with local black hole physics. Several continuations of this work suggest themselves: In \cite{Smulders:2026bya} we investigate black hole ringdown and the associated observational signatures around the hairy black holes identified here. It would clearly also be interesting to explore whether viable black hole solutions can be constructed within other general dynamical dark energy theories along the lines of the cubic Galileon solution discussed here. In addition to exploring the existence of stable background solutions in specific covariant theories, it should be very fruitful to explore the interplay of this with both recent insights from effective field theory approaches to black hole perturbations \cite{Franciolini:2018uyq,Noller:2019chl,Hui:2021cpm,Mukohyama:2022enj,Khoury:2022zor,Mukohyama:2022skk,Mukohyama:2023xyf,Mukohyama:2024pqe,Barura:2024uog,Mukohyama:2025jzk,Mukohyama:2025owu} as well as other `parametrised hair' approaches to understanding such perturbations, see e.g. \cite{Tattersall:2019nmh,Sirera:2023pbs}. Finally, it is noticeable that most black hole solutions in the presence of a timelike scalar that have been discussed in the literature are for cases where an exact GR solution can be found for the metric background, where such solutions are often referred to as `stealth' solutions. The solution we have found here instead generically (for $\approxcon \neq 1$) yields a departure from such a standard GR solution -- c.f. the metric solution \eqref{eq:ShortRangeSolutions}. In addition we have found that obtaining a stable black hole solution on short distance scales in such scalar-tensor theories also requires going beyond purely static solutions, closely related to the results found in \cite{DeFelice:2022qaz, Rosen:2017dvn, Lara:2025hqh}. It will be interesting to better understand whether one or both of these assumptions will be required to obtain well-behaved black hole solutions in analogous dynamical dark energy theories beyond the cubic Galileon. 

\section*{Acknowledgments}

We thank Gilles Esposito-Far\`ese, Pedro Ferreira, Shinji Mukohyama, Sergi Sirera, and Leonardo Trombetta for useful discussions. LS is supported by STFC. JN is supported by an STFC Ernest Rutherford Fellowship (ST/S004572/1). In deriving the results of this paper, we have used \texttt{xAct}~\cite{xAct} and the \texttt{ringdown calculations} repository~\cite{ringdown-calculations}. For the purpose of open access, the authors have applied a Creative Commons Attribution (CC BY) licence to any Author Accepted Manuscript version arising from this work.
\\

\noindent{\bf Data availability} \red{Details on the calculations supporting the results in this paper can be found in the accompanying GitHub repository~\cite{stable-black-holes-cosmological-hair}. Further supporting research data are available on reasonable request from the authors.}

\appendix
\section{Accuracy of the approximate solutions} \label{sec:AppendixShortRangeLimitAccuracy}
In this appendix we consider the accuracy of the approximate solutions in equation \eqref{eq:ShortRangeSolutions} for $x\gg1$.
For the approximate solutions to be accurate, we require that the relative errors in $h$, $f$ and $\Xi$ are small. We can write some set of exact solutions as
\begin{align}
    \label{eq:SolutionsErrors}
    h(x)&=\bar{h}(x)+\delta_h(x),
    \nonumber\\
    f(x)&=\bar{f}(x)+\delta_f(x),
    \nonumber\\
    \Xi(x)&=\bar{\Xi}(x)+\delta_\Xi(x),
\end{align}
where $\bar{h}$, $\bar{f}$ and $\bar{\Xi}$ are the approximate solutions found in section \ref{sec:ShortRangeLimit}. Our accuracy requirement is then
\begin{align}
    \label{eq:AccuracyRequirement}
    \frac{\delta_{f_i}^{(n)}}{\bar{f}_i^{(n)}}\ll1
\end{align}
for all the functions $f_i\in\{f,h,\Xi\}$ and their $n^{th}$ derivatives. We can write each of the equations of motion \eqref{eq:EoMsDimensionless1}-\eqref{eq:EoMsDimensionless3} as $E+\mathcal{E}=0$, where $E$ represents the dominant terms in $r_s/r_c$ that we kept in calculating our approximate solutions and $\mathcal{E}$ represents the remaining terms that we discarded. Therefore $\bar{E}=0$ since the approximate solutions solve this part of the equation. Expanding about our approximate solutions the equations then become
\begin{align}
    \sum_{i,n}\overline{\frac{\partial E}{\partial f_i^{(n)}}f_i^{(n)}}\frac{\delta_{f_i}^{(n)}}{\bar{f}_i^{(n)}}
    +\bar{\mathcal{E}}+\dots
    =0
\end{align}
where the bars indicate evaluation at the approximate solutions and we have omitted sub-leading terms in the $\delta_{f_i}$. Since $\bar{\mathcal{E}}$ does not generically vanish on its own, an appropriate choice for the $\delta_{f_i}$ that is consistent with the accuracy requirement \eqref{eq:AccuracyRequirement} and solves this equation can only exist if
\begin{align}
    \label{eq:AccuracyTest}
    \sum_{i,n}\left|\overline{\frac{\partial E}{\partial f_i^{(n)}}f_i^{(n)}}\right|\gg\left|\bar{\mathcal{E}}\right|.
\end{align}

For example, for equation \eqref{eq:EoMsDimensionless1} we have 
\begin{align}
    \sum_{i,n}\left|\overline{\frac{\partial E}{\partial f_i^{(n)}}f_i^{(n)}}\right|&=4\alpha_1\approxcon x^2,
    \nonumber\\
    \left|\bar{\mathcal{E}}\right|&=2\approxcon^{3/2}x^3\frac{x-1}{\sqrt{4x-3}}.
\end{align}
For $x\sim\order{1}$ \eqref{eq:AccuracyTest} is satisfied due to the large size of $\alpha_1$ as expected. For $x\gg1$ this requirement becomes
\begin{align}
    x\ll\alpha_1^{2/3}\sim(r_c/r_s)^{2/3}.
\end{align}
Including the bounds from equations \eqref{eq:EoMsDimensionless2} and \eqref{eq:EoMsDimensionless3} we find the same overall bound on $x$. Therefore, the approximate solutions \eqref{eq:ShortRangeSolutions} are accurate as long as $x\ll(r_c/r_s)^{2/3}$. We find the same limit when applying this to the approximate solutions in the quasi-stationary case (section \ref{sec:QuasiStationaryBlackHoles} and appendix \ref{sec:AppendixTimeDependenceSuppression}).

\section{The branch of the cosmological horizon} \label{sec:AppendixCosmologicalHorizonBranch}
In section \ref{sec:BranchingStructure}, we saw that the long-range solutions must follow the $\Xi_+$ branch. In this appendix we show that for any cosmological horizon where $f$ and $h$ cross zero from above, the scalar solution must follow the $\Xi_+$ branch.

Suppose there is a horizon at $x=x_0$. We can Taylor expand about this point and write
\begin{align}
    f(x)=\sum_{n=m_f}^{\infty}\frac{a_n^{(f)}}{n!}(x-x_0)^n,
    \quad
    h(x)=\sum_{n=m_h}^{\infty}\frac{a_n^{(h)}}{n!}(x-x_0)^n,
\end{align}
where $m_f$ and $m_h$ are the multiplicities of the zeros. For horizons, we require that these are odd positive numbers.
For positive $a_{m_f}^{(f)}$ and $a_{m_h}^{(h)}$, $f$ and $h$ cross from below and we have a horizon like the black hole horizon, while in the opposite case the horizon is of cosmological type.

We then expand $A$, $B$ and $C$ from equation \eqref{eq: A, B, C defs} to next-to-leading order and find
\begin{align}
    A
    &=\alpha_1(x^4h)'\frac{f}{h}
    =\alpha_1 x_0^4 m_h a_{m_f}^{(f)}(x-x_0)^{m_f-1}
    \nonumber\\
    &\times\left[
    1
    +\left(\frac{4(m_h+1)}{m_h x_0}+\frac{a_{m_h+1}^{(h)}}{m_h a_{m_h}^{(h)}}+\frac{a_{m_f+1}^{(f)}}{a_{m_f}^{(f)}}\right)(x-x_0)
    \right.
    \nonumber\\
    \Biggl.
    &\quad\quad\quad
    +\order{x-x_0}^2
    \Biggr],
    \nonumber\\\nonumber\\
    B
    &=2x^4h
    =2x_0^4a_{m_h}^{(h)}(x-x_0)^{m_h}
    \nonumber\\
    &\times
    \left[
    1
    +\left(\frac{4}{x_0}+\frac{a_{m_h+1}^{(h)}}{a_{m_h}^{(h)}}\right)(x-x_0)
    +\order{x-x_0}^2
    \right],
    \nonumber\\\nonumber\\
    C
    &=-\alpha_1 x^4 h'
    =-\alpha_1 m_h a_{m_h}^{(h)}x_0^4(x-x_0)^{m_h-1}
    \nonumber\\
    &\times\left[
    1
    +\left(\frac{4}{x_0}+\frac{(m_h+1)a_{m_h+1}^{(h)}}{m_h a_{m_h}^{(h)}}\right)(x-x_0)
    \right.
    \nonumber\\
    \Biggl.
    &\quad\quad\quad+\order{x-x_0}^2
    \Biggr].
\end{align}
We can also write down the solution for $\Xi$ in terms of $f$ and $h$ from equation \eqref{eq:EoMsDimensionless2} and expand to next-to-leading order. This gives
\begin{align}
    \Xi
    &=\pm\sqrt{\frac{a_{m_h}^{(h)}}{a_{m_f}^{(f)}}}(x-x_0)^\frac{m_h-m_f}{2}
    \left[
    1
    +\left(\frac{a_{m_h+1}^{(h)}}{2a_{m_h}^{(h)}}-\frac{a_{m_f+1}^{(f)}}{2a_{m_f}^{(f)}}\right)(x-x_0)
    \right.
    \nonumber\\
    &\left.
    -\frac{a_{m_h}^{(h)}}{\alpha_2 x_0^2}(x-x_0)^{m_h}+\frac{m_h a_{m_f}^{(f)}a_{m_h}^{(h)}}{\alpha_2 x_0}(x-x_0)^{m_f+m_h-1}
    \right.
    \nonumber\\
    &\Biggl.
    +\order{x-x_0}^2
    \Biggr],
\end{align}
where the last two terms in the expansion should only be included in the next-to-leading order for $m_h=1$ and $m_h=m_f=1$, respectively. We can substitute this result into equation \eqref{eq: A, B, C defs} and expand to leading order to find
\begin{align}
    &4 \alpha_1 a_{m_h}^{(h)} x_0^3 (x-x_0)^{m_h}
    \Biggl[
    1
    -\frac{m_h}{2\alpha_2 x_0}a_{m_h}^{(h)}(x-x_0)^{m_h-1}
    \Biggr.
    \nonumber\\
    &\left.
    +\frac{m_h^2}{2\alpha_2}a_{m_f}^{(f)}a_{m_h}^{(h)}(x-x_0)^{m_f+m_h-2}
    +\order{x-x_0}
    \right]
    \nonumber\\
    &\pm 2 x_0^4 a_{m_h}^{(h)}\sqrt{\frac{a_{m_h}^{(h)}}{a_{m_f}^{(f)}}}(x-x_0)^{\frac{3m_h-m_f}{2}}
    \left[1+\order{x-x_0}\right]=0,
\end{align}
where again, the last two terms in the first expansion should only be included in the leading order for $m_h=1$ and $m_h=m_f=1$, respectively. For $m_h>1$ and $m_f>1$ we then have
\begin{align}
    \left(4 \alpha_1 a_{m_h}^{(h)} x_0^3 (x-x_0)^{m_h}
    \pm2x_0^4 a_{m_h}\sqrt{\frac{a_{m_h}^{(h)}}{a_{m_f}^{(f)}}}(x-x_0)^\frac{3m_h-m_f}{2}\right)
    \nonumber\\
    \times
    \left(1+\order{x-x_0}\right)=0.
\end{align}
This only has solutions at the leading order for $m_h=(3m_h-m_f)/2\Rightarrow m_h=m_f$. For $m_h=1$ and $m_f>1$ we find
\begin{align}
    \pm 2 x_0^4 a_{m_h}^{(h)}\sqrt{\frac{a_{m_h}^{(h)}}{a_{m_f}^{(f)}}}(x-x_0)^\frac{3-m_f}{2}\left(1+\order{x-x_0}\right)=0.
\end{align}
This has no solutions. Since the only remaining option is $m_h=m_f=1$ we conclude that we must have $m_h=m_f=m$, with $m$ some odd positive integer. We then calculate the branches for $\Xi$ to next to leading order from equation \eqref{eq: A, B, C defs}:
\begin{align}
    \label{eq:XiExpansionEqn1}
    \Xi_\pm
    &=\pm\text{sign}(\alpha_1)\text{sign}\left(a_m^{(f)}\right)
    \Biggl[
        1
        +\Biggl(
            -\frac{2}{m x_0}
            +\frac{a_{m+1}^{(h)}}{2a_m^{(h)}}
            -\frac{a_{m+1}^{(f)}}{2a_m^{(f)}}
            \Biggr.
            \nonumber\\
            &\left.
            \mp\frac{\text{sign}\left(a_m^{(h)}\right)}{m|\alpha_1|}\sqrt{\frac{a_m^{(h)}}{a_m^{(f)}}}
        \right)(x-x_0)
        +\order{x-x_0}^2
    \Biggr].
\end{align}
However, we can also calculate $\Xi_\pm$ to next-to-leading order from equation \eqref{eq:EoMsDimensionless2}. Matching the leading order to the expansion above we find
\begin{align}
    \label{eq:XiExpansionEqn2}
    \Xi_\pm&=\pm\text{sign}(\alpha_1)\text{sign}\left(a_m^{(f)}\right)\sqrt{\frac{a_m^{(h)}}{a_m^{(f)}}}
    \nonumber\\
    &\times
    \left[
        1
        +\left(\frac{a_{m+1}^{(h)}}{2a_m^{(h)}}-   \frac{a_{m+1}^{(f)}}{2a_m^{(f)}}\right)(x-x_0)
        -\frac{a_m^{(h)}}{\alpha_2x_0^2}(x-x_0)^m
        \right.
        \nonumber\\
        &\left.
        +\frac{m a_m^{(f)}a_m^{(h)}}{\alpha_2 x_0}(x-x_0)^{2m-1}
        +\order{x-x_0}^2
    \right],
\end{align}
where the last two terms in the expansion should only be included in the next-to-leading order for $m=1$. It is immediately clear from these expansions that the two branches do not overlap. Therefore, we cannot have a branch point at a horizon. Furthermore, by matching the next-to-leading orders of \eqref{eq:XiExpansionEqn1} and \eqref{eq:XiExpansionEqn2} we find for $m>1$
\begin{align}
    -\frac{2}{m x_0}\mp\frac{\text{sign}\left(a_m^{(h)}\right)}{m|\alpha_1|}\sqrt{\frac{a_m^{(h)}}{a_m^{(f)}}}=0.
\end{align}
For $a_m^{(h)}<0$ (i.e. a cosmological-type horizon) this can only be solved for the upper sign and therefore for $\Xi$ following the $\Xi_+$ branch.
Doing the same for $m=1$ we find
\begin{align}
    \mp\frac{\text{sign}\left(a_m^{(h)}\right)}{|\alpha_1|}\sqrt{\frac{a_m^{(h)}}{a_m^{(f)}}}
    =\frac{2}{x_0}
    +\frac{a_m^{(f)}a_m^{(h)}}{\alpha_2 x_0}-\frac{a_m^{(h)}}{\alpha_2 x_0^2}.
\end{align}
Again, we see that for $a_m^{(h)}<0$, $\Xi$ must follow the $\Xi_+$ branch. Therefore, at any cosmological horizon the solution for $\Xi$ must follow the $\Xi_+$ branch.

\section{Scalar decoupling at leading order} \label{sec:AppendixScalarDecoupling}

In section \ref{sec:QuadraticActionsSchematic} we used a schematic representation of the quadratic action near the black hole to argue why we can ignore terms mixing the $\delta g$ and $\delta \varphi$ perturbations. In this appendix we will show this more rigourously by directly demixing the scalar and metric terms in the kinetic part of the action at the covariant level through a shift in the metric perturbation. For the cubic Galileon this is possible for generic backgrounds \cite{Babichev:2012re}. Using integration by parts we can find the kinetic part of the quadratic action as:
\begin{align}
    \frac{\mathcal L^{(2)}_{\text{kin}}}{2\sqrt{-g}}
    &=-\frac{r_s^2}{8}\nabla_\mu\delta g_{\alpha_\beta}\left(\frac{1}{2}g^{\alpha\gamma}g^{\beta\delta}-\frac{1}{4}g^{\alpha\beta}g^{\gamma\delta}\right)\nabla^\mu\delta g_{\gamma\delta}
    \nonumber\\
    &+\frac{r_s^2}{8}\left(\nabla_\nu\delta g_\mu^\nu-\frac{1}{2}\partial_\mu\delta g\right)^2
    +S^{\mu\nu}\partial_\mu\delta\varphi\partial_\nu\delta\varphi
    \nonumber\\
    &+\text{sign}(\alpha_1)\text{sign}(q)\frac{\sqrt{|\alpha_1\alpha_2|}}{2\sqrt{2}q^2}\partial^\mu\bar{\varphi}\partial^\nu\bar{\varphi}
    \nonumber\\
    &\times\left[
        \partial^\kappa\delta\varphi\nabla_\kappa\delta g_{\mu\nu}
        -2\partial_\nu\delta\varphi\left(\nabla_\kappa\delta g_\mu^\kappa-\frac{1}{2}\partial_\mu\delta g\right)   
    \right],
    \nonumber\\\nonumber\\
    S^{\mu\nu}&=\text{sign}(\alpha_1)\left(\frac{\Box\bar{\varphi}}{qr_s}+\frac{1}{2\alpha_1r_s^2}\right)\bar{g}^{\mu\nu}
    \nonumber\\
    &-\text{sign}(\alpha_1)\frac{1}{qr_s}\nabla^\mu\partial^\nu\bar{\varphi},
\end{align}
where we have used the perturbations from equation \eqref{eq:NormalisedPerturbations} and $\delta g=\bar{g}^{\mu\nu}\delta g_{\mu\nu}$. This is demixed by defining a new spin 2 perturbation:
\begin{align}
    \delta g'_{\mu\nu}=\delta g_{\mu\nu}-\text{sign}(\alpha_1)\text{sign}(q)\frac{2\sqrt{2|\alpha_1\alpha_2|}}{q^2r_s^2}&\biggl[\partial_\mu\bar{\varphi}\partial_\nu\bar{\varphi}
    \biggr.\nonumber\\\left.
    -\frac{1}{2}(\partial\bar{\varphi})^2\bar{g}_{\mu\nu}\right]&\delta\varphi.
\end{align}
Recalling that $\partial\bar{\varphi}\sim qr_s$ near the black hole, we see that the shift in the metric perturbation is suppressed by $\sqrt{|\alpha_1\alpha_2|}\sim\sqrt{r_s/r_c}$. Therefore, near the black hole $\delta g'_{\mu\nu}=\delta g_{\mu\nu}$ to leading order in $r_s/r_c$.

We have now shown that demixing happens at leading order in $r_s/r_c$ for $r\sim r_s$. However, since our approximate background solutions are accurate over a much larger range, we can consider whether this demixing also happens over larger ranges. We can consider the kinetic part of the Lagrangian near $r=r_0$ by using coordinates $x=r/r_0$ and $y=t/r_0$, but now with $r_0\gg r_s$. Firstly, since the background functions will not be $\order{1}$ anymore (e.g. derivatives of the metric functions will be suppressed by factors of $r_s/r_0$), we will need to use a different normalisation of the perturbations to obtain a canonical normalisation of the kinetic terms, i.e. one where the prefactors of the $(\partial\delta g)^2$ and $(\partial\delta\varphi)^2$ terms are $\order{1}$. The correct normalisation near $r=r_0$ is
\begin{align}
    \label{eq:NormalisedPerturbationsAllScales}
    g_{\mu\nu}=\bar{g}_{\mu\nu}+\frac{r_0}{\mpl}\delta g_{\mu\nu},&&
    \varphi=\bar{\varphi}+\sqrt{\frac{\Lambda_3^3}{|\conthree q|r_0}}\left(\frac{r_0}{r_s}\right)^{1/4}\delta\varphi.
\end{align}
We then find for the kinetic part of the quadratic action
\begin{align}
        \frac{\mathcal L^{(2)}_{\text{kin}}}{2\sqrt{-g}}
    &=-\frac{r_0^2}{8}\nabla_\mu\delta g_{\alpha\beta}\left(\frac{1}{2}g^{\alpha\gamma}g^{\beta\delta}-\frac{1}{4}g^{\alpha\beta}g^{\gamma\delta}\right)\nabla^\mu\delta g_{\gamma\delta}
    \nonumber\\
    &+\frac{r_0^2}{8}\left(\nabla_\nu\delta g_\mu^\nu-\frac{1}{2}\partial_\mu\delta g\right)^2
    +\sqrt{\frac{r_0}{r_s}}S^{\mu\nu}\partial_\mu\delta\varphi\partial_\nu\delta\varphi
    \nonumber\\
    &+\text{sign}(\alpha_1)\text{sign}(q)\frac{\sqrt{|\alpha_1\alpha_2|}}{2\sqrt{2}q^2}\left(\frac{r_0}{r_s}\right)^{1/4}\partial^\mu\bar{\varphi}\partial^\nu\bar{\varphi}
    \nonumber\\
    &\times\left[
        \partial^\kappa\delta\varphi\nabla_\kappa\delta g_{\mu\nu}
        -2\partial_\nu\delta\varphi\left(\nabla_\kappa\delta g_\mu^\kappa-\frac{1}{2}\partial_\mu\delta g\right)   
    \right],
    \nonumber\\\nonumber\\
    S^{\mu\nu}&=\text{sign}(\alpha_1)\left(\frac{\Box\bar{\varphi}}{qr_0}+\frac{1}{2\alpha_1r_0^2}\right)\bar{g}^{\mu\nu}
    \nonumber\\
    &-\text{sign}(\alpha_1)\frac{1}{qr_0}\nabla^\mu\partial^\nu\bar{\varphi},
\end{align}
where the $\alpha_i$ parameters are defined as in \eqref{eq:DimensionlessParameters}, i.e. with $r_0\neq r_s$. By substituting in our background solutions we find that $S^{\mu\nu}\sim\sqrt{r_s/r_0}$, so the perturbations are canonically normalised. We then again demix the $\delta g$ and $\delta \varphi$ perturbations by introducing
\begin{align}
    \delta g_{\mu\nu}'=\delta g'_{\mu\nu}=\delta g_{\mu\nu}&-\text{sign}(\alpha_1)\text{sign}(q)\frac{2\sqrt{2|\alpha_1\alpha_2|}}{q^2r_0^2}\left(\frac{r_0}{r_s}\right)^{1/4}
    \nonumber\\
    &\times\left[\partial_\mu\bar{\varphi}\partial_\nu\bar{\varphi}
    -\frac{1}{2}(\partial\bar{\varphi})^2\bar{g}_{\mu\nu}\right]\delta\varphi.
\end{align}
Through using the background solutions we find that the shift is of the order $\order{\frac{r_0^3}{r_c^2r_s}}^{1/4}$. We are therefore justified in dropping the mixing terms at leading order as long as
\begin{align}
    \left(\frac{r_0^3}{r_c^2r_s}\right)^{1/4}\ll1
    \Rightarrow \frac{r_0}{r_s}\ll(r_c/r_s)^{2/3}.
\end{align}
Note that this is exactly the same range as the one for which the approximate solutions are accurate. Going back to our usual dimensionless coordinates $x=r/r_s$ and $y=t/r_s$ we find that the perturbation analysis described in section \ref{sec:BlackHolePerturbations}, which uses the demixing of scalar and metric perturbations and the approximate background solutions \eqref{eq:ShortRangeSolutions} is applicable for $x\ll (r_c/r_s)^{2/3}$. Note that the same arguments are valid for the decoupling in the quasi-stationary case described in section \ref{sec:QuasiStationaryBlackHoles}. 

\section{Regge-Wheeler gauge} \label{sec:AppendixReggeWheelerGauge}
For the derivation of the Regge-Wheeler gauge, we mostly follow \cite{Zerilli:1970wzz} and \cite{maggiore_12_2018}, with the inclusion of scalar perturbations.

Firstly, we expand the scalar perturbation in spherical harmonics and the metric perturbation in terms of Zerilli tensor harmonics:
\begin{align}
    \delta g_{\mu\nu}^{\text{even}}&=\sum_{\ell=0}^{\infty}\sum_{m=-\ell}^{\ell}\left[
    h_{\ell m}^{tt}(\textbf{t}_{\ell m}^{tt})_{\mu\nu}
    +h_{\ell m}^{Rt}(\textbf{t}_{\ell m}^{Rt})_{\mu\nu}
    \right.\nonumber\\&\left.
    +h_{\ell m}^{T0}(\textbf{t}_{\ell m}^{T0})_{\mu\nu}
    +h_{\ell m}^{L0}(\textbf{t}_{\ell m}^{L0})_{\mu\nu}
    \right]
    \nonumber\\
    &+\sum_{\ell=1}^{\infty}\sum_{m=-\ell}^{\ell}\left[
    h_{\ell m}^{Et}(\textbf{t}_{\ell m}^{Et})_{\mu\nu}
    +h_{\ell m}^{E1}(\textbf{t}_{\ell m}^{E1})_{\mu\nu}
    \right]
    \nonumber\\
    &+\sum_{\ell=2}^{\infty}\sum_{m=-\ell}^{\ell}h_{\ell m}^{E2}(\textbf{t}_{\ell m}^{E2})_{\mu\nu},
    \nonumber\\\nonumber\\
    \delta g_{\mu\nu}^{\text{odd}}&=\sum_{\ell=1}^{\infty}\sum_{m=-\ell}^{\ell}\left[
    h_{\ell m}^{Bt}(\textbf{t}_{\ell m}^{Bt})_{\mu\nu}
    +h_{\ell m}^{B1}(\textbf{t}_{\ell m}^{B1})_{\mu\nu}
    \right]
    \nonumber\\
    &+\sum_{\ell=2}^{\infty}\sum_{m=-\ell}^{\ell}h_{\ell m}^{B2}(\textbf{t}_{\ell m}^{B2})_{\mu\nu},
    \nonumber\\\nonumber\\
    \delta\phi&=\sum_{\ell=0}^{\infty}\sum_{m=-\ell}^{\ell}\delta\varphi_{\ell m}Y_{\ell m},
\end{align}
where we have also separated into even and odd modes. All the spherical and tensor harmonics carry angular dependence, while the $h_{\ell m}$ and $\delta\varphi_{\ell m}$ functions carry temporal and radial dependence. The Zerilli tensors are given by
\begin{align}
    (\textbf{t}_{\ell m}^{tt})_{\mu\nu}
    &=\begin{pmatrix}
        1&0&0&0\\
        0&0&0&0\\
        0&0&0&0\\
        0&0&0&0
    \end{pmatrix}Y_{\ell m},
    \nonumber\\\nonumber\\
    (\textbf{t}_{\ell m}^{Rt})_{\mu\nu}
    &=\begin{pmatrix}
        0&1&0&0\\
        1&0&0&0\\
        0&0&0&0\\
        0&0&0&0
    \end{pmatrix}Y_{\ell m},
    \nonumber\\\nonumber\\
    (\textbf{t}_{\ell m}^{L0})_{\mu\nu}
    &=\begin{pmatrix}
        0&0&0&0\\
        0&1&0&0\\
        0&0&0&0\\
        0&0&0&0
    \end{pmatrix}Y_{\ell m},
    \nonumber\\\nonumber\\
    (\textbf{t}_{\ell m}^{T0})_{\mu\nu}
    &=\begin{pmatrix}
        0&0&0&0\\
        0&0&0&0\\
        0&0&1&0\\
        0&0&0&\sin^2\theta
    \end{pmatrix}Y_{\ell m},
    \nonumber\\\nonumber\\
    (\textbf{t}_{\ell m}^{Et})_{\mu\nu}
    &=\begin{pmatrix}
        0&0&\partial_\theta&\partial_\phi\\
        0&0&0&0\\
        \partial_\theta&0&0&0\\
        \partial_\phi&0&0&0
    \end{pmatrix}Y_{\ell m},
    \nonumber\\\nonumber\\
    (\textbf{t}_{\ell m}^{E1})_{\mu\nu}
    &=\begin{pmatrix}
        0&0&0&0\\
        0&0&\partial_\theta&\partial_\phi\\
        0&\partial_\theta&0&0\\
        0&\partial_\phi&0&0
    \end{pmatrix}Y_{\ell m},
    \nonumber\\\nonumber\\
    (\textbf{t}_{\ell m}^{E2})_{\mu\nu}
    &=\begin{pmatrix}
        0&0&0&0\\
        0&0&0&0\\
        0&0&W&X\\
        0&0&X&-\sin^2\theta W
    \end{pmatrix}Y_{\ell m},
    \nonumber\\\nonumber\\
    (\textbf{t}_{\ell m}^{Bt})_{\mu\nu}
    &=\begin{pmatrix}
        0&0&(1/\sin\theta)\partial_\phi&-\sin\theta\partial_\theta\\
        0&0&0&0\\
        (1/\sin\theta)\partial_\phi&0&0&0\\
        -\sin\theta\partial_\theta&0&0&0
    \end{pmatrix}Y_{\ell m},
    \nonumber\\\nonumber\\
    (\textbf{t}_{\ell m}^{B1})_{\mu\nu}
    &=\begin{pmatrix}
        &0&0&0\\
        0&0&(1/\sin\theta)\partial_\phi&-\sin\theta\partial_\theta\\
        0&(1/\sin\theta)\partial_\phi&0&0\\
        0&-\sin\theta\partial_\theta&0&0
    \end{pmatrix}Y_{\ell m},
    \nonumber\\\nonumber\\
    (\textbf{t}_{\ell m}^{B2})_{\mu\nu}
    &=\begin{pmatrix}
        1&0&0&0\\
        0&0&0&0\\
        0&0&-(1/\sin\theta)X&\sin\theta W\\
        0&0&\sin\theta W&\sin\theta X
    \end{pmatrix}Y_{\ell m},
    \nonumber\\\nonumber\\
    X &= 2\partial_\theta\partial_\phi-2\cot\theta\partial_\phi,
    \nonumber\\
    W &= \partial_\theta^2-\cot\theta\partial_\theta-\frac{1}{\sin^2\theta}\partial_\phi^2.
\end{align}
To study gauge transformations we then define the infinitesimal coordinate transformation
\begin{align}
    x^\mu\rightarrow x'^\mu=x^\mu+\frac{r_s}{\mpl}\xi^\mu,
\end{align}
which leads to transformations on the metric and scalar:
\begin{align}
    \delta g'_{\mu\nu}&=\delta g_{\mu\nu}-2\bar{\nabla}_{(\mu}\xi_{\nu)},
    \nonumber\\
    \delta\varphi'&=\delta\varphi-\sqrt\frac{\left|\conthree q\right|r_s^3}{\mpl^2\Lambda_3^3}\xi^\mu\partial_\mu\bar{\varphi}.
\end{align}
We next expand $\xi_\mu$ in terms of vector harmonics:
\begin{align}
    \xi_y&=\sum_{\ell=0}^{\infty}\sum_{m=-\ell}^{\ell}\xi_{\ell m}^{(t)}Y_{\ell m},
    \nonumber\\
    \xi_x&=\sum_{\ell=0}^{\infty}\sum_{m=-\ell}^{\ell}\xi_{\ell m}^{(R)}Y_{\ell m},
    \nonumber\\
    \xi_\theta&=\sum_{\ell=1}^{\infty}\sum_{m=-\ell}^{\ell}\left[
        \xi_{\ell m}^{(E)}\partial_\theta Y_{\ell m}
        -\xi_{\ell m}^{(B)}\frac{1}{\sin\theta}\partial_\phi Y_{\ell m}
    \right],
    \nonumber\\
    \xi_\phi&=\sum_{\ell=1}^{\infty}\sum_{m=-\ell}^{\ell}\left[
        \xi_{\ell m}^{(E)}\partial_\phi Y_{\ell m}
        +\xi_{\ell m}^{(B)}\sin\theta\partial_\theta Y_{\ell m}
    \right].
\end{align}
Substituting this in and using the properties of the spherical harmonics we find for the gauge transformation on the $h_{\ell m}$ and $\delta\varphi_{\ell m}$ functions, in dimensionless coordinates,
\begin{align}
    h_{\ell m}^{tt}&\rightarrow h_{\ell m}^{tt}+fh'\xi_{\ell m}^{(R)}-2\partial_y\xi_{\ell m}^{(t)},
    \nonumber\\
    h_{\ell m}^{Rt}&\rightarrow h_{\ell m}^{Rt}+\frac{h'}{h}\xi_{\ell m}^{(t)}-\partial_y\xi_{\ell m}^{(R)},
    \nonumber\\
    h_{\ell m}^{L0}&\rightarrow h_{\ell m}^{L0}-\frac{f'}{f}\xi_{\ell m}^{(R)}-2\partial_x\xi_{\ell m}^{(R)},
    \nonumber\\
    h_{\ell m}^{T0}&\rightarrow h_{\ell m}^{T0}+\ell(\ell+1)\xi_{\ell m}^{(E)}-2xf\xi_{\ell m}^{(R)},
    \nonumber\\
    h_{\ell m}^{Et}&\rightarrow h_{\ell m}^{Et}-\xi_{\ell m}^{(t)}-\partial_y\xi_{\ell m}^{(E)},
    \nonumber\\
    h_{\ell m}^{E1}&\rightarrow h_{\ell m}^{E1}+\frac{2}{x}\xi_{\ell m}^{(E)}-\partial_x\xi_{\ell m}^{(E)}-\xi_{\ell m}^{(R)},
    \nonumber\\
    h_{\ell m}^{E2}&\rightarrow h_{\ell m}^{E2}-\xi_{\ell m}^{(E)},
    \nonumber\\
    h_{\ell m}^{Bt}&\rightarrow h_{\ell m}^{Bt}+\partial_y\xi_{\ell m}^{(B)},
    \nonumber\\
    h_{\ell m}^{B1}&\rightarrow h_{\ell m}^{B1}-\frac{2}{x}\xi_{\ell m}^{(B)}+\partial_x\xi_{\ell m}^{(B)},
    \nonumber\\
    h_{\ell m}^{B2}&\rightarrow h_{\ell m}^{B2}-\xi_{\ell m}^{(B)},
    \nonumber\\
    \delta\varphi_{\ell m}&\rightarrow \delta\varphi_{\ell m}-\text{sign}(q)\sqrt{\frac{|\alpha_1\alpha_2|}{2}}\left(-\frac{1}{h}\xi_{\ell m}^{(t)}+\frac{f}{h}\Xi\xi_{\ell m}^{(R)}\right),
\end{align}
where we need to keep in mind that $\xi_{00}^{(E)}=\xi_{00}^{(B)}=h_{00}^{Et}=h_{00}^{E1}=h_{00}^{Bt}=h_{00}^{B1}=h_{1m}^{E2}=h_{1m}^{B2}=0$. Starting from some arbitrary gauge, we can then choose the $\xi_{\ell m}$ functions in a way that sets some of the perturbation functions to zero. For $\ell\geq2$, we can choose $\xi$ to set $h_{\ell m}^{Et}$, $h_{\ell m}^{E1}$, $h_{\ell m}^{E2}$ and $h_{\ell m}^{B2}$ to zero. This fully fixes the gauge for $\ell\geq2$.

For $\ell=1$, $h_{\ell m}^{E2}$ and $h_{\ell m}^{B2}$ are already zero, so we can choose $\xi_{1m}^{(E)}$ and $\xi_{1m}^{(B)}$ to also set $h_{\ell m}^{T0}$ and $h_{\ell m}^{B1}$ to zero. However, there is no unique choice to do this and we will still be able to perform further gauge transformations that do not shift $h_{1m}^{T0}$, $h_{1m}^{Et}$, $h_{1m}^{E1}$ or $h_{1m}^{B1}$. These will be of the form
\begin{align}
    \xi_{1m}^{(E)}&=x^2D(y)\exp\left[-\int\frac{dx}{xf}\right],
    \nonumber\\
    \xi_{1m}^{(R)}&=\frac{x}{f}D(y)\exp\left[-\int\frac{dx}{xf}\right],
    \nonumber\\
    \xi_{1m}^{(t)}&=-x^2\dot{D}(y)\exp\left[-\int\frac{dx}{xf}\right],
    \nonumber\\
    \xi_{1m}^{(B)}&=C(y)x^2,
\end{align}
where $C$ and $D$ are arbitrary functions of the temporal coordinate $y$.

Similarly, for $\ell=0$ we can use the gauge freedom to set $h_{00}^{Rt}$ and $h_{00}^{T0}$ to zero. Again there is a residual gauge freedom of the form
\begin{align}
    \xi_{00}^{(t)}=E(y)h,&&\xi_{00}^{(R)}=0.
\end{align}
We then redefine the fields to correspond to more conventional choices:
\begin{align}
    h_{\ell m}^{tt}(y,x)=h(x)H_{\ell m}^{(0)}(y,x),&&
    h_{\ell m}^{Rt}(y,x)=H_{\ell m}^{(1)}(y,x),
    \nonumber\\
    h_{\ell m}^{L0}(y,x)=\frac{1}{f(x)}H_{\ell m}^{(2)}(y,x),&&
    h_{\ell m}^{T0}(y,x)=x^2K_{\ell m}(y,x),
    \nonumber\\
    h_{\ell m}^{Bt}(y,x)=-h_{\ell m}^{(0)}(y,x),&&
    h_{\ell m}^{B1}(y,x)=-h_{\ell m}^{(1)}(y,x).
\end{align}
Using these definitions and the gauge choices described above gives the perturbations in the Regge-Wheeler gauge as given in equation \eqref{eq:ReggeWheelerGauge}. There are then residual gauge freedoms for $\ell = 1$ and $\ell = 0$:
\begin{align}
    \label{eq:ResidualGaugeFreedoms}
    H_{1m}^{(0)}&\rightarrow H_{1m}^{(0)}+\left(\frac{xh'}{h}D+2\frac{x^2}{h}\ddot{D}\right)\exp\left(-\int\frac{dx}{xf}\right),
    \nonumber\\
    H_{1m}^{(1)}&\rightarrow H_{1m}^{(1)}+\left(2-\frac{2}{f}-\frac{xh'}{h}\right)x\dot{D}\exp\left(-\int\frac{dx}{xf}\right),
    \nonumber\\
    H_{1m}^{(2)}&\rightarrow H_{1m}^{(2)}-\left(2-\frac{2}{f}-\frac{xf'}{f}\right)D\exp\left(-\int\frac{dx}{xf}\right),
    \nonumber\\
    h_{1m}^{(0)}&\rightarrow h_{1m}^{(0)}-\dot{C}x^2,
    \nonumber\\
    \delta\varphi_{1m}&\rightarrow\delta\varphi_{1m}-\text{sign}{(q)}\sqrt{\frac{\left|\alpha_1\alpha_2\right|}{2}}\left(\frac{x^2}{h}\dot{D}+\frac{x}{h}\Xi D\right)
    \nonumber\\
    &\times\exp\left(-\int\frac{dx}{xf}\right),
    \nonumber\\
    H_{00}^{(0)}&\rightarrow H_{00}^{(0)}-2\dot{E},
    \nonumber\\
    \delta\varphi_{00}&\rightarrow\delta\varphi_{00}+\text{sign}{(q)}\sqrt{\frac{\left|\alpha_1\alpha_2\right|}{2}}E,
\end{align}
where $C$, $D$ and $E$ are arbitrary functions of $y$.

\section{Modified quadratic actions} \label{sec:AppendixQuadraticActions}

Here we describe in more detail how we arrive at the quadratic actions shown in equation \eqref{eq:QuadraticActionsFull}.

Firstly, the odd part of the quadratic action that we find after substituting in the Regge-Wheeler functions and doing some integrations by parts is
\begin{align}
    \label{eq:OddQuadraticActionInitial}
    \frac{4\sqrt{\approxcon}}{\ell(\ell+1)}\mathcal L_{\text{odd}}^{(2)}
    &=\frac{\ell(\ell+1)x-2}{x^2(x-1)}h_0^2
    \nonumber\\
    &-\approxcon\frac{(\ell+2)(\ell-1)(x-1)}{x^3}h_1^2
    \nonumber\\
    &+\frac{4}{x}h_0\dot{h_1}
    +h_0'^2
    -2h_0'\dot{h_1}
    +\dot{h_1}^2,
\end{align}
where we have suppressed the $(y,x)$ dependence of the fields, as well as the sum over angular indices.
We have written $h^{(0)}_{\ell m}(y,x)=h_0$, etc. Since derivatives of both fields appear, we cannot simply integrate out one of the fields. However, we can use a trick \cite{Kobayashi:2012kh} which involves first completing the square in the derivatives and subsequently introducing an auxiliary field $w(y,x)$:
\begin{align}
    \frac{4\sqrt{\approxcon}}{\ell(\ell+1)}\mathcal L_{\text{odd}}^{(2)}
    &=\frac{(\ell+2)(\ell-1)}{x(x-1)}h_0^2
    \nonumber\\
    &-\approxcon\frac{(\ell+2)(\ell-1)(x-1)}{x^3}h_1^2
    \nonumber\\
    &+2w\left(\dot{h_1}-h_0'+\frac{2}{x}h_0\right)
    -w^2.
\end{align}
Varying this action with respect to $h_0$ and $h_1$ gives two equations of motion that are algebraic in both of these fields. We can therefore use these equations to express $h_0$ and $h_1$ in terms of $w$ and its derivatives. Substituting this back into the action and integrating by parts gives the action in terms of $w$ given in equation \eqref{eq:QuadraticActionsFull}.

Secondly, for the even metric modes we find, using the same notation as in equation \eqref{eq:OddQuadraticActionInitial},
\begin{align}
    \mathcal L_{\text{even}}^{(2)}
    &=a_0H_1^2+H_1\left(a_1\dot{H_2}+a_2\dot{K}+a_3\dot{K}'\right)
    \nonumber\\
    &+H_0\left(a_4H_2+a_5K+a_6H_2'+a_7K'+a_8K''\right)
    \nonumber\\
    &+b_0H_2^2+b_1K'^2+b_2\dot{K}^2+b_3H_2K+b_4H_2K'
    \nonumber\\
    &+b_5\dot{H_2}\dot{K},
\end{align}
where the $a_i$ and $b_i$ are $x$-dependent coefficients. No derivatives of $H_0$ and $H_1$ appear, which means they act as auxiliary fields. We can use the equation of motion resulting from variation with respect to $H_1$ to integrate out this field. For $H_0$ there is no quadratic term, which means it acts as a Lagrange multiplier providing a constraint on the remaining two fields:
\begin{align}
    a_4H_2+a_5K+a_6H_2'+a_7K'+a_8K''=0.
\end{align}
Since this constraint equation is not algebraic in either of the fields, we cannot easily use it to remove one of them. We overcome this by defining a new field $\psi(y,x)$, with
\begin{align}
    H_2=\psi+\gamma_1(x)K+\gamma_2(x)K'.
\end{align}
$\gamma_1$ and $\gamma_2$ are chosen in such a way that, after replacing $H_2$, the derivatives of $K$ are removed from the constraint equation. We can then solve the equation algebraically for $K$ and substitute this back into the quadratic action, leaving us with an action solely in terms of the new field $\psi$. After performing some integrations by parts we arrive at the $\mathcal L_{\text{even}}^{(2)}$ given in equation \eqref{eq:QuadraticActionsFull}.

Lastly, for the scalar part of the quadratic action there is already only a single field describing the scalar degree of freedom. Therefore, we only need to do some straight-forward integrations by parts to arrive at the form of $\mathcal L_{\text{scalar}}^{(2)}$ given in equation \eqref{eq:QuadraticActionsFull}.

\section{Monopole and dipole perturbations} \label{sec:AppendixMonopoleDipole}
Due to the additional gauge freedoms for $\ell=0,1$, the discussion of the quadratic actions is slightly different from the one in the main text for $\ell\geq2$. However, we will find that no monopole or dipole dynamical degrees of freedom exist in the metric perturbations. For the scalar perturbations they do exist and behave exactly as the $\ell\geq2$ perturbations.

We will start by considering the dipole perturbations for the odd modes. As described in appendix \ref{sec:AppendixReggeWheelerGauge}, we use part of the extra gauge freedom here to set $h_{1m}^{(1)}=0$. Setting $\ell=1$ in the quadratic action, calculating the equations of motion and taking $h^{(1)}_{1m}\rightarrow0$, we find
\begin{align}
    h_0''-\frac{2}{x^2}h_0&=0,
    \nonumber\\
    \dot{h}_0'-\frac{2}{x}\dot{h}_0&=0.
\end{align}
The first equation gives
\begin{align}
    h_0(y,x)=\frac{d_1(y)}{x}+d_2(y)x^2,
\end{align}
where $d_1$ and $d_2$ are arbitrary functions of time. From equation \eqref{eq:ResidualGaugeFreedoms}, we see that the second term is pure gauge. Therefore we can use our last remaining residual gauge freedom in the odd sector to remove this term. Substituting the solution into the second equation, we find $d_1(y)=const.$. Rewriting the integration constant in a suggestive way and expressing $h_{1m}^{(0)}$ in terms of our original dimensionful  coordinates we finally find
\begin{align}
    h_{1m}^{(0)}(t,r)=\frac{4\sqrt{\pi}\mpl J}{3r}.
\end{align}
We see that there is no odd dipole dynamical degree of freedom. The perturbation corresponds to what we would expect for the Kerr correction to a slowly rotating black hole. This is exactly analogous to the result found in \cite{Kobayashi:2012kh}, where a purely radial scalar background is used in a general Horndeski theory.

Subsequently, we can consider the dipole perturbation for the even metric modes. As described in appendix \ref{sec:AppendixReggeWheelerGauge}, $K_{1m}=0$. In the same way as for $\ell\geq2$, we can integrate out $H_1$ as an auxiliary field. Again, $H_0$ acts as a Lagrange multiplier. The constraint resulting from this is then, after setting $K=0$,
\begin{align}
    &(x-1)H_2'+2H_2=0
    \nonumber\\
    &\Rightarrow H_2=\frac{e_1(y)}{(x-1)^2},
\end{align}
where $e_1$ is an arbitrary integration function of time. From equation \eqref{eq:ResidualGaugeFreedoms}, we see that this solution for $H_2$ is pure gauge. Therefore we can use our remaining gauge freedom to set it to zero. The equations of motion resulting from variation of $H_2$ and $K$ then yield $H_0=0$ as well. There are therefore also no even metric dipole degrees of freedom. This corresponds to the result found in \cite{Kobayashi:2014wsa}.

Finally, we consider the even monopole perturbations. In this case, we can use the gauge freedom to set $H_1=0$ as well. For $\ell=0$, we find that quadratic terms appear for neither $H_0$ nor $H_1$ in the quadratic action. Therefore they both act as Lagrange multipliers, providing constraints on $H_2$:
\begin{align}
    (x-1)H_2'+H_2&=0,
    \nonumber\\
    \dot{H}_2&=0
    \nonumber\\
    \Rightarrow H_2&=\frac{f_1}{x-1},
\end{align}
where $f_1$ is an arbitrary integration constant.
The equations of motion resulting from variation with respect to $H_2$ and $K$ will then be dependent equations yielding
\begin{align}
    H_0=\frac{f_1}{x-1}+f_2(y),
\end{align}
with $f_2$ an arbitrary integration function of time. From equation \eqref{eq:ResidualGaugeFreedoms} we see that this $f_2$ term can be removed by using our last residual gauge freedom. We then finally find
\begin{align}
    H_{00}^{(2)}=H_{00}^{(0)}=\frac{f_1}{x-1},
\end{align}
which corresponds to a simple shift of the horizon in the background solutions as in \cite{Kobayashi:2014wsa}.

Note that monopole perturbations in the odd sector vanish by definition. Since we have fully gauge fixed now, there is no more gauge freedom that will change anything about the monopole and dipole perturbations in the scalar sector. Therefore, the monopole and dipole scalar perturbations are dynamical and the discussion for $\ell\geq2$ in the main text still applies for these perturbations.

\section{Stability and Hamiltonian boundedness} \label{sec:AppendixHamiltonianBoundedness}

In \cite{Babichev:2018uiw}, the important point is made that while a Hamiltonian density that is bounded from below guarantees stability, the converse is not true: a Hamiltonian density that is unbounded from below does not necessarily imply the presence of an instability. This is because the Hamiltonian density depends on the coordinate system in which it is calculated while stability is a coordinate-independent property. In naively unstable cases it might still be possible to find a different coordinate system in which the Hamiltonian \'is bounded from below. Therefore we need to be more careful in cases where the Hamiltonian we find initially is unbounded from below. See also the detailed discussion in \cite{Sawicki:2024ryt}.

For scalar fields that are minimally coupled to the metric, (e.g. standard model fields), the kinetic part of the Lagrangian will have the form
\begin{align}
    \label{eq:KineticTermMinimallyCoupledField}
    -\frac{1}{2}g^{\mu\nu}\partial_\mu\varphi\partial_\nu\varphi.
\end{align}
However, for fields that couple to the metric differently, the kinetic part will generically look like
\begin{align}
    \label{eq:KineticTermEffectiveMetric}
    -\frac{1}{2}S^{\mu\nu}\partial_\mu\psi\partial_\nu\psi,
\end{align}
which defines an effective metric $S^{-1}_{\mu\nu}$. Note that here the indices have not been lowered with $g_{\mu\nu}$, but it is simply the inverse of $S^{\mu\nu}$. We will then typically solve a Cauchy problem, defining initial conditions on some spatial hypersurface and propagating these along some timelike direction. However, note that in the case \eqref{eq:KineticTermEffectiveMetric}, we need to use a hypersurface and direction that are spacelike and timelike respectively with respect to the \textit{effective} metric $S_{\mu\nu}^{-1}$. Therefore, if we want to solve for multiple fields we require the existence of a spacelike hypersurface and timelike direction that are spacelike and timelike with respect to all effective metrics simultaneously. Since standard model fields will always be present, even if we do not explicitly consider them, the effective metrics need to at least be consistent with the regular metric $g_{\mu\nu}$. This is a different condition for stability that is independent of the choice of coordinates.

We can study this in a bit more detail and see how it relates to the requirement of Hamiltonian boundedness. In particular, we can show that if there exist suitable coordinates in which the Hamiltonian density is bounded from below, the above conditions must be satisfied. We will consider a 2D case and suppose that the Hamiltonian is bounded from below in some coordinates $(\tilde{t},\tilde{r})$, where $\partial_{\tilde{t}}$ is timelike and $\partial_{\tilde{r}}$ is spacelike with respect to the regular metric $g_{\mu\nu}$. The kinetic part of the Lagrangian will have the form
\begin{align}
    \mathcal L_{\text{kin}}&=\frac{1}{2}a_0(\partial_{\tilde{t}}\psi)^2-a_1\partial_{\tilde{t}}\psi\partial_{\tilde{r}}\psi-\frac{1}{2}a_2(\partial_{\tilde{r}}\psi)^2
    \nonumber\\
    &=-\frac{1}{2}S^{\mu\nu}\partial_\mu\psi\partial_\nu\psi,
    \nonumber\\\nonumber\\
    S^{\mu\nu}&=\begin{pmatrix}
        -a_0&a_1\\
        a_1&a_2
    \end{pmatrix},
\end{align}
where the boundedness from below implies $a_0>0$ and $a_2>0$.
We then have an effective metric
\begin{align}
    S_{\mu\nu}^{-1}=-\frac{1}{a_0a_2+a_1^2}\begin{pmatrix}
        a_2&-a_1\\
        -a_1&-a_0
    \end{pmatrix}.
\end{align}
For this to define a sensible metric it needs to satisfy the \textit{hyperbolicity} condition:
\begin{align}
    \det S_{\mu\nu}^{-1}=-\frac{1}{a_0a_2+a_1^2}<0.
\end{align}
This is automatically satisfied for $a_0, a_2>0$. We can finally show that the vectors $\partial_{\tilde{t}}$ and $\partial_{\tilde{r}}$, which are timelike and spacelike with respect to $g_{\mu\nu}$ are also timelike and spacelike with respect to $S_{\mu\nu}^{-1}$:
\begin{align}
    S_{\mu\nu}^{-1}(\partial_{\tilde{t}})^\mu(\partial_{\tilde{t}})^\nu&=S^{-1}_{00}=\frac{-a_2}{a_0a_2+a_1^2}<0,
    \nonumber\\
    S_{\mu\nu}^{-1}(\partial_{\tilde{r}})^\mu(\partial_{\tilde{r}})^\nu&=S^{-1}_{11}=\frac{a_0}{a_0a_2+a_1^2}>0.
\end{align}
Therefore, if some coordinates $(\tilde{t},\tilde{r})$ exist in which the Hamiltonian is bounded from below, there must exist vectors which are timelike and spacelike with respect to the effective as well as the regular metric.

\begin{figure}
    \centering
    \includegraphics[width=\linewidth]{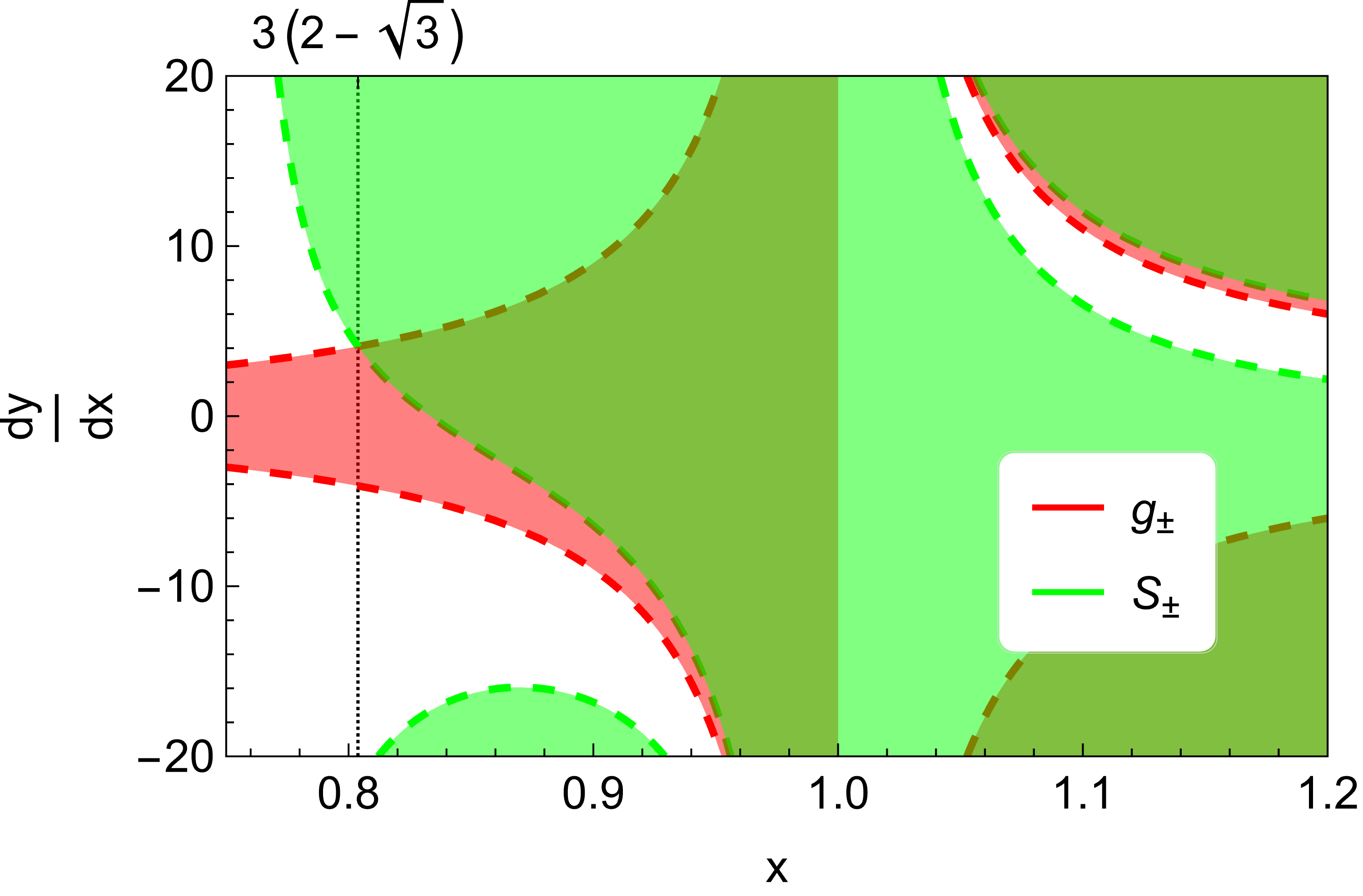}
    \caption{An example of the lightcone structure for the $\Xi_-$ background and $\text{sign}(\alpha_1)=+1$. The dashed lines show the $dy/dx$ derivatives of null curves with respect to the spacetime metric $g_{\mu\nu}$ as well as the effective metric for scalar perturbations $S^{-1}_{\mu\nu}$ for different radii. Worldlines with derivatives in the shaded region are timelike with respect to the corresponding metric, while wordlines outside the shaded region are spacelike. To find `good' coordinates that are time/space-like with respect to both metrics, the lightcones must have overlapping interiors as well as overlapping exteriors for all radii. In the plot, this means that for every $x$ there must be a value of $dy/dx$ lying in both shaded regions simultaneously, as well as a value not lying in either shaded region. Crucially, for $x<3\left(2-\sqrt{3}\right)$, there is no overlap between the shaded regions. Therefore, a common spacelike coordinate cannot be found below this radius, indicating a ghost instability.}
    \label{fig:LightconeStructure}
\end{figure}

We can now apply this to our case. We will study the stability over the range where our local solutions \eqref{eq:ShortRangeSolutions} and quadratic actions \eqref{eq:QuadraticActionsFull} are accurate, i.e. the region where $x\ll(r_c/r_s)^{2/3}$. Note that we also investigate the region from $x=3/4$ to $x=1$. Although this is behind the black hole horizon, the scalar perturbations can escape this horizon and we therefore need to study their stability in the black hole interior as well -- see section \ref{sec:Stability}. We have for our fields $w$, $\psi$ and $\delta\varphi$:
\begin{align}
    \mathcal L^{(2)}_{\text{kin}}
    =-\frac{1}{2} U^{\mu\nu}\partial_\mu w\partial_\nu w
    -\frac{1}{2} T^{\mu\nu}\partial_\mu\psi\partial_\nu\psi
    \nonumber\\
    -\frac{1}{2} S^{\mu\nu}\partial_\mu\delta\varphi\partial_\nu\delta\varphi,
\end{align}
where $U^{\mu\nu}$, $T^{\mu\nu}$ and $S^{\mu\nu}$ can be deduced from equation \eqref{eq:QuadraticActionsFull}. We then find effective metrics
\begin{align}
    U_{\mu\nu}^{-1}&=\frac{2\sqrt{\approxcon}(\ell+2)(\ell-1)}{\ell(\ell+1)x^2}g_{\mu\nu},
    \nonumber\\\nonumber\\
    T^{-1}_{\mu\nu}&=\frac{\ell(\ell+1)\left((\ell+2)(\ell-1)x+3\right)^2}{2\sqrt{\approxcon}(\ell+2)(\ell-1)x^2(x-1)^2}g_{\mu\nu},
    \nonumber\\\nonumber\\
    S_{\mu\nu}^{-1}&=\frac{\approxcon(4x-3)}{24x^2}\begin{pmatrix}
        \zeta\sqrt{4x-3}&\frac{\lambda x}{\sqrt{\approxcon}(x-1)}\\
        \frac{\lambda x}{\sqrt{\approxcon}(x-1)}&-\zeta\frac{x^2(6x-5)(2x-3)}{\approxcon(x-1)^2(4x-3)^{3/2}}
    \end{pmatrix},
\end{align}
where we have used $\zeta=\pm1$ to indicate the $\Xi_\zeta$ branch. Since $U_{\mu\nu}^{-1}$ and $T_{\mu\nu}^{-1}$ are proportional to the the regular metric (with positive proportionality constants), any vector that is time- or spacelike with respect to $g_{\mu\nu}$ will also be time- or spacelike with respect to both of these effective metrics. We can therefore easily see that the perturbations they describe are stable. For $S_{\mu\nu}^{-1}$ on the other hand the situation is more complicated. We firstly have the hyperbolicity condition:
\begin{align}
    \det S_{\mu\nu}^{-1}=-\frac{\approxcon(4x-3)}{48x^2}<0.
\end{align}
This is satisfied in the regime where our solutions are valid ($x>3/4$). Since a positive determinant would have led to complex propagation speeds, this would correspond to a gradient instability. Therefore, the perturbations are gradient stable. Subsequently, we need to identify the lightcone. This is defined by
\begin{align}
    S_{\mu\nu}^{-1}dx^\mu dx^\nu=S_{00}^{-1}dy^2+2S_{01}^{-1}dydx+S_{11}^{-1}dx^2<0.
\end{align}
The lightcone will therefore be bounded by null-curves which must have derivatives
\begin{align}
    \frac{dy}{dx}&=-\frac{S_{01}^{-1}}{S_{00}^{-1}}\pm\frac{1}{S_{00}^{-1}}\sqrt{-\det S_{\mu\nu}^{-1}},
    \nonumber\\
    \Rightarrow \left(\frac{dy}{dx}\right)_{S\pm}&=-\zeta\frac{\lambda x}{\sqrt{\approxcon}(x-1)\sqrt{4x-3}}\pm\frac{2\sqrt{3}x}{\sqrt{\approxcon}(4x-3)}.
\end{align}
Then for $S_{00}^{-1}<0$, the lightcone will be the region
\begin{align}
    \frac{dy}{dx}<\left(\frac{dy}{dx}\right)_{S-}\lor\frac{dy}{dx}>\left(\frac{dy}{dx}\right)_{S+},
\end{align}
while for $S_{00}^{-1}>0$, it will be
\begin{align}
    \left(\frac{dy}{dx}\right)_{S-}<\frac{dy}{dx}<\left(\frac{dy}{dx}\right)_{S+}.
\end{align}
Similarly, for the regular metric we have null curves
\begin{align}
    \left(\frac{dy}{dx}\right)_{g\pm}=\pm\frac{x}{\sqrt{\approxcon}|x-1|},
\end{align}
and a lightcone
\begin{align}
    \frac{dy}{dx}<\left(\frac{dy}{dx}\right)_{g-}\lor\frac{dy}{dx}>\left(\frac{dy}{dx}\right)_{g+},
    &&x>1,
    \nonumber\\\nonumber\\
    \left(\frac{dy}{dx}\right)_{g-}<\frac{dy}{dx}<\left(\frac{dy}{dx}\right)_{g+},
    &&x<1.
\end{align}
For ghost stability, these lightcones need to have common exteriors as well as common interiors over the whole range of radii we are studying. Using the expressions above, we find that the interiors of the lightcones have no overlap for $x<3(2-\sqrt{3})$ when using the $\Xi_-$ branch -- see figure \ref{fig:LightconeStructure}. Similarly, when using the $\Xi_+$ branch, the interiors do not overlap for $x>3(2+\sqrt{3})$. We therefore find that both branches display ghost instabilities, with the instability appearing behind the black hole horizon for the $\Xi_-$ branch and outside it for the $\Xi_+$ branch.

\section{Time dependence in the scalar ansatz} \label{sec:AppendixScalarAnsatzTimeDependence}

In section \ref{sec:TimeDependentEoMs} we mentioned that using a simpler ansatz for the scalar field with only linear time dependence as in the static case is inconsistent with the resulting equations of motion if we aim to find solutions with a nonzero time dependence resembling the GR Schwarzschild solution. We will demonstrate this here in more detail.

Firstly note that this simpler ansatz is a special case of the ansatz with general time dependence \eqref{eq:TimeDependentAnsatz} so we can describe it using the full time-dependent equations of motion \eqref{eq:EoMsTimeDependentFull} with the additional constraint:
\begin{align}
    \label{eq:TimeIndependentAnsatzConstraint}
    \partial_t\Psi(t,r)=0\Rightarrow\partial_t\frac{\Xi(t,r)}{h(t,r)}=0.
\end{align}
Since we require that the solutions resemble the static ones, we take $\alpha_{BH}\lesssim1$ -- see section \ref{sec:QuasiStationaryBlackHoles}. We can therefore use the approximate solutions \eqref{eq:ShortRangeSolutions} and \eqref{eq:XiShortRangeQuasiStationarySolution} to write out this constraint in terms of $\dot{r_s}$ and $\dot{\beta}$. In addition one could allow for a small time dependence in $\alpha_{BH}$ to consider the most general scenario. We must choose $\dot{r}_s$, $\dot{\beta}$ and $\dot{\alpha}_{BH}$ such that \eqref{eq:TimeIndependentAnsatzConstraint} is satisfied for all $r$ (in the range of validity of the approximate solutions). Since $r_s$, $\beta$ and $\alpha_{BH}$ cannot have any radial dependence -- see \eqref{eq:ShortRangeSimpleEOM} and \eqref{eq:ConservedCurrent} -- this is nontrivial. In particular when expanding about $r=r_s$ we find
\begin{align}
    \partial_t \frac{\Xi(t,r)}{h(t,r)}=\pm \text{sign}(\alpha_1)\sqrt{\frac{r_s(t)}{\beta(t)}}\dot{r}_s(t)\left(r-r_s(t)\right)^{-2}+\dots.
\end{align}
\eqref{eq:TimeIndependentAnsatzConstraint} can therefore only be satisfied arbitrarily close to the black hole horizon for $\dot{r}_s=0$, which from equation \eqref{eq:RadiusEvolution} gives $\alpha_{BH}=0$. In this case there is no accretion or time dependence at all and we recover the static case. Therefore if we require a nonzero time dependence and solutions resembling the GR Schwarzschild one, we must use the more general ansatz for the scalar \eqref{eq:TimeDependentAnsatz}. In other words, the extra time dependence in the metric functions necessarily induces an additional time dependence in the scalar field on top of the purely linear time dependence that is already present in the static case. However, recall from section \ref{sec:BlackHoleSolutionsQuasiStationary} that this time dependence is suppressed in the quasi-stationary solutions we find.

\section{Stability on a quasi-stationary background} \label{sec:AppendixStabilityQuasiStationary}

In this appendix we show the detailed analysis of stability of perturbations about the quasi-stationary background described in section \ref{sec:QuasiStationaryBlackHoles}. We focus on scalar perturbations about the $\Xi_+$ branch since this is the branch required to connect to the correct cosmological asymptotes. From the quadratic action \eqref{eq:ScalarQuadraticActionQuasiStationary} we find the effective metric governing the propagation of these perturbations
\begin{align}
    S_{\mu\nu}^{-1}
    &=\frac{\approxcon(4x-3)}{24x^2}
    \begin{pmatrix}
        \frac{\sqrt{4x-3}}{\xi(x)}
        &
        \frac{\lambda x}{\sqrt{\approxcon}(x-1)\xi^2(x)}
        \\
        \bullet
        &
        -\frac{x^2(6x-5)(2x-3)+12x(x-1)^3\frac{\alpha_4\sqrt{\approxcon}}{\alpha_1}}{\approxcon(x-1)^2(4x-3)^{3/2}\xi^3(x)}
    \end{pmatrix},
    \nonumber\\\nonumber\\
    \xi(x)&=\sqrt{1+\frac{\alpha_4\sqrt{\approxcon}}{\alpha_1}\frac{x-1}{x}}.
\end{align}
The hyperbolicity condition gives
\begin{align}
    \det{S_{\mu\nu}^{-1}}=-\frac{\approxcon(4x-3)}{48x^2\xi^2}<0,
\end{align}
which is satisfied for $x>3/4$. The corresponding lightcone will be bounded by the null curves
\begin{align}
\left(\frac{dy}{dx}\right)_{S\pm}
    =-\frac{\lambda x}{\sqrt{\approxcon}(x-1)\sqrt{4x-3}\xi}
    \pm\frac{2\sqrt{3}x}{\sqrt{\approxcon}(4x-3)}.
\end{align}

Using the expressions above we find that the lightcones do not have overlapping interiors for any $x$ if $\alpha_4\sqrt{\approxcon}/\alpha_1>3$. However, for $\alpha_4\sqrt{\approxcon}/\alpha_1\leq3$, the interiors of the lightcones overlap up to
\begin{align}
    x_{\text{max}}=\frac{(3+2\sqrt{3})\sqrt{3-\frac{\alpha_4\sqrt{\approxcon}}{\alpha_1}}+\sqrt{3}\left(3+2\sqrt{3}-\frac{\alpha_4\sqrt{\approxcon}}{\alpha_1}\right)}{2\left(1+\frac{\alpha_4\sqrt{\approxcon}}{\alpha_1}\right)}.
\end{align}
By choosing $\alpha_4$ appropriately we can find a solution that is stable against scalar perturbations over the whole range where the approximate solutions are accurate, as described in section \ref{sec:QuasiStationaryBlackHoles}.

By transforming our coordinates as in \eqref{eq:ExplicitlyStableCoordinatesQuasiStationary} we find the quadratic Hamiltonian \eqref{eq:ExplicitlyStableHamiltonianQuasiStationary}. Here we give the full forms of the $a_i$ coefficients.
\begin{widetext}
\begin{align}
    a_0=\frac{x^2}{4\approxcon\xi(x-1)^2(4x-3)^{3/2}}
    &\Biggl[
    4x-3
    +2\xi\sqrt{4x-3}\left(-3-2\sqrt{3}+2\left(2+\sqrt{3}\right)x\right)
    \Biggr.
    \nonumber\\
    &\Biggl.
    +\xi^2\left(3\left(-9+4\sqrt{3}\right)+4x\left(12-7\sqrt{3}+\left(-5+4\sqrt{3}\right)x\right)\right)
    \Biggr]
    \nonumber\\\nonumber\\
    a_1 = -\frac{x^2}{4\approxcon\xi(x-1)^2(4x-3)^{3/2}}
    &\Biggl[
    4x-3
    +2\xi\sqrt{4x-3}\left(3+2\sqrt{3}-2\left(2+\sqrt{3}\right)x\right)
    \Biggr.
    \nonumber\\
    &\Biggl.
    +\xi^2\left(3\left(-9+4\sqrt{3}\right)+4x\left(12-7\sqrt{3}+\left(-5+4\sqrt{3}\right)x\right)\right)
    \Biggr]
    \nonumber\\\nonumber\\
    a_2 = -\frac{\ell(\ell+1)}{\xi x^2(4x-3)^{3/2}}
    &\left[
    4\left(1+\frac{\alpha_4\sqrt{\approxcon}}{\alpha_1}\right)x^2
    -6\left(1+\frac{\alpha_4\sqrt{\approxcon}}{\alpha_1}\right)x
    +3\frac{\alpha_4\sqrt{\approxcon}}{\alpha_1}
    \right]
\end{align}
\end{widetext}

\section{Suppression of time dependence} \label{sec:AppendixTimeDependenceSuppression}
The numerical solution found in section \ref{sec:FullStableQuasiStationaryBlackHoleSolution} solves the quasi-stationary equations of motion \eqref{eq:EoMsQuasiStationary}. In this appendix we will explicitly show that this solution is also a good approximation to the full time-dependent equations of motion and that we were therefore justified in our assumption that the time dependence is suppressed across all $x$.

Firstly, we assume a characteristic time-scale $\Gamma^{-1}$ over which the functions $f$, $h$ and $\Psi$ vary. This time-scale can differ at different $x$. We then replace all temporal derivatives in the equations of motion as $\partial_y\rightarrow \Gamma r_s$ and substitute in our numerical solutions. $\Gamma$ must have some minimal value to solve the full time-dependent $(tr)$ equation (the first equation in \eqref{eq:EoMsTimeDependentFull}). We substitute this minimal value into the other equations and use the accuracy requirement \eqref{eq:AccuracyTest} to test whether the time-dependent terms are sufficiently suppressed in these equations for a range of $x$ values.

In figure \ref{fig:TimeDependenceSuppression} we plot the suppression factor -- see appendix \ref{sec:AppendixShortRangeLimitAccuracy}:
\begin{align}
    \label{eq:SuppresionFactor}
    \frac{\left|\bar{\mathcal{E}}\right|}{\sum_{i,n}\left|\overline{\frac{\partial E}{\partial f_i^{(n)}}f_i^{(n)}}\right|}.
\end{align}
The figure shows that in the vicinity of the black hole the time dependence is suppressed by a factor of $\sim r_s/r_c$ as expected. This holds up to the cosmological horizon, after which the suppression becomes stronger.
\begin{figure}
    \centering
    \includegraphics[width=\linewidth]{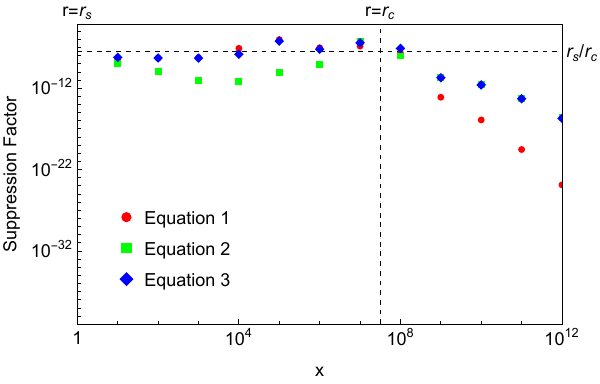}
    \caption{The factor by which the time-dependent equations are suppressed as defined in \eqref{eq:SuppresionFactor} for the different quasi-stationary equations \eqref{eq:EoMsQuasiStationary}. In the vicinity of the black hole the time dependence is suppressed by a factor of $\sim r_s/r_c$ as expected. This holds up to the cosmological horizon, after which the suppression becomes even stronger.}
    \label{fig:TimeDependenceSuppression}
\end{figure}

\bibliographystyle{utphys}
\bibliography{BH_CubicGal}

@article{Charmousis:2025xug,
    author = "Charmousis, Christos and Iteanu, Simon and Langlois, David and Noui, Karim",
    title = "{Axial perturbations of black holes with primary scalar hair}",
    eprint = "2503.22348",
    archivePrefix = "arXiv",
    primaryClass = "gr-qc",
    reportNumber = "CERN-TH-2025-058",
    doi = "10.1088/1475-7516/2025/05/102",
    journal = "JCAP",
    volume = "05",
    pages = "102",
    year = "2025"
}

@article{Sawicki:2024ryt,
    author = "Sawicki, Ignacy and Trenkler, Georg and Vikman, Alexander",
    title = "{Causality and stability from acoustic geometry}",
    eprint = "2412.21169",
    archivePrefix = "arXiv",
    primaryClass = "gr-qc",
    doi = "10.1007/JHEP10(2025)227",
    journal = "JHEP",
    volume = "10",
    pages = "227",
    year = "2025"
}

@article{Lara:2025hqh,
    author = "Lara, Guillermo and Trenkler, Georg and Trombetta, Leonardo G.",
    title = "{Primary black-hole scalar charges and kinetic screening in $K$-essence-Gauss-Bonnet gravity}",
    eprint = "2512.23683",
    archivePrefix = "arXiv",
    primaryClass = "gr-qc",
    month = "12",
    year = "2025"
}

@article{Kobayashi:2009wr,
    author = "Kobayashi, Tsutomu and Tashiro, Hiroyuki and Suzuki, Daichi",
    title = "{Evolution of linear cosmological perturbations and its observational implications in Galileon-type modified gravity}",
    eprint = "0912.4641",
    archivePrefix = "arXiv",
    primaryClass = "astro-ph.CO",
    reportNumber = "WU-AP-305-09",
    doi = "10.1103/PhysRevD.81.063513",
    journal = "Phys. Rev. D",
    volume = "81",
    pages = "063513",
    year = "2010"
}

@article{Smulders:2026bya,
    author = "Smulders, Laurens and Noller, Johannes and Sirera, Sergi",
    title = "{Testing Dark Energy with Black Hole Ringdown}",
    eprint = "2603.23634",
    archivePrefix = "arXiv",
    primaryClass = "gr-qc",
    month = "3",
    year = "2026"
}

@article{Bekenstein:1995un,
    author = "Bekenstein, J. D.",
    title = "{Novel {\textquoteleft}{\textquoteleft}no-scalar-hair{\textquoteright}{\textquoteright} theorem for black holes}",
    doi = "10.1103/PhysRevD.51.R6608",
    journal = "Phys. Rev. D",
    volume = "51",
    number = "12",
    pages = "R6608",
    year = "1995"
}

@article{Tattersall:QNMsHorn,
    author = "Tattersall, Oliver J. and Ferreira, Pedro G.",
    title = "{Quasinormal modes of black holes in Horndeski gravity}",
    eprint = "1804.08950",
    archivePrefix = "arXiv",
    primaryClass = "gr-qc",
    doi = "10.1103/PhysRevD.97.104047",
    journal = "Phys. Rev. D",
    volume = "97",
    number = "10",
    pages = "104047",
    year = "2018"
}

@article{Sotiriou:2013qea,
    author = "Sotiriou, Thomas P. and Zhou, Shuang-Yong",
    title = "{Black hole hair in generalized scalar-tensor gravity}",
    eprint = "1312.3622",
    archivePrefix = "arXiv",
    primaryClass = "gr-qc",
    doi = "10.1103/PhysRevLett.112.251102",
    journal = "Phys. Rev. Lett.",
    volume = "112",
    pages = "251102",
    year = "2014"
}

@article{Noller:2019chl,
    author = "Noller, Johannes and Santoni, Luca and Trincherini, Enrico and Trombetta, Leonardo G.",
    title = "{Black Hole Ringdown as a Probe for Dark Energy}",
    eprint = "1911.11671",
    archivePrefix = "arXiv",
    primaryClass = "gr-qc",
    doi = "10.1103/PhysRevD.101.084049",
    journal = "Phys. Rev. D",
    volume = "101",
    pages = "084049",
    year = "2020"
}

@book{Maggiore:2018sht,
    author = "Maggiore, Michele",
    title = "{Gravitational Waves. Vol. 2: Astrophysics and Cosmology}",
    isbn = "978-0-19-857089-9",
    publisher = "Oxford University Press",
    month = "3",
    year = "2018"
}

@misc{xAct,
 author = {"Mart\'in-Garc\'ia, Jose M"},
 title = {{xAct}},
 url = {http://www.xact.es/},
 howpublished = {}
}

@article{deRham:2014wfa,
      author         = "de Rham, Claudia and Ribeiro, Raquel H.",
      title          = "{Riding on irrelevant operators}",
      journal        = "JCAP",
      volume         = "1411",
      year           = "2014",
      number         = "11",
      pages          = "016",
      doi            = "10.1088/1475-7516/2014/11/016",
      eprint         = "1405.5213",
      archivePrefix  = "arXiv",
      primaryClass   = "hep-th",
      SLACcitation   = "%%CITATION = ARXIV:1405.5213;%%"
}

@article{Nicolis:2004qq,
      author         = "Nicolis, Alberto and Rattazzi, Riccardo",
      title          = "{Classical and quantum consistency of the DGP model}",
      journal        = "JHEP",
      volume         = "06",
      year           = "2004",
      pages          = "059",
      doi            = "10.1088/1126-6708/2004/06/059",
      eprint         = "hep-th/0404159",
      archivePrefix  = "arXiv",
      primaryClass   = "hep-th",
      reportNumber   = "IFT-UAM-CSIC-04-13, CERN-PH-TH-2004-063",
      SLACcitation   = "%%CITATION = HEP-TH/0404159;%%"
}

@article{Horndeski:1974wa,
      author         = "Horndeski, Gregory Walter",
      title          = "{Second-order scalar-tensor field equations in a
                        four-dimensional space}",
      journal        = "Int. J. Theor. Phys.",
      volume         = "10",
      year           = "1974",
      pages          = "363-384",
      doi            = "10.1007/BF01807638",
      SLACcitation   = "%%CITATION = IJTPB,10,363;%%"
}

@article{deRham:2010eu,
    author = "de Rham, Claudia and Tolley, Andrew J.",
    title = "{DBI and the Galileon reunited}",
    eprint = "1003.5917",
    archivePrefix = "arXiv",
    primaryClass = "hep-th",
    doi = "10.1088/1475-7516/2010/05/015",
    journal = "JCAP",
    volume = "05",
    pages = "015",
    year = "2010"
}

@article{Burrage:2010cu,
    author = "Burrage, Clare and de Rham, Claudia and Seery, David and Tolley, Andrew J.",
    title = "{Galileon inflation}",
    eprint = "1009.2497",
    archivePrefix = "arXiv",
    primaryClass = "hep-th",
    reportNumber = "DESY-10-132",
    doi = "10.1088/1475-7516/2011/01/014",
    journal = "JCAP",
    volume = "01",
    pages = "014",
    year = "2011"
}

@article{Clifton:2011jh,
      author         = "Clifton, Timothy and Ferreira, Pedro G. and Padilla,
                        Antonio and Skordis, Constantinos",
      title          = "{Modified Gravity and Cosmology}",
      journal        = "Phys. Rept.",
      volume         = "513",
      year           = "2012",
      pages          = "1-189",
      doi            = "10.1016/j.physrep.2012.01.001",
      eprint         = "1106.2476",
      archivePrefix  = "arXiv",
      primaryClass   = "astro-ph.CO",
      SLACcitation   = "%%CITATION = ARXIV:1106.2476;%%"
}

@article{Traykova:2021hbr,
    author = "Traykova, Dina and Bellini, Emilio and Ferreira, Pedro G. and Garc\'\i{}a-Garc\'\i{}a, Carlos and Noller, Johannes and Zumalac\'arregui, Miguel",
    title = "{Theoretical priors in scalar-tensor cosmologies: Shift-symmetric Horndeski models}",
    eprint = "2103.11195",
    archivePrefix = "arXiv",
    primaryClass = "astro-ph.CO",
    doi = "10.1103/PhysRevD.104.083502",
    journal = "Phys. Rev. D",
    volume = "104",
    number = "8",
    pages = "083502",
    year = "2021"
}

@article{Kobayashi:2012kh,
    author = "Kobayashi, Tsutomu and Motohashi, Hayato and Suyama, Teruaki",
    title = "{Black hole perturbation in the most general scalar-tensor theory with second-order field equations I: the odd-parity sector}",
    eprint = "1202.4893",
    archivePrefix = "arXiv",
    primaryClass = "gr-qc",
    reportNumber = "RESCEU-3-12, KUNS-2385",
    doi = "10.1103/PhysRevD.85.084025",
    journal = "Phys. Rev. D",
    volume = "85",
    pages = "084025",
    year = "2012",
    note = "[Erratum: Phys.Rev.D 96, 109903 (2017)]"
}

@article{Luty:2003vm,
	author         = "Luty, Markus A. and Porrati, Massimo and Rattazzi,
	Riccardo",
	title          = "{Strong interactions and stability in the DGP model}",
	journal        = "JHEP",
	volume         = "09",
	year           = "2003",
	pages          = "029",
	doi            = "10.1088/1126-6708/2003/09/029",
	eprint         = "hep-th/0303116",
	archivePrefix  = "arXiv",
	primaryClass   = "hep-th",
	reportNumber   = "CERN-TH-2003-044, UMD-PP-03-041",
	SLACcitation   = "%%CITATION = HEP-TH/0303116;%%"
}

@article{Nicolis:2008in,
	author         = "Nicolis, Alberto and Rattazzi, Riccardo and Trincherini,
	Enrico",
	title          = "{The Galileon as a local modification of gravity}",
	journal        = "Phys. Rev.",
	volume         = "D79",
	year           = "2009",
	pages          = "064036",
	doi            = "10.1103/PhysRevD.79.064036",
	eprint         = "0811.2197",
	archivePrefix  = "arXiv",
	primaryClass   = "hep-th",
	SLACcitation   = "%%CITATION = ARXIV:0811.2197;%%"
}

@article{Pirtskhalava:2015nla,
	author         = "Pirtskhalava, David and Santoni, Luca and Trincherini,
	Enrico and Vernizzi, Filippo",
	title          = "{Weakly Broken Galileon Symmetry}",
	journal        = "JCAP",
	volume         = "1509",
	year           = "2015",
	number         = "09",
	pages          = "007",
	doi            = "10.1088/1475-7516/2015/09/007",
	eprint         = "1505.00007",
	archivePrefix  = "arXiv",
	primaryClass   = "hep-th",
	SLACcitation   = "%%CITATION = ARXIV:1505.00007;%%"
}

@article{Deffayet:2011gz,
	author         = "Deffayet, C. and Gao, Xian and Steer, D. A. and
	Zahariade, G.",
	title          = "{From k-essence to generalised Galileons}",
	journal        = "Phys. Rev.",
	volume         = "D84",
	year           = "2011",
	pages          = "064039",
	doi            = "10.1103/PhysRevD.84.064039",
	eprint         = "1103.3260",
	archivePrefix  = "arXiv",
	primaryClass   = "hep-th",
	SLACcitation   = "%%CITATION = ARXIV:1103.3260;%%"
}

@article{DeFelice:2011bh,
	author         = "De Felice, Antonio and Tsujikawa, Shinji",
	title          = "{Conditions for the cosmological viability of the most
	general scalar-tensor theories and their applications to
	extended Galileon dark energy models}",
	journal        = "JCAP",
	volume         = "1202",
	year           = "2012",
	pages          = "007",
	doi            = "10.1088/1475-7516/2012/02/007",
	eprint         = "1110.3878",
	archivePrefix  = "arXiv",
	primaryClass   = "gr-qc",
	SLACcitation   = "%%CITATION = ARXIV:1110.3878;%%"
}

@article{Barreira:2013jma,
    author = "Barreira, Alexandre and Li, Baojiu and Sanchez, Ariel and Baugh, Carlton M. and Pascoli, Silvia",
    title = "{Parameter space in Galileon gravity models}",
    eprint = "1302.6241",
    archivePrefix = "arXiv",
    primaryClass = "astro-ph.CO",
    doi = "10.1103/PhysRevD.87.103511",
    journal = "Phys. Rev. D",
    volume = "87",
    pages = "103511",
    year = "2013"
}

@article{Babichev:2025ric,
    author = "Babichev, Eugeny and Esposito-Far{\`e}se, Gilles and Sawicki, Ignacy and Trombetta, Leonardo G.",
    title = "{Large black-hole scalar charges induced by cosmology in Horndeski theories}",
    eprint = "2504.07882",
    archivePrefix = "arXiv",
    primaryClass = "gr-qc",
    doi = "10.1103/bv8z-qbfj",
    journal = "Phys. Rev. D",
    volume = "112",
    number = "2",
    pages = "024043",
    year = "2025"
}

@article{Goon:2020myi,
    author = "Goon, Garrett and Melville, Scott and Noller, Johannes",
    title = "{Quantum corrections to generic branes: DBI, NLSM, and more}",
    eprint = "2010.05913",
    archivePrefix = "arXiv",
    primaryClass = "hep-th",
    doi = "10.1007/JHEP01(2021)159",
    journal = "JHEP",
    volume = "01",
    pages = "159",
    year = "2021"
}

@article{Heisenberg:2020cyi,
    author = "Heisenberg, Lavinia and Noller, Johannes and Zosso, Jann",
    title = "{Horndeski under the quantum loupe}",
    eprint = "2004.11655",
    archivePrefix = "arXiv",
    primaryClass = "hep-th",
    doi = "10.1088/1475-7516/2020/10/010",
    journal = "JCAP",
    volume = "10",
    pages = "010",
    year = "2020"
}

@article{Noller:2018eht,
    author = "Noller, Johannes and Nicola, Andrina",
    title = "{Radiative stability and observational constraints on dark energy and modified gravity}",
    eprint = "1811.03082",
    archivePrefix = "arXiv",
    primaryClass = "astro-ph.CO",
    doi = "10.1103/PhysRevD.102.104045",
    journal = "Phys. Rev. D",
    volume = "102",
    number = "10",
    pages = "104045",
    year = "2020"
}

@article{deRham:2012ew,
    author = "de Rham, Claudia and Gabadadze, Gregory and Heisenberg, Lavinia and Pirtskhalava, David",
    title = "{Nonrenormalization and naturalness in a class of scalar-tensor theories}",
    eprint = "1212.4128",
    archivePrefix = "arXiv",
    primaryClass = "hep-th",
    reportNumber = "UCSD-PTH-12-20, NYU-TH-11-11-12",
    doi = "10.1103/PhysRevD.87.085017",
    journal = "Phys. Rev. D",
    volume = "87",
    number = "8",
    pages = "085017",
    year = "2013"
}

@article{Heisenberg:2019wjv,
    author = "Heisenberg, Lavinia and Steinwachs, Christian F.",
    title = "{Geometrized quantum Galileons}",
    eprint = "1909.07111",
    archivePrefix = "arXiv",
    primaryClass = "hep-th",
    reportNumber = "FR-PHENO-2019-015",
    doi = "10.1088/1475-7516/2020/02/031",
    journal = "JCAP",
    volume = "02",
    pages = "031",
    year = "2020"
}

@article{Saltas:2016nkg,
    author = "Saltas, Ippocratis D. and Vitagliano, Vincenzo",
    title = "{Covariantly Quantum Galileon}",
    eprint = "1611.07984",
    archivePrefix = "arXiv",
    primaryClass = "hep-th",
    doi = "10.1103/PhysRevD.95.105002",
    journal = "Phys. Rev. D",
    volume = "95",
    number = "10",
    pages = "105002",
    year = "2017"
}

@article{Goon:2016ihr,
    author = "Goon, Garrett and Hinterbichler, Kurt and Joyce, Austin and Trodden, Mark",
    title = "{Aspects of Galileon Non-Renormalization}",
    eprint = "1606.02295",
    archivePrefix = "arXiv",
    primaryClass = "hep-th",
    doi = "10.1007/JHEP11(2016)100",
    journal = "JHEP",
    volume = "11",
    pages = "100",
    year = "2016"
}

@article{ReggeWheeler,
    author = "Regge, Tullio and Wheeler, John A.",
    title = "{Stability of a Schwarzschild singularity}",
    doi = "10.1103/PhysRev.108.1063",
    journal = "Phys. Rev.",
    volume = "108",
    pages = "1063--1069",
    year = "1957"
}

@article{Kobayashi:2014wsa,
    author = "Kobayashi, Tsutomu and Motohashi, Hayato and Suyama, Teruaki",
    title = "{Black hole perturbation in the most general scalar-tensor theory with second-order field equations II: the even-parity sector}",
    eprint = "1402.6740",
    archivePrefix = "arXiv",
    primaryClass = "gr-qc",
    reportNumber = "RESCEU-5-14, RUP-14-4, RESCEU-5/14",
    doi = "10.1103/PhysRevD.89.084042",
    journal = "Phys. Rev. D",
    volume = "89",
    number = "8",
    pages = "084042",
    year = "2014"
}

@article{Franciolini:2018uyq,
    author = "Franciolini, Gabriele and Hui, Lam and Penco, Riccardo and Santoni, Luca and Trincherini, Enrico",
    title = "{Effective Field Theory of Black Hole Quasinormal Modes in Scalar-Tensor Theories}",
    eprint = "1810.07706",
    archivePrefix = "arXiv",
    primaryClass = "hep-th",
    doi = "10.1007/JHEP02(2019)127",
    journal = "JHEP",
    volume = "02",
    pages = "127",
    year = "2019"
}

@article{deRham:2019gha,
    author = "de Rham, Claudia and Zhang, Jun",
    title = "{Perturbations of stealth black holes in degenerate higher-order scalar-tensor theories}",
    eprint = "1907.00699",
    archivePrefix = "arXiv",
    primaryClass = "hep-th",
    reportNumber = "Imperial/TP/2019/CdR/03",
    doi = "10.1103/PhysRevD.100.124023",
    journal = "Phys. Rev. D",
    volume = "100",
    number = "12",
    pages = "124023",
    year = "2019"
}

@article{Khoury:2020aya,
    author = "Khoury, Justin and Trodden, Mark and Wong, Sam S. C.",
    title = "{Existence and instability of hairy black holes in shift-symmetric Horndeski theories}",
    eprint = "2007.01320",
    archivePrefix = "arXiv",
    primaryClass = "astro-ph.CO",
    doi = "10.1088/1475-7516/2020/11/044",
    journal = "JCAP",
    volume = "11",
    pages = "044",
    year = "2020"
}

@article{Hui:2021cpm,
    author = "Hui, Lam and Podo, Alessandro and Santoni, Luca and Trincherini, Enrico",
    title = "{Effective Field Theory for the perturbations of a slowly rotating black hole}",
    eprint = "2111.02072",
    archivePrefix = "arXiv",
    primaryClass = "hep-th",
    doi = "10.1007/JHEP12(2021)183",
    journal = "JHEP",
    volume = "12",
    pages = "183",
    year = "2021"
}

@software{ringdown-calculations,
author = {Sirera, Sergi},
title = {{ringdown-calculations}},
url = {https://github.com/sergisl/ringdown-calculations},
howpublished = {}
}

@software{stable-black-holes-cosmological-hair,
author = {Smulders, Laurens},
title = {{stable-black-holes-cosmological-hair}},
url = {https://github.com/laurenssmulders/stable_black_holes_cosmological_hair},
howpublished = {}
}

@article{Bakopoulos:2023fmv,
    author = "Bakopoulos, Athanasios and Charmousis, Christos and Kanti, Panagiota and Lecoeur, Nicolas and Nakas, Theodoros",
    title = "{Black holes with primary scalar hair}",
    eprint = "2310.11919",
    archivePrefix = "arXiv",
    primaryClass = "gr-qc",
    month = "10",
    year = "2023"
}

@article{Koyama:2015vza,
    author = "Koyama, Kazuya",
    title = "{Cosmological Tests of Modified Gravity}",
    eprint = "1504.04623",
    archivePrefix = "arXiv",
    primaryClass = "astro-ph.CO",
    doi = "10.1088/0034-4885/79/4/046902",
    journal = "Rept. Prog. Phys.",
    volume = "79",
    number = "4",
    pages = "046902",
    year = "2016"
}

@article{Babichev:2013cya,
    author = "Babichev, Eugeny and Charmousis, Christos",
    title = "{Dressing a black hole with a time-dependent Galileon}",
    eprint = "1312.3204",
    archivePrefix = "arXiv",
    primaryClass = "gr-qc",
    reportNumber = "LPT-ORSAY-13-105",
    doi = "10.1007/JHEP08(2014)106",
    journal = "JHEP",
    volume = "08",
    pages = "106",
    year = "2014"
}

@article{Kobayashi:2014eva,
    author = "Kobayashi, Tsutomu and Tanahashi, Norihiro",
    title = "{Exact black hole solutions in shift symmetric scalar\textendash{}tensor theories}",
    eprint = "1403.4364",
    archivePrefix = "arXiv",
    primaryClass = "gr-qc",
    reportNumber = "RUP-14-7, IPMU14-0057",
    doi = "10.1093/ptep/ptu096",
    journal = "PTEP",
    volume = "2014",
    pages = "073E02",
    year = "2014"
}

@article{Babichev:2016fbg,
    author = "Babichev, Eugeny and Charmousis, Christos and Leh\'ebel, Antoine and Moskalets, Tetiana",
    title = "{Black holes in a cubic Galileon universe}",
    eprint = "1605.07438",
    archivePrefix = "arXiv",
    primaryClass = "gr-qc",
    reportNumber = "LPT-ORSAY-16-42",
    doi = "10.1088/1475-7516/2016/09/011",
    journal = "JCAP",
    volume = "09",
    pages = "011",
    year = "2016"
}

@article{Babichev:2018uiw,
    author = "Babichev, Eugeny and Charmousis, Christos and Esposito-Far\`ese, Gilles and Leh\'ebel, Antoine",
    title = "{Hamiltonian unboundedness vs stability with an application to Horndeski theory}",
    eprint = "1803.11444",
    archivePrefix = "arXiv",
    primaryClass = "gr-qc",
    reportNumber = "LPT-ORSAY-18-57, LPT-Orsay-18-57",
    doi = "10.1103/PhysRevD.98.104050",
    journal = "Phys. Rev. D",
    volume = "98",
    number = "10",
    pages = "104050",
    year = "2018"
}

@article{Babichev:2016kdt,
    author = "Babichev, Eugeny and Esposito-Farese, Gilles",
    title = "{Cosmological self-tuning and local solutions in generalized Horndeski theories}",
    eprint = "1609.09798",
    archivePrefix = "arXiv",
    primaryClass = "gr-qc",
    reportNumber = "LPT-ORSAY-16-87",
    doi = "10.1103/PhysRevD.95.024020",
    journal = "Phys. Rev. D",
    volume = "95",
    number = "2",
    pages = "024020",
    year = "2017"
}

@article{BenAchour:2018dap,
    author = "Ben Achour, Jibril and Liu, Hongguang",
    title = "{Hairy Schwarzschild-(A)dS black hole solutions in degenerate higher order scalar-tensor theories beyond shift symmetry}",
    eprint = "1811.05369",
    archivePrefix = "arXiv",
    primaryClass = "gr-qc",
    doi = "10.1103/PhysRevD.99.064042",
    journal = "Phys. Rev. D",
    volume = "99",
    number = "6",
    pages = "064042",
    year = "2019"
}

@article{Motohashi:2019sen,
    author = "Motohashi, Hayato and Minamitsuji, Masato",
    title = "{Exact black hole solutions in shift-symmetric quadratic degenerate higher-order scalar-tensor theories}",
    eprint = "1901.04658",
    archivePrefix = "arXiv",
    primaryClass = "gr-qc",
    reportNumber = "YITP-19-01",
    doi = "10.1103/PhysRevD.99.064040",
    journal = "Phys. Rev. D",
    volume = "99",
    number = "6",
    pages = "064040",
    year = "2019"
}

@article{Takahashi:2020hso,
    author = "Takahashi, Kazufumi and Motohashi, Hayato",
    title = "{General Relativity solutions with stealth scalar hair in quadratic higher-order scalar-tensor theories}",
    eprint = "2004.03883",
    archivePrefix = "arXiv",
    primaryClass = "gr-qc",
    reportNumber = "KOBE-COSMO-20-05, YITP-20-35",
    doi = "10.1088/1475-7516/2020/06/034",
    journal = "JCAP",
    volume = "06",
    pages = "034",
    year = "2020"
}

@article{Babichev:2012re,
    author = "Babichev, Eugeny and Esposito-Far\`ese, Gilles",
    title = "{Time-Dependent Spherically Symmetric Covariant Galileons}",
    eprint = "1212.1394",
    archivePrefix = "arXiv",
    primaryClass = "gr-qc",
    doi = "10.1103/PhysRevD.87.044032",
    journal = "Phys. Rev. D",
    volume = "87",
    pages = "044032",
    year = "2013"
}

@article{Mukohyama:2022enj,
    author = "Mukohyama, Shinji and Yingcharoenrat, Vicharit",
    title = "{Effective field theory of black hole perturbations with timelike scalar profile: formulation}",
    eprint = "2204.00228",
    archivePrefix = "arXiv",
    primaryClass = "hep-th",
    reportNumber = "YITP-22-27, IPMU22-0012",
    doi = "10.1088/1475-7516/2022/09/010",
    journal = "JCAP",
    volume = "09",
    pages = "010",
    year = "2022"
}

@article{Khoury:2022zor,
    author = "Khoury, Justin and Noumi, Toshifumi and Trodden, Mark and Wong, Sam S. C.",
    title = "{Stability of hairy black holes in shift-symmetric scalar-tensor theories via the effective field theory approach}",
    eprint = "2208.02823",
    archivePrefix = "arXiv",
    primaryClass = "hep-th",
    doi = "10.1088/1475-7516/2023/04/035",
    journal = "JCAP",
    volume = "04",
    pages = "035",
    year = "2023"
}

@article{Mukohyama:2022skk,
    author = "Mukohyama, Shinji and Takahashi, Kazufumi and Yingcharoenrat, Vicharit",
    title = "{Generalized Regge-Wheeler equation from Effective Field Theory of black hole perturbations with a timelike scalar profile}",
    eprint = "2208.02943",
    archivePrefix = "arXiv",
    primaryClass = "gr-qc",
    reportNumber = "YITP-22-78, IPMU22-0039",
    doi = "10.1088/1475-7516/2022/10/050",
    journal = "JCAP",
    volume = "10",
    pages = "050",
    year = "2022"
}

@article{Mukohyama:2023xyf,
    author = "Mukohyama, Shinji and Takahashi, Kazufumi and Tomikawa, Keitaro and Yingcharoenrat, Vicharit",
    title = "{Quasinormal modes from EFT of black hole perturbations with timelike scalar profile}",
    eprint = "2304.14304",
    archivePrefix = "arXiv",
    primaryClass = "gr-qc",
    reportNumber = "YITP-23-45, IPMU23-0007, RUP-23-8",
    doi = "10.1088/1475-7516/2023/07/050",
    journal = "JCAP",
    volume = "07",
    pages = "050",
    year = "2023"
}

@article{Barura:2024uog,
    author = "Barura, Chams Gharib Ali and Kobayashi, Hajime and Mukohyama, Shinji and Oshita, Naritaka and Takahashi, Kazufumi and Yingcharoenrat, Vicharit",
    title = "{Tidal Love Numbers from EFT of Black Hole Perturbations with Timelike Scalar Profile}",
    eprint = "2405.10813",
    archivePrefix = "arXiv",
    primaryClass = "gr-qc",
    reportNumber = "YITP-24-62, IPMU24-0018, RIKEN-iTHEMS-Report-24",
    month = "5",
    year = "2024"
}

@article{Motohashi:2019ymr,
    author = "Motohashi, Hayato and Mukohyama, Shinji",
    title = "{Weakly-coupled stealth solution in scordatura degenerate theory}",
    eprint = "1912.00378",
    archivePrefix = "arXiv",
    primaryClass = "gr-qc",
    reportNumber = "YITP-19-98, IPMU19-0156",
    doi = "10.1088/1475-7516/2020/01/030",
    journal = "JCAP",
    volume = "01",
    pages = "030",
    year = "2020"
}

@article{DeFelice:2022xvq,
    author = "De Felice, Antonio and Mukohyama, Shinji and Takahashi, Kazufumi",
    title = "{Avoidance of Strong Coupling in General Relativity Solutions with a Timelike Scalar Profile in a Class of Ghost-Free Scalar-Tensor Theories}",
    eprint = "2204.02032",
    archivePrefix = "arXiv",
    primaryClass = "gr-qc",
    reportNumber = "YITP-22-25, IPMU22-0014",
    doi = "10.1103/PhysRevLett.129.031103",
    journal = "Phys. Rev. Lett.",
    volume = "129",
    number = "3",
    pages = "031103",
    year = "2022"
}

@article{Charmousis:2019vnf,
    author = "Charmousis, Christos and Crisostomi, Marco and Gregory, Ruth and Stergioulas, Nikolaos",
    title = "{Rotating Black Holes in Higher Order Gravity}",
    eprint = "1903.05519",
    archivePrefix = "arXiv",
    primaryClass = "hep-th",
    doi = "10.1103/PhysRevD.100.084020",
    journal = "Phys. Rev. D",
    volume = "100",
    number = "8",
    pages = "084020",
    year = "2019"
}

@article{Takahashi:2021bml,
    author = "Takahashi, Kazufumi and Motohashi, Hayato",
    title = "{Black hole perturbations in DHOST theories: master variables, gradient instability, and strong coupling}",
    eprint = "2106.07128",
    archivePrefix = "arXiv",
    primaryClass = "gr-qc",
    reportNumber = "YITP-21-38",
    doi = "10.1088/1475-7516/2021/08/013",
    journal = "JCAP",
    volume = "08",
    pages = "013",
    year = "2021"
}

@article{Hui:2012qt,
	title        = {{No-Hair Theorem for the Galileon}},
	author       = {Hui, Lam and Nicolis, Alberto},
	year         = 2013,
	journal      = {Phys. Rev. Lett.},
	volume       = 110,
	pages        = 241104,
	doi          = {10.1103/PhysRevLett.110.241104},
	archiveprefix = {arXiv},
	eprint       = {1202.1296},
	primaryclass = {hep-th}
}

@article{Zerilli:1970wzz,
    author = "Zerilli, F. J.",
    title = "{Gravitational field of a particle falling in a schwarzschild geometry analyzed in tensor harmonics}",
    doi = "10.1103/PhysRevD.2.2141",
    journal = "Phys. Rev. D",
    volume = "2",
    pages = "2141--2160",
    year = "1970"
}

@incollection{maggiore_12_2018,
    title = {12 {Black}-hole perturbation theory},
    isbn = {978-0-19-857089-9},
    url = {https://doi.org/10.1093/oso/9780198570899.003.0003},
    urldate = {2024-09-25},
    booktitle = {Gravitational {Waves}: {Volume} 2: {Astrophysics} and {Cosmology}},
    publisher = {Oxford University Press},
    author = {Maggiore, Michele},
    editor = {Maggiore, Michele},
    month = mar,
    year = {2018},
    doi = {10.1093/oso/9780198570899.003.0003},
    pages = {0},
}

@article{Emond:2019myx,
    author = "Emond, William T. and Leh{\'e}bel, Antoine and Saffin, Paul M.",
    title = "{Black holes in self-tuning cubic Horndeski cosmology}",
    eprint = "1912.09199",
    archivePrefix = "arXiv",
    primaryClass = "gr-qc",
    doi = "10.1103/PhysRevD.101.084008",
    journal = "Phys. Rev. D",
    volume = "101",
    number = "8",
    pages = "084008",
    year = "2020"
}

@book{doi:10.1137/1.9781611971224,
author = {Brenan, K. E. and Campbell, S. L. and Petzold, L. R.},
title = {Numerical Solution of Initial-Value Problems in Differential-Algebraic Equations},
publisher = {Society for Industrial and Applied Mathematics},
year = {1995},
doi = {10.1137/1.9781611971224},
address = {},
edition   = {}
}

@book{hairer_numerical_1989,
	address = {Berlin, Heidelberg},
	series = {Lecture {Notes} in {Mathematics}},
	title = {The {Numerical} {Solution} of {Differential}-{Algebraic} {Systems} by {Runge}-{Kutta} {Methods}},
	volume = {1409},
	copyright = {http://www.springer.com/tdm},
	isbn = {978-3-540-51860-0 978-3-540-46832-5},
	url = {http://link.springer.com/10.1007/BFb0093947},
	urldate = {2025-10-28},
	publisher = {Springer},
	author = {Hairer, Ernst and Roche, Michel and Lubich, Christian},
	year = {1989},
	doi = {10.1007/BFb0093947},
}

@article{Wald:1979lth,
    author = "Wald, Robert M.",
    title = "{Note on the stability of the Schwarzschild metric}",
    doi = "10.1063/1.524181",
    journal = "J. Math. Phys.",
    volume = "20",
    number = "6",
    pages = "1056",
    year = "1979"
}

@article{Copeland:2006wr,
    author = "Copeland, Edmund J. and Sami, M. and Tsujikawa, Shinji",
    title = "{Dynamics of dark energy}",
    eprint = "hep-th/0603057",
    archivePrefix = "arXiv",
    doi = "10.1142/S021827180600942X",
    journal = "Int. J. Mod. Phys. D",
    volume = "15",
    pages = "1753--1936",
    year = "2006"
}

@article{Kobayashi:2019hrl,
    author = "Kobayashi, Tsutomu",
    title = "{Horndeski theory and beyond: a review}",
    eprint = "1901.07183",
    archivePrefix = "arXiv",
    primaryClass = "gr-qc",
    reportNumber = "RUP-19-3",
    doi = "10.1088/1361-6633/ab2429",
    journal = "Rept. Prog. Phys.",
    volume = "82",
    number = "8",
    pages = "086901",
    year = "2019"
}

@article{Joyce:2014kja,
    author = "Joyce, Austin and Jain, Bhuvnesh and Khoury, Justin and Trodden, Mark",
    title = "{Beyond the Cosmological Standard Model}",
    eprint = "1407.0059",
    archivePrefix = "arXiv",
    primaryClass = "astro-ph.CO",
    doi = "10.1016/j.physrep.2014.12.002",
    journal = "Phys. Rept.",
    volume = "568",
    pages = "1--98",
    year = "2015"
}

@inproceedings{Bekenstein:1996pn,
    author = "Bekenstein, Jacob D.",
    title = "{Black hole hair: 25 - years after}",
    booktitle = "{2nd International Sakharov Conference on Physics}",
    eprint = "gr-qc/9605059",
    archivePrefix = "arXiv",
    pages = "216--219",
    month = "5",
    year = "1996"
}

@article{Yazadjiev:2025ezx,
    author = "Yazadjiev, Stoytcho S. and Doneva, Daniela D.",
    title = "{No-hair theorems in general relativity and scalar{\textendash}tensor theories}",
    eprint = "2505.01038",
    archivePrefix = "arXiv",
    primaryClass = "gr-qc",
    doi = "10.1142/S0218271825300046",
    journal = "Int. J. Mod. Phys. D",
    volume = "34",
    number = "09",
    pages = "2530004",
    year = "2025"
}

@article{Maselli:2015yva,
    author = "Maselli, Andrea and Silva, Hector O. and Minamitsuji, Masato and Berti, Emanuele",
    title = "{Slowly rotating black hole solutions in Horndeski gravity}",
    eprint = "1508.03044",
    archivePrefix = "arXiv",
    primaryClass = "gr-qc",
    doi = "10.1103/PhysRevD.92.104049",
    journal = "Phys. Rev. D",
    volume = "92",
    number = "10",
    pages = "104049",
    year = "2015"
}

@article{Babichev:2013vji,
    author = "Babichev, E. O. and Dokuchaev, V. I. and Eroshenko, Yu N.",
    title = "{Black holes in the presence of dark energy}",
    eprint = "1406.0841",
    archivePrefix = "arXiv",
    primaryClass = "gr-qc",
    reportNumber = "LPT-ORSAY-14-33",
    doi = "10.3367/UFNe.0183.201312a.1257",
    journal = "Phys. Usp.",
    volume = "56",
    pages = "1155--1175",
    year = "2013"
}

@article{Sirera:2023pbs,
    author = "Sirera, Sergi and Noller, Johannes",
    title = "{Testing the speed of gravity with black hole ringdowns}",
    eprint = "2301.10272",
    archivePrefix = "arXiv",
    primaryClass = "gr-qc",
    doi = "10.1103/PhysRevD.107.124054",
    journal = "Phys. Rev. D",
    volume = "107",
    number = "12",
    pages = "124054",
    year = "2023"
}

@article{Sirera:2024ghv,
    author = "Sirera, Sergi and Noller, Johannes",
    title = "{Stability and quasinormal modes for black holes with time-dependent scalar hair}",
    eprint = "2408.01720",
    archivePrefix = "arXiv",
    primaryClass = "gr-qc",
    doi = "10.1103/PhysRevD.111.044067",
    journal = "Phys. Rev. D",
    volume = "111",
    number = "4",
    pages = "044067",
    year = "2025"
}

@article{Kobayashi:2025evr,
    author = "Kobayashi, Hajime and Mukohyama, Shinji and Noller, Johannes and Sirera, Sergi and Takahashi, Kazufumi and Yingcharoenrat, Vicharit",
    title = "{Inverting no-hair theorems: How requiring general relativity solutions restricts scalar-tensor theories}",
    eprint = "2503.05651",
    archivePrefix = "arXiv",
    primaryClass = "gr-qc",
    reportNumber = "YITP-25-34, IPMU25-0010",
    doi = "10.1103/75sw-4f7n",
    journal = "Phys. Rev. D",
    volume = "111",
    number = "12",
    pages = "124022",
    year = "2025"
}

@article{Ruffini:1971bza,
    author = "Ruffini, Remo and Wheeler, John A.",
    title = "{Introducing the black hole}",
    doi = "10.1063/1.3022513",
    journal = "Phys. Today",
    volume = "24",
    number = "1",
    pages = "30",
    year = "1971"
}

@article{Israel:1967wq,
    author = "Israel, Werner",
    title = "{Event horizons in static vacuum space-times}",
    doi = "10.1103/PhysRev.164.1776",
    journal = "Phys. Rev.",
    volume = "164",
    pages = "1776--1779",
    year = "1967"
}

@article{Robinson:1975bv,
    author = "Robinson, D. C.",
    title = "{Uniqueness of the Kerr black hole}",
    doi = "10.1103/PhysRevLett.34.905",
    journal = "Phys. Rev. Lett.",
    volume = "34",
    pages = "905--906",
    year = "1975"
}

@article{Carter:1971zc,
    author = "Carter, B.",
    title = "{Axisymmetric Black Hole Has Only Two Degrees of Freedom}",
    doi = "10.1103/PhysRevLett.26.331",
    journal = "Phys. Rev. Lett.",
    volume = "26",
    pages = "331--333",
    year = "1971"
}

@article{Bekenstein:1971hc,
    author = "Bekenstein, Jacob D.",
    title = "{Nonexistence of baryon number for static black holes}",
    doi = "10.1103/PhysRevD.5.1239",
    journal = "Phys. Rev. D",
    volume = "5",
    pages = "1239--1246",
    year = "1972"
}

@article{Bekenstein:1972ky,
    author = "Bekenstein, J. D.",
    title = "{Nonexistence of baryon number for black holes. ii}",
    doi = "10.1103/PhysRevD.5.2403",
    journal = "Phys. Rev. D",
    volume = "5",
    pages = "2403--2412",
    year = "1972"
}

@article{Jacobson:1999vr,
    author = "Jacobson, Ted",
    title = "{Primordial black hole evolution in tensor scalar cosmology}",
    eprint = "astro-ph/9905303",
    archivePrefix = "arXiv",
    reportNumber = "NSF-ITP-99-35",
    doi = "10.1103/PhysRevLett.83.2699",
    journal = "Phys. Rev. Lett.",
    volume = "83",
    pages = "2699--2702",
    year = "1999"
}

@article{DeFelice:2010pv,
    author = "De Felice, Antonio and Tsujikawa, Shinji",
    title = "{Cosmology of a covariant Galileon field}",
    eprint = "1007.2700",
    archivePrefix = "arXiv",
    primaryClass = "astro-ph.CO",
    doi = "10.1103/PhysRevLett.105.111301",
    journal = "Phys. Rev. Lett.",
    volume = "105",
    pages = "111301",
    year = "2010"
}

@article{Silva:2009km,
    author = "Silva, Fabio P and Koyama, Kazuya",
    title = "{Self-Accelerating Universe in Galileon Cosmology}",
    eprint = "0909.4538",
    archivePrefix = "arXiv",
    primaryClass = "astro-ph.CO",
    doi = "10.1103/PhysRevD.80.121301",
    journal = "Phys. Rev. D",
    volume = "80",
    pages = "121301",
    year = "2009"
}

@article{Babichev:2013usa,
    author = "Babichev, Eugeny and Deffayet, C{\'e}dric",
    title = "{An introduction to the Vainshtein mechanism}",
    eprint = "1304.7240",
    archivePrefix = "arXiv",
    primaryClass = "gr-qc",
    reportNumber = "LPT-ORSAY-13-45",
    doi = "10.1088/0264-9381/30/18/184001",
    journal = "Class. Quant. Grav.",
    volume = "30",
    pages = "184001",
    year = "2013"
}

@article{Burrage:2010rs,
    author = "Burrage, Clare and Seery, David",
    title = "{Revisiting fifth forces in the Galileon model}",
    eprint = "1005.1927",
    archivePrefix = "arXiv",
    primaryClass = "astro-ph.CO",
    reportNumber = "DESY-10-062",
    doi = "10.1088/1475-7516/2010/08/011",
    journal = "JCAP",
    volume = "08",
    pages = "011",
    year = "2010"
}

@article{Burrage:2011bt,
    author = "Burrage, Clare and de Rham, Claudia and Heisenberg, Lavinia",
    title = "{de Sitter Galileon}",
    eprint = "1104.0155",
    archivePrefix = "arXiv",
    primaryClass = "hep-th",
    doi = "10.1088/1475-7516/2011/05/025",
    journal = "JCAP",
    volume = "05",
    pages = "025",
    year = "2011"
}

@article{Mukohyama:2024pqe,
    author = "Mukohyama, Shinji and Seraille, Emeric and Takahashi, Kazufumi and Yingcharoenrat, Vicharit",
    title = "{Bridging dark energy and black holes with EFT: frame transformation and gravitational wave speed}",
    eprint = "2407.15123",
    archivePrefix = "arXiv",
    primaryClass = "gr-qc",
    reportNumber = "YITP-24-86, IPMU24-0030",
    doi = "10.1088/1475-7516/2025/01/085",
    journal = "JCAP",
    volume = "01",
    pages = "085",
    year = "2025"
}

@article{Mukohyama:2025owu,
    author = "Mukohyama, Shinji and Takahashi, Kazufumi and Tomikawa, Keitaro and Yingcharoenrat, Vicharit",
    title = "{Spherical black hole perturbations in EFT of scalar-tensor gravity with timelike scalar profile}",
    eprint = "2503.00520",
    archivePrefix = "arXiv",
    primaryClass = "gr-qc",
    reportNumber = "YITP-25-30, IPMU25-0009",
    doi = "10.1088/1475-7516/2025/05/084",
    journal = "JCAP",
    volume = "05",
    pages = "084",
    year = "2025"
}

@article{Mukohyama:2025jzk,
    author = "Mukohyama, Shinji and Seraille, Emeric and Takahashi, Kazufumi and Yingcharoenrat, Vicharit",
    title = "{Effective field theory of perturbations on arbitrary black hole backgrounds with spacelike scalar profile}",
    eprint = "2507.02066",
    archivePrefix = "arXiv",
    primaryClass = "gr-qc",
    reportNumber = "YITP-25-98, IPMU25-0036",
    doi = "10.1007/JHEP10(2025)128",
    journal = "JHEP",
    volume = "10",
    pages = "128",
    year = "2025"
}

@article{Capuano:2023yyh,
    author = "Capuano, Lodovico and Santoni, Luca and Barausse, Enrico",
    title = "{Black hole hairs in scalar-tensor gravity and the lack thereof}",
    eprint = "2304.12750",
    archivePrefix = "arXiv",
    primaryClass = "gr-qc",
    reportNumber = "DE13253",
    doi = "10.1103/PhysRevD.108.064058",
    journal = "Phys. Rev. D",
    volume = "108",
    number = "6",
    pages = "064058",
    year = "2023"
}

@article{Tattersall:2019nmh,
    author = "Tattersall, Oliver J.",
    title = "{Quasi-Normal Modes of Hairy Scalar Tensor Black Holes: Odd Parity}",
    eprint = "1911.07593",
    archivePrefix = "arXiv",
    primaryClass = "gr-qc",
    doi = "10.1088/1361-6382/ab839b",
    journal = "Class. Quant. Grav.",
    volume = "37",
    number = "11",
    pages = "115007",
    year = "2020"
}

@article{DeFelice:2022qaz,
    author = "De Felice, Antonio and Mukohyama, Shinji and Takahashi, Kazufumi",
    title = "{Approximately stealth black hole in higher-order scalar-tensor theories}",
    eprint = "2212.13031",
    archivePrefix = "arXiv",
    primaryClass = "gr-qc",
    reportNumber = "YITP-22-161, IPMU22-0071",
    doi = "10.1088/1475-7516/2023/03/050",
    journal = "JCAP",
    volume = "03",
    pages = "050",
    year = "2023"
}

@article{Rosen:2017dvn,
    author = "Rosen, Rachel A",
    title = "{Non-Singular Black Holes in Massive Gravity: Time-Dependent Solutions}",
    eprint = "1702.06543",
    archivePrefix = "arXiv",
    primaryClass = "hep-th",
    doi = "10.1007/JHEP10(2017)206",
    journal = "JHEP",
    volume = "10",
    pages = "206",
    year = "2017"
}

\end{document}